\documentclass[final,3p,times]{elsarticle}
\usepackage{amssymb,amsmath,amsthm,mathtools}
\usepackage{lineno,natbib}
\usepackage{xcolor,soul}
\usepackage[colorlinks]{hyperref}
\usepackage{cases}
\usepackage{enumitem}
\usepackage{algorithm}
\usepackage{algpseudocode}
\usepackage{standalone}
\usepackage{booktabs}
\usepackage[OT1]{fontenc} 
\usepackage{xpatch}

\newtheorem{theorem}{Theorem}[section]

\newtheorem{remark}[theorem]{Remark}

\graphicspath{ {./Images/} }
\usepackage{graphicx}
\usepackage{tikz}
\usepackage{caption}
\usepackage{subcaption}
\usepackage{multirow}

\begin{document}
	\makeatletter
	\def\ps@pprintTitle{%
		\let\@oddhead\@empty
		\let\@evenhead\@empty
		\let\@oddfoot\@empty
		\let\@evenfoot\@oddfoot}
	\makeatother
	
	\begin{frontmatter}
		
		\title{Integrating Physics of the Problem into Data-Driven Methods  to Enhance  Elastic Full-Waveform Inversion with Uncertainty Quantification }

		\author[main]{Vahid Negahdari\corref{cor}}
		\ead{vahid\_negahdari@outlook.com}

		\author[main]{Seyed Reza Moghadasi}
		\ead{moghadasi@sharif.edu}
		
		\author[main]{Mohammad Reza Razvan}
		\ead{razvan@sharif.edu}

		\cortext[cor]{Corresponding author}	
		\address[main]{Department of Mathematical Sciences, Sharif University of Technology, Tehran, 11155-9415, Iran}

		\begin{abstract}
			
			Full-Waveform Inversion (FWI) is a nonlinear iterative seismic imaging technique that, by reducing the misfit between recorded and predicted seismic waveforms, can produce detailed estimates of subsurface geophysical properties. Nevertheless, the strong nonlinearity of FWI  can trap the optimization in local minima. This issue arises due to factors such as improper initial values, the absence of low frequencies in the measurements, noise, and other related considerations. To address this challenge and with the advent of advanced machine-learning techniques, data-driven methods, such as deep learning, have attracted significantly increasing attention in the geophysical community.  Furthermore,  the elastic wave equation should be included in FWI to  represent elastic effects accurately. The intersection of data-driven techniques and elastic scattering theories presents opportunities and challenges. In this paper, by using the knowledge of elastic scattering (physics of the problem) and integrating it with machine learning techniques, we propose methods for the solution of time-harmonic FWI to enhance accuracy compared to pure data-driven and physics-based approaches. Moreover, to address uncertainty quantification, by modifying the structure of the Variational Autoencoder, we introduce a probabilistic deep learning method based on the physics of the problem that enables us to  explore the uncertainties of the solution. According to the limited availability of datasets in this field and to assess the performance and accuracy of the proposed methods, we create a comprehensive dataset  close to reality and conduct a comparative analysis of the presented approaches to it. 

		\end{abstract}
		
	\end{frontmatter}

	\section{Introduction}		
	
	Inverse wave scattering is the problem of determining the intrinsic characteristics of an object based on waves scattered from it under the illumination of an incident wave. Generally, scattering problems include various types, such as acoustic wave scattering, electromagnetic wave scattering, and elastic wave scattering. In particular, in this study,  we concentrate on elastic wave scattering, which is more complex due to the presence of shear and compressional waves. The inverse elastic scattering problems have attracted much attention lately, driven by applications in nondestructive testing \cite{gupta, rose}, medical imaging \cite{ormachea, li4}, and seismic imaging in geophysics exploration \cite{li5,  wang3}. 
	
	Motivated by the applications mentioned above, extensive studies have been conducted on the elastic inverse problems, and many mathematical results are available in this area, with a particular focus on the existence and uniqueness of the solution \cite{bao, hahner, hahner2, ammar}.
	
	In the context of the inverse elastic scattering problem within seismic imaging, main focus is on the determination of properties related to either a penetrable inhomogeneous medium (specifically, the reconstruction of Lame parameters, density, etc.), which is the subject of this paper, or a bounded impenetrable obstacle. Seismic imaging can capture detailed subsurface images through the analysis of seismic data. This data is acquired by deploying seismic sources either on or near the surface and recording the responses using an extensive array of receivers positioned on the specific locations like the surface.
	
	Seismic imaging inversion methods can be solved in two main ways  based on how complex the forward modeling is. Although travel-time inversion \cite{tarantola2} is a simpler method with a linear forward operator, it yields results with lower accuracy and resolution. Conversely, the forward operator in Full-Waveform Inversion (FWI) techniques  is nonlinear and computationally expensive. However,  it offers better solutions by simulating wave propagation in the subsurface. FWI \cite{tarantola, virieux2} is an optimization problem with PDE constraints. By iteratively minimizing the difference between predicted and observed seismic waveforms in a least-squares sense, FWI uses the PDE solver to find optimal values for seismic waves, density, and Lame parameters in the acoustic or elastic approximation \cite{lailly, xue}. 
	
	From a physics-based perspective, numerous local optimization methods have been developed over recent decades to minimize the FWI objective function \cite{choi,  yan, pratt, he, brossier, epanomeritakis, metivier}.  Applying local minimization methods to the conventional FWI objective function remains challenging because they are usually hindered by multiple local minima, resulting from its high nonlinearity and ill-posedness \cite{virieux}. This implies that when employing local minimization techniques, an appropriate initial model or low-frequency data are needed \cite{bozdaug,  zhang}. Regularization methods \cite{haung},  wavefield reconstruction inversion \cite{van}, adaptive waveform inversion \cite{warner},  prior information-based methods \cite{zhang2, ma2, ma3}, multiscale inversion approaches \cite{bunks, tran}, preconditioning methods \cite{tang},  tomographic FWI inversion \cite{biondi}, model reduction \cite{barnier}, model reparameterization \cite{guitton} and extended modeling methods  \cite{fu, fu2} are some of the strategies developed to address the multiple minima issue.  Despite all these solutions, utilizing physics-based methods is challenging as there are still significant weaknesses,  such as high computational cost and high dependence on the initial value. Data-driven approaches have been suggested as a way to improve these limitations.
	
	Data-driven methods, particularly deep learning techniques, have shown a remarkable capacity to approximate nonlinear mapping functions between different data domains, including  natural language processing \cite{bharadiya}, speech recognition \cite{mehrish}, and many other fields \cite{sharifani, jiang}, especially for inverse problems like image reconstruction \cite{szczyk}, image super-resolution \cite{wang,  dong},  etc. These cutting-edge innovations offer new possibilities for Inverse Scattering Problem (ISP) and consequently seismic imaging. Deep learning approaches provide a forthcoming imaging technique by avoiding bottlenecks mentioned earlier \cite{ovcharenko, alregib}. One main topic in this research area is to build a direct inverse mapping from observations to subsurface structure by training neural networks on paired data of seismic waveforms and desired parameters of the model  \cite{kazei, li2, wu5,  yang3}. Actually, instead of solving the wave equation, this method considers FWI as a simple deep-learning problem, the same as image recognition.   Some research efforts have been conducted in this field, such as VAE, GAN, Unet, and RNN, as detailed in the comprehensive review articles {\cite{yu, review2}}. The techniques as mentioned above, can achieve exceptional efficiency  once fully trained. However, the generalizability of data-driven approaches is constrained by the size and diversity of the training set. Furthermore, the availability of data is relatively limited in this field due to the high acquisition cost. According to the high commercial value of this data, a very limited amount of it is accessible to the public. Moreover, since these models learn from data directly without incorporating domain knowledge, they are susceptible to generating unreasonable or unrealistic predictions that may not adapt to the physical mechanism. As a result, these observations have prompted researchers to integrate physics-based and data-driven methods.
	
	The simultaneous use of the physics-based method and data-driven approach is an idea that has recently gained attention, which is called the Physics-Guided Machine Learning (PGML) approach. Numerous studies have shown how crucial prior knowledge is to enhancing the effectiveness of the physics-based FWI optimization technique. The existing parameter  distribution \cite{zhu, ren, he3}, the low-frequency component of the measured signals \cite{hu,jin2}, assistance in resolving the cycle skipping problem, and other things are examples of this prior knowledge. On the other hand, Physics-Informed Neural Networks (PINNs)  \cite{raissi, rasht}, a subset of PGML,  become an effective tool for resolving inverse problems, mainly when insufficient scatter measurements are available, and no complete system knowledge is known.  PINNs attempt to incorporate physics into the trained network, commonly achieved by augmenting the loss function with a residual term derived from the governing equation. This approach effectively regularizes the solution space during training. Specifically, incorporating the wave equation in data-driven methods has been associated with some factors, such as improved network accuracy, enhanced generalization, increased robustness, a reduction in the amount of required training data, and, in certain instances, the elimination of the need for labeled data {\cite{review-PINN, PINN3, Gan-PINN}}. Like PINN, several works have been completed by applying deep learning and PDEs simultaneously to seismic FWI {\cite{sun, jin, dhara, FWI-ML1, FWI-ML2}}. Estimating the parameters of interest using a learned representation of the underlying distribution of the physical parameters is an impressive method of integrating physics into deep learning. More specifically, a widely used network architecture utilized to describe data distribution is the generative adversarial network (GAN). This category is illustrated by several recent studies \cite{mosser, richardson}. Other methods within PGML include Data Augmentation \cite{feng, rojas}, Unrolling-Based Techniques, which help to reduce computation efficiently \cite{wu3}, and the Learning Loss Function, which chooses an appropriate loss function \cite{zhang4, sun3}. A comprehensive overview of works in this field is provided in the article \cite{review}.
	
	%%%%%%%%%%%%%%%%%%%%%%%%%%%%%%%%%%%%%%%%%%%%%%%%%%%%%%%%%%%%%%%%%%%%%%%%%%%%%%%%%%
	%%%%%%%%%%%%%%%%%%%%%%%%%%%%%%%%%%%%%%%%%%%%%%%%%%%%%%%%%%%%%%%%%%%%%%%%%%%%%%%%%%

	Probabilistic methods within seismic inversion, including FWI and elastic FWI, allow for a more thorough exploration of the model space and quantifying uncertainties associated with the inversion results. These methods are typically implemented through Bayesian frameworks, where the posterior distribution of subsurface parameters is estimated, taking into account prior knowledge and observational data. Such methods can reveal the most likely model and provide a quantitative understanding of uncertainty within the inversion results. Among the relevant papers in this field, we can mention \cite{Bayesian1, Bayesian2, Bayesian3, Bayesian4}.	In practice, these probabilistic methods often utilize Monte Carlo sampling, such as Hamiltonian Monte Carlo {\cite{Hamilton} and Markov Chain Monte Carlo (MCMC)  {\cite{MCMC1, MCMC2, MCMC3} techniques, to explore the high-dimensional model space. Though these techniques approximate the posterior distribution, they are computationally intensive due to the requirement for numerous forward simulations. Some recent advancements aim to mitigate these computational demands by using techniques like low-rank approximations of the Hessian matrix and quasi-Newton methods, which reduce the number of necessary samples and computational resources  \cite{low1, low2}. Advanced probabilistic methods, including ensemble-based approaches \cite{Ensemble} and variational inference  \cite{Inference1, Inference2}, have also been explored, even though challenges related to computational cost and convergence remain significant hurdles. Additionally, machine learning and deep learning tools are increasingly being integrated with probabilistic FWI and elastic FWI, leveraging data-driven insights and reducing computational costs  \cite{deep1, deep2, zhu}. This evolving landscape not only enhances the accuracy of inversion results but also broadens the application of these methods in complex geological settings, making them valuable tools in the fields of seismic exploration and reservoir characterization. Furthermore, PINNs have attracted significant attention in probabilistic seismic inversion due to their ability to incorporate physics directly into the neural network training process. PINNs' probabilistic formulation aids in capturing the inherent uncertainty in seismic inversion, yielding solutions that are both physically consistent and statistically reliable \cite{PINN1, PINN2}}.

		%%%%%%%%%%%%%%%%%%%%%%%%%%%%%%%%%%%%%%%%%%%%%%%%%%%%%%%%%%%%%%%%%%%%%%%%%%%%%%%%%%
		%%%%%%%%%%%%%%%%%%%%%%%%%%%%%%%%%%%%%%%%%%%%%%%%%%%%%%%%%%%%%%%%%%%%%%%%%%%%%%%%%%
		
		This paper introduces approaches to solving time-harmonic elastic FWI through data-driven techniques and their combination within physical constraints. We subsequently conduct a comparative analysis of these methods  with pure data-driven and physics-based methods. A consequence of the inherent ill-conditioning of the problem is that small noises in the data can lead to significant errors, which is evident in the proposed methods. Some tricks are employed to mitigate this issue. Another challenge arises from the fact that numerous plausible reconstructions may fit with the noisy measurements. This variety of solutions recommends employing techniques that recover multiple reconstructions. A probabilistic framework in characterizing solutions enhances the reliability of interpretation for reconstructions, particularly when the problem becomes more ill-posed. In this paper, we introduce a novel probabilistic method based on Variational Autoencoder (VAE) for uncertainty quantification.  In all the methods to be discussed, we make the assumption that the Lame parameters are constant and known. Additionally, we assume that the receivers have collected information (displacement field) on the surface of the field. From this collected data, we want to acquire the density of the field in the entire desired area.
		
		Section \ref{Function setting} contains the physics of the problem and the relations between variables. To use data-driven techniques to solve the problem, we create a dataset. In section \ref{Dataset}, we aim to produce more complex and comprehensive models by making artificial fields. These artificial fields can be good simulators for natural fields. So, we use them and their related displacement field (derived from the governed elastic wave equation) in the learning process. In the first method \ref{First},  which is a pure data-driven approach, we
		employ a Convolutional Neural Network (CNN) to solve the inverse problem. This CNN takes surface information as input and produces the density of the subsurface as output. In the second method \ref{Second}, which is a PGML approach, we utilize CNN to derive the displacement field over the entire area from the information on the surface. Subsequently, using different techniques and leveraging the physics of the problem, we determine the desired density. The third method \ref{Third}, which is a probabilistic deep learning inverse approach, inspired by the structure of VAE, we aim to introduce a method that not only generates solutions but also can provide a probabilistic solution for uncertainty quantification.
		Additionally, this method can  utilize the physics of the problem to enhance and refine its results. In the fourth method {\ref{Fourth}}, two traditional physics-based techniques are implemented to facilitate comparison with the approaches presented in this paper.

		\section{Governing equations and main results}\label{Function setting}
		\label{Section2}
		The behavior of time-harmonic waves as they propagate within an isotropic inhomogeneous medium characterized by a specified density $\rho$ and Lame constants $\lambda$ and $\mu$  is described by the Navier equation:
		\begin{equation}\label{base}
			\mu \Delta \boldsymbol{u} + (\lambda + \mu) \nabla \, div\, \boldsymbol{u} + \rho \omega^2 \boldsymbol{u} = 0 \qquad \text{in} \,\, \mathbb{R}^2
		\end{equation}
		where $ \boldsymbol{u} :\, \mathbb{R}^2 \to \mathbb{C}^2 $ and $\omega $ denote the displacement field and the circular frequency, respectively. In this paper, it is assumed that the density $\rho \equiv 1$  outside of the region $\Omega \subseteq \mathbb{R}^2 $. The assumptions on the Lame constants $ (\mu > 0, 2\mu +\lambda > 0) $ ensure that the Navier system is strongly elliptic.
		
		The total field $\boldsymbol{u}$ comprises two distinct components, the incident field $\boldsymbol{u}^{inc} $ and the scattered field $\boldsymbol{u}^{sc}$. $\boldsymbol{u}^{inc} $  is a plane wave of the following form:  
		\begin{equation}\label{incident}
			\boldsymbol{u}^{inc}(\boldsymbol{x}) = \alpha \boldsymbol{u}^{inc}_p(\boldsymbol{x}) + \beta \boldsymbol{u}^{inc}_s(\boldsymbol{x}) \quad  \alpha , \beta \in \mathbb{R}	
		\end{equation}
		Here, $ \boldsymbol{u}^{inc}_p := \boldsymbol{\theta} \exp (i\kappa_p\,\boldsymbol{x}\cdot\boldsymbol{\theta}) $ is the compressional plane wave, and $ \boldsymbol{u}^{inc}_s := \boldsymbol{\theta}^{\perp} \exp (i\kappa_s\,\boldsymbol{x}\cdot\boldsymbol{\theta}) $ denotes the shear plane wave, where $\boldsymbol{\theta} \in \mathbb{S}^1 := \Big\{\boldsymbol{y} \in \mathbb{R}^2 : |\boldsymbol{y}| = 1\Big\}$ represents the unit propagation direction, $\boldsymbol{\theta}^{\perp} \in  \mathbb{S}^1 $ is a vector orthogonal to $\boldsymbol{\theta} $, and $\kappa_p := \dfrac{\omega}{\sqrt{\lambda + 2\mu}} $ and $\kappa_s := \dfrac{\omega}{\sqrt{\mu}} $ 
		denote the compressional and shear wave numbers, respectively. It can be verified that the incident field $\boldsymbol{u}^{inc}(\boldsymbol{x}) $ satisfies:
		$$  \mu \Delta \boldsymbol{u}^{inc} + (\lambda + \mu) \nabla \, div\, \boldsymbol{u}^{inc} + \omega^2 \boldsymbol{u}^{inc} = 0 \qquad \text{in} \,\, \mathbb{R}^2$$
		
		Based on the Helmholtz decomposition \cite{lai}, the scattered field $ \boldsymbol{u}^{sc} $ in $\Omega^{c}$ can be decomposed into the compressional wave component $ \boldsymbol{u}^{sc}_p $ and the shear wave component  $ \boldsymbol{u}^{sc}_s $:
		$$ \boldsymbol{u}^{sc}_p = -\dfrac{1}{\kappa_{p}^2} \nabla \, div\,\boldsymbol{u}^{sc} \quad , \quad \boldsymbol{u}^{sc}_s = \dfrac{1}{\kappa_{s}^2} curl \, \boldsymbol{curl} (\boldsymbol{u}^{sc})   \quad , \quad \boldsymbol{u}^{sc} = \boldsymbol{u}^{sc}_p + \boldsymbol{u}^{sc}_s $$
		
		The definitions of the two-dimensional $curl$ and $\boldsymbol{curl}$ operators are as follows:
		$$   \boldsymbol{curl} (\boldsymbol{w}) =  \dfrac{\partial w_2}{\partial x_1}  -\dfrac{\partial w_1}{\partial x_2} 
		\quad, \quad  curl (w) =  [\dfrac{\partial w}{\partial x_2} , -\dfrac{\partial w}{\partial x_1} ]^{\intercal} $$	
		where $w$ is a scalar function and $\boldsymbol{w}=(w_1,w_2)^{\intercal} $ denotes a vector function.
		
		Since the problem is formulated in the whole space $\mathbb{R}^2$, an appropriate radiation condition is needed to ensure the uniqueness of the solution. Kupradze proposed a radiation condition for  $\boldsymbol{u}^{sc} $,  which states that $\boldsymbol{u}^{sc}_p$ and $\boldsymbol{u}^{sc}_s$ should both satisfy the Sommerfeld radiation condition:

		\begin{equation}\label{sommer}
			\displaystyle\lim_{\boldsymbol{x}\rightarrow \infty} |\boldsymbol{x}|\Big(\partial_{|\boldsymbol{x}|}\boldsymbol{u}^{sc}_p - i\kappa_p \boldsymbol{u}^{sc}_p\Big) =0  \quad , \quad \displaystyle\lim_{\boldsymbol{x}\rightarrow \infty} |\boldsymbol{x}|\Big(\partial_{|\boldsymbol{x}|}\boldsymbol{u}^{sc}_s - i\kappa_s \boldsymbol{u}^{sc}_s \Big) =0 
		\end{equation}
		
		Following  Green tensor $\mathbf{G}(\boldsymbol{x},\boldsymbol{y},\omega) \in \mathbb{C}^{2\times 2} $  can be defined for the Navier equation \ref{base} with $\rho\equiv 1 $:
		\begin{equation}\label{Green}
			\mathbf{G}\big(\boldsymbol{x},\boldsymbol{y},\omega\big) =\dfrac{1}{\mu}\Phi\big(\boldsymbol{x},\boldsymbol{y},\kappa_s\big)\mathbf{I} + \dfrac{1}{\omega ^{2}}\nabla_x \nabla_{x}^{\perp}\Big(\Phi\big(\boldsymbol{x},\boldsymbol{y},\kappa_s\big)-\Phi\big(\boldsymbol{x},\boldsymbol{y},\kappa_p\big)\Big)
		\end{equation}
		where $\mathbf{I}$ is the $2 \times 2$ identity matrix, $\Phi\big(\boldsymbol{x},\boldsymbol{y},\kappa\big) = \dfrac{i}{4}H_{0}^{1}\big(\kappa|\boldsymbol{x}-\boldsymbol{y}|\big) $
		is the fundamental solution for the two-dimensional Helmholtz equation, and $ \nabla_x \nabla_{x}^{\perp} $ is Hessian matrix. It is easy to see that the Green tensor $\mathbf{G}\big(\boldsymbol{x},\boldsymbol{y},\omega\big) $ is symmetric with respect to the variables $\boldsymbol{x}$ and $\boldsymbol{y}$. 
		The following  theorem shows that the solution to the above scattering problem is also
		the solution to a Lippmann-Schwinger-type integral equation and vice versa.
		\begin{theorem}\label{Lippman-Theorem}
			\textbf {(The Lippmann Schwinger integral equation)	} Given a real valued function
			$\rho \in C^{1,\gamma} (\mathbb{R}^2) (0<\gamma<1) $ with supp $ (1-\rho)\subseteq \Omega  $. If there exists a $\boldsymbol{u}^{sc} \in C^2(\mathbb{R}^2)$ that satisfies the boundary condition  \ref{sommer} and $ \boldsymbol{u} = \boldsymbol{u}^{sc} + \boldsymbol{u}^{inc} $  is the solution of the Navier equation \ref{base}, then $\boldsymbol{u} $ is the solution to: 
			\begin{equation}\label{Lippman}
				\boldsymbol{u}(\boldsymbol{x}) = \boldsymbol{u}^{inc}(\boldsymbol{x}) -\omega^2 \displaystyle\int_\Omega  \mathbf{G}\big(\boldsymbol{x},\boldsymbol{y},\omega\big)\boldsymbol{u}(\boldsymbol{y})\Big(1-\rho(\boldsymbol{y})\Big) d\boldsymbol{y}
			\end{equation}
			conversely, if there exists a function $\boldsymbol{u} \in C^0(\mathbb{R}^2)$  that satisfies the  integral equation \ref{Lippman}, then $u$ is the solution to the Navier equation with Kupradze radiation condition.
			Furthermore $\boldsymbol{u} $ is unique.
		\end{theorem}
		The details and proof of this theorem are given in  \cite{bao, hahner, hahner2}.
		%%%%%%%%%%%%%%%%%%%%%%%%%%%%%%%%%%%%%%%%%%%%%%%%%%%%%%%%%%%%%%%%%%%%%%%%%%%%%%%%%%%%%%%%%%%%%%%%%%%%%%%%%%%%%%%%%%%%%%%%%%%%%%%%%%%%%%%%%%%%%%%%%%%%%%%%%%%%%%%%%%%%%%%%%%%%%%%%%%%%%%%%%%%%%%%%%%%%%%%%%%%%%%%%%%%%%%%%%
		\begin{remark}
			In the full waveform inversion problem, we usually seek three parameters: density, P-wave, and S-wave. The Green's function introduced in equation {\ref{Green}}, which is used in Theorem {\ref{Lippman-Theorem}}, is valid for constant Lame parameters. 
			If we find a Green's function for an elastic scattering problem with non-constant Lame parameters, either analytically or numerically, the results presented in this paper will be generalized.
			paper \cite{green} suggests approaches to drive Green's function for the heterogeneous elastic equations in specific cases, which can further generalize this study.
			Consequently, in this paper, we focus on finding the density and omit P-wave and S-wave.
		\end{remark}		
		In the following sections, we will explain the methodology for time-harmonic elastic FWI, which includes data-driven techniques and the integration of principles of physics related to the problem with it. Our goal is to  recover the subsurface using the data we have collected from the surface of the field. To achieve our objective, we have created a synthetic dataset, which is thoroughly described in section \ref{Dataset} using equation \ref{Lippman}. We divided the region $\Omega$ into $N = n\times n$ parts, assuming that the displacement field $\boldsymbol{u}=(u_1 , u_2)$   is measured at $n$ receivers on the Earth's surface, yielding to $2n$ values for each incident wave. Additionally, the number of incident waves (sources) is assumed to be $k$ with a single frequency for all of them.

		\section{First method (Direct Deep Learning Inversion)}\label{First}
		
		Our objective is to determine the unknown subsurface density values represented by $\rho$ by utilizing the data we have collected from the displacement field at the surface of the earth. The forward modeling of elastic scattering  problem can be expressed as:
		\begin{equation}
			\boldsymbol{f} (\rho) = \boldsymbol{u}|_{\partial \Omega} 
		\end{equation}
		
		We will train a network which is known as InversionNet in FWI. This network is designed to approximate the inverse mapping, denoted as $\boldsymbol{f}^{-1}$, from the values of the displacement field $\boldsymbol{u}|_{\partial \Omega}$ to the density $\rho$. 
		
		We intend to use a CNN to predict the density values in the area $\Omega$ by taking the surface displacement fields corresponding to different incident waves as its input. The structure of this network is illustrated in  Figure \ref{Dens_Cnn}.

		\begin{figure}[H]
			\centerline{\includegraphics[scale=0.8]{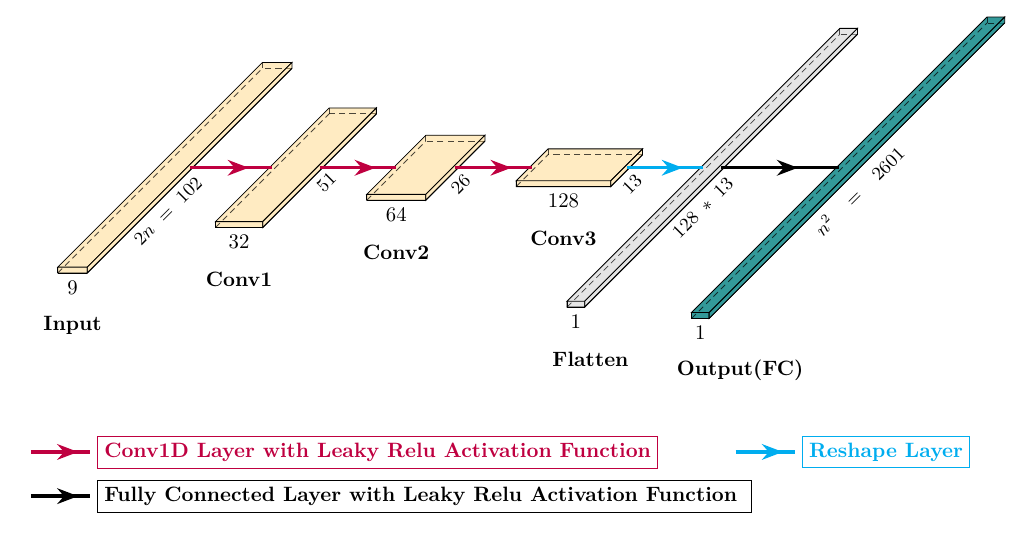}}
			\caption{The above network is specifically configured to take the real components of surface displacement fields (for two separate horizontal and vertical sections, each of size $51$) corresponding to $k=9$ incident waves. The network's objective is to produce an output that represents the density, denoted as $\overline{\rho}$, with a total size of $N=n^2=2601$. We train this network using the dataset we made in section \ref{Dataset} and  the $L_2$ loss function $\|\rho-\overline{\rho}\|^2_2$ between $\overline{\rho}$ and orginal $\rho$.}
			\label{Dens_Cnn}
		\end{figure}

		\section{Second method}\label{Second}
		As we know, the time-harmonic elastic FWI problem is nonlinear (in integral relation {\ref{Lippman}}, $\rho$ and $\boldsymbol{u} $ are multiplied). If we have the values of $\boldsymbol{u} $ over the entire region $\Omega $, the problem becomes linear, although it is highly ill-conditioned. In this section, we will present methods for solving this problem by estimating $\boldsymbol{u} $ over the entire region.
		
		In this section, initially, we train  a CNN to obtain the displacement field $\boldsymbol{u} $ within region $\Omega$. This network takes the same input as the one used in the First method  {\ref{First}} and provides the displacement fields within region $\Omega$ as its output. These displacement fields are collectively denoted as $\overline{U}=\{ \overline{U}_1 , ... ,\overline{U}_k\}$ and correspond to all $k$ incident waves.
		
		Since the input and output of this network are of the same type, and the output is  a continuation of the input, this process achieves very high accuracy, as demonstrated in Table \ref{table1} in the Results section. The structure of this network is illustrated in Figure \ref{Disp_Cnn}.
		
		\begin{figure}[H]
			\centerline{\includegraphics[scale=0.74]{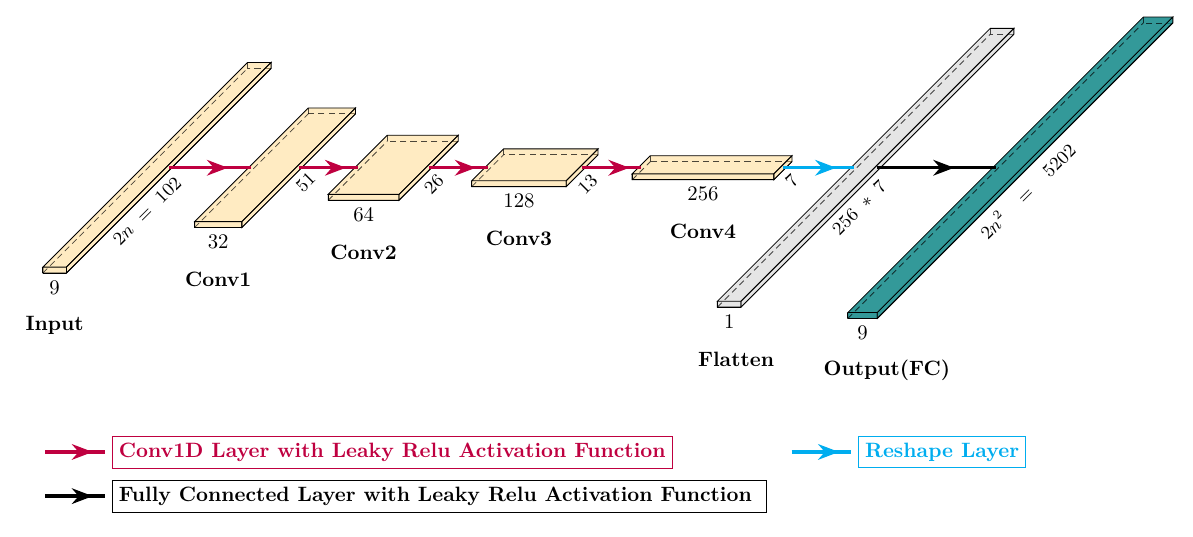}}
			\caption{ 	 The above network is designed to take the real part of surface displacement fields (for $k=9$ incident waves) from two horizontal and vertical segments, each with a size of n = 51. This input is directed to achieve an estimation of the real part of field $U$ (called $\overline{ U} $) in whole area $\Omega$ with a size of $k * 2 n^2$ as an output (because for each incident wave, we have a horizontal and vertical part for displacement field). We train this network with the dataset we made in section \ref{Dataset}. The loss function for this network is defined as $||U - \overline{ U}||^2_2$, representing the $L_2$ norm between $\overline{ U}$ and the original $U$. We also train another network with the same structure to estimate the imaginary part of $U$. }
			\label{Disp_Cnn}
		\end{figure}	
		
		Using the obtained displacement field and the integral equation \ref{Lippman}, we want to find the density of the desired area. We aim to simultaneously apply deep learning and the problem's physics to improve accuracy through some techniques. It should be noted that in this section, for convenience, we use $\rho $ instead of $1-\rho$ in formula \ref{Lippman}.

		\subsection{Least Square}\label{Least Square}
		Considering that we have the displacement field across the entire desired area and the use of relation \ref{Lippman}, the initial approach involves solving the problem using the least squares method.
		So, it is enough to collect all the information and write it as a linear system.

		By discretizing the integral equation \ref{Lippman}, it can be written as follows:
		\begin{equation}\label{linear_system}
			(U_j - U_j^{inc})_{2n^2 \times 1} = A_{2n^2 \times 2n^2} \,(\hat{\rho} U_j)_{2n^2 \times 1} 
		\end{equation}
		where $U_j = [u_1^j(x_1),...,u_1^j(x_N),\, u_2^j(x_1),...,u_2^j(x_N)]^{\intercal} $ , 
		$U^{inc}_j = [u^{inc_j}_1(x_1),...,u^{inc_j}_1(x_N),\, u^{inc_j}_2(x_1),...,u^{inc_j}_2(x_N)]^{\intercal} $ and $\hat{\rho} =  [\rho(x_1),...,\rho(x_N),\, \rho(x_1),...,\rho(x_N)]^{\intercal} $.
		
		Having a displacement field $\overline{U}_j$, which is an approximation of $U_j$ that was obtained at the beginning of this section, the linear equation described above can be written as $A_{u_j}\, \hat{\rho} = \overline{U}_j- U^{inc}_j$. To provide a well-defined representation of the problem, we can rewrite it as follows:
		\begin{equation}\label{least_square1}
			\begin{cases}
				B_{2n^2 \times n^2} := A_{u_j}[: , 0:n^2] + A_{u_j}[: , n^2:2n^2] \\
				b:=(\overline{U}_j - U^{inc}_j)_{2n^2 \times 1}  		 
			\end{cases}	
			\hspace{0.5 cm}\Longrightarrow \hspace{0.35 cm}  b_{2n^2 \times 1}= B_{2n^2 \times n^2} \rho_{n^2 \times 1} \\
		\end{equation}	
		where $\rho =  [\rho(x_1),...,\rho(x_N)]^{\intercal} $  and $ B(h_1,h_2) = A_{u_j}(h_1,h_2) + A_{u_j}(h_1,h_2+n^2) $.
		
		Then, for every $U^{inc}_j$ and $\overline{U}_j$ dependent on it, a linear equation \ref{least_square1}  can be formed to obtain $\rho$.
		To do this operation, since we have several $U^{inc}_j$ and $\overline{U}_j$ dependent on a single density $\rho $, we can consider all of them as a system of linear equations and obtain the desired density.
		Suppose we have k incident waves, consequently: 
		\begin{equation}\label{least_square2}
			\begin{cases}
				b_{u_1} = B_{u_1}\,\rho  \\
				\vdots \hspace{2 cm} \Longrightarrow  \\
				b_{u_k} = B_{u_k}\,\rho
			\end{cases}	
			\begin{bmatrix}
				b_{u_1}            \\ 
				\vdots       \\
				b_{u_k}      
			\end{bmatrix}_{2kn^2\times 1} = 
			\begin{bmatrix}
				B_{u_1}            \\ 
				\vdots       \\
				B_{u_k}      
			\end{bmatrix}_{2kn^2 \times n^2}
			\begin{bmatrix}
				\rho(x_1)            \\ 
				\vdots       \\
				\rho(x_N)      
			\end{bmatrix}_{n^2 \times 1}	
		\end{equation}	
		
		A suitable technique for solving the equation mentioned above is the least-squares method with $L_2$  regularization. One notable challenge when dealing with equation \ref{least_square2} is its considerable size. With an increment in the number of incident waves and $N$, which leads to an increase in matrix size, using this method can become quite challenging. The main difficulty in this method is that matrix $A$ is ill-posed in relation \ref{linear_system}. When this matrix is combined with the displacement vector $\overline{U} $,  which may contain a small error, this ill-posedness leads to a significant error in the solution of the problem.
		
		\subsection{Inverse Convolution} \label{Inverse Convolution}
		The integral form of relation \ref{Lippman} and the Green's tensor $\mathbf{G}$ in it, evokes a convolution-like relation in our mind. By defining this particular type of convolution, we want to obtain the inverse of $\mathbf{G}$ with this convolution.
		
		Using equation \ref{Lippman} and discretizing it, this equation  can be written as follows:
		\begin{equation}\label{Lippman_conv}
			\begin{bmatrix}
				u_1 - u^{inc}_1     \\ 
				u_2 - u^{inc}_2
			\end{bmatrix}
			=
			\begin{bmatrix}
				G_1      & G_2  \\ 
				G_2      & G_4 
			\end{bmatrix}
			\circledast
			\begin{bmatrix}
				u_1.\,\rho      \\ 
				u_2.\,\rho 
			\end{bmatrix}
		\end{equation}
		where $ u_1 $ and $ u_2$ are $n\times n $ complex matrices, dependent on a single $\boldsymbol{u}^{inc}=(u^{inc}_1 , u^{inc}_2)$, that represent the components of the displacement field, and $ \rho$ is a $n\times n $ density matrix within region $\Omega$. In the above equation, "." is the  matrix elementwise multiplication, and
		"$\circledast $"  is defined as follows:
		\begin{equation}\label{matconv}
			\begin{bmatrix}
				a      & b  \\ 
				c      & d 
			\end{bmatrix}
			\circledast
			\begin{bmatrix}
				e      & f  \\ 
				g      & h 
			\end{bmatrix}
			:=
			\begin{bmatrix}
				a\ast e + b\ast g      &   a\ast f + b\ast h  \\ 
				c\ast e + d\ast g      &   c\ast f + d\ast h
			\end{bmatrix}
		\end{equation}
		Furthermore, Multiplication $\ast$ in  equation \ref{matconv} is ordinary matrix convolution.
		
		We aim to find the matrix $\overline{\mathbf{G}} $ such that:
		\begin{equation}\label{kron}
			\begin{bmatrix}
				\overline{G}_1      & \overline{G}_2  \\ 
				\overline{G}_2      & \overline{G}_4 
			\end{bmatrix}
			\circledast
			\begin{bmatrix}
				G_1      & G_2  \\ 
				G_2      & G_4 
			\end{bmatrix}
			=
			\begin{bmatrix}
				\delta      &   0  \\ 
				0      &   \delta
			\end{bmatrix}
		\end{equation}
		where $\delta$ is a Kronecker matrix with its central element equal to 1, while all other elements are set to zero.
		
		If we find matrix $\overline{\mathbf{G}} $ or an approximation of it, we can multiply  $\overline{\mathbf{G}} $ on both sides of the equation \ref{Lippman_conv} using the  $ \circledast$ multiplication operator. This operation can be performed on the approximation we obtained for $\boldsymbol{u}=(u_1,u_2)$ at the beginning of this section, which we denoted as $\overline{\boldsymbol{u}}=(\overline{u}_1,\overline{u}_2)$:
		\begin{equation}\label{conv_inverse}
			\begin{bmatrix}
				v_1     \\ 
				v_2
			\end{bmatrix}	:=
			\begin{bmatrix}
				\overline{G}_1      & \overline{G}_2  \\ 
				\overline{G}_2      & \overline{G}_4 
			\end{bmatrix}
			\circledast
			\begin{bmatrix}
				\overline{u}_1 - u^{inc}_1     \\ 
				\overline{u}_2 - u^{inc}_2
			\end{bmatrix}
			\simeq
			\begin{bmatrix}
				\overline{u}_1.\,\rho      \\ 
				\overline{u}_2.\,\rho 
			\end{bmatrix}
			\hspace{0.5 cm}
			\Longrightarrow
			\hspace{0.5 cm}
			\begin{bmatrix}
				\rho     \\ 
				\rho
			\end{bmatrix}
			\simeq
			\begin{bmatrix}
				\Re(v_1)     \\ 
				\Re(v_2)
			\end{bmatrix}
			/
			\begin{bmatrix}
				\Re(\overline{u}_1)      \\ 
				\Re(\overline{u}_2) 
			\end{bmatrix}	
			\simeq			
			\begin{bmatrix}
				\Im(v_1)     \\ 
				\Im(v_2)
			\end{bmatrix}
			/
			\begin{bmatrix}
				\Im(\overline{u}_1)      \\ 
				\Im(\overline{u}_2) 
			\end{bmatrix}		
		\end{equation}
		where $/ $ refers to elementwise division. As a result, the density can be extracted using the above equation.	
		
		Two techniques can be employed to determine the matrix $\tilde{\mathbf{G}}$. The first one is to write the equation \ref{kron}   as a linear equation and solving it. However, this technique encounters challenges when $\mathbf{G}$ becomes large. To solve this problem, we offer another technique in which $\tilde{\mathbf{G}} $ is learned through a process. This procedure is formally presented in Algorithm  \ref{alg2}.

		\begin{algorithm}
			\caption{(Optimization algorithm to find $\tilde{G} $)}
			\label{alg2}
			\hspace*{\algorithmicindent} \textbf{Input : } Matrix $G_1 , G_2 , G_4 $ \\
			\hspace*{\algorithmicindent} \textbf{Output :} Matrix $\tilde{G}_1 , \tilde{G}_2 , \tilde{G}_4 $
			\begin{algorithmic}[1]
				\State  Initialize the complex matrix $\tilde{G}_1 , \tilde{G}_2 , \tilde{G}_4 $ with random normal distributions for the real and imaginary components.
				\While{Loss $\geq$ Threshold}
				\State  Loss = $||\tilde{G}_1*G_1 + \tilde{G}_2*G_2 - \delta ||^2_2 + ||\tilde{G}_1*G_2 + \tilde{G}_2*G_4  ||^2_2  +
				||\tilde{G}_2*G_1 + \tilde{G}_4*G_2  ||^2_2 +
				||\tilde{G}_2*G_2 + \tilde{G}_4*G_4 - \delta ||^2_2$ 				
				\State Update matrix $\tilde{G}_i$ with the Adam optimizer to minimize Loss 
				\EndWhile \\
				\Return  $\tilde{G_1} , \tilde{G}_2 , \tilde{G}_4 $ 
				
			\end{algorithmic}
		\end{algorithm}
		
		Note that in relation \ref{conv_inverse}, we drive four approximated densities. Furthermore, for each $\boldsymbol{u}^{inc}$ and its corresponding  $\overline{\boldsymbol{u}}$, we can form the relation \ref{conv_inverse} to obtain $\rho$. While each of these densities can be utilized to estimate the original density, deciding which one to choose is  challenging. It is crucial to have a framework that can provide a unique estimate by incorporating  these approximations. Employing the Unet network structure presents a promising solution. Since we will encounter a similar situation in the upcoming subsection, we will elaborate on the application of the Unet structure and provide more comprehensive details on this problem in method \ref{Linear_to_nonlinear}.

		\subsection{Linear-to-Nonlinear}\label{Linear_to_nonlinear}
		While the accuracy of the estimated displacement field $\boldsymbol{u} $ is notably high,  methods \ref{Least Square} and \ref{Inverse Convolution} do not yield satisfactory results.
		This issue arises because the mentioned linear device is ill-conditioned, and even the slightest noise in the linear system can result in significant errors, undermining the entire process. However, the mentioned techniques can be effective for well-conditioned problems. In the previous method \ref{Inverse Convolution}, we sought to perform an inverse convolution, effectively aiming to approximate the inverse of this linear equation by employing a specific structure. In this method, we aim to approximate the inverse of the linear system using  a nonlinear transformation. Inspired by the previous method \ref{Inverse Convolution}, we seek a nonlinear function $\hat{G} : \mathbb{C}^{2n^2} \to \mathbb{C}^{2n^2}$ such that:
		\begin{equation}\label{rhou1}
			\hat{G}(U_j - U^{inc}_j) = \hat{\rho} U_j
		\end{equation}
		where $U_j , U^{inc}_j $ and $  \hat{\rho}$ are the same as defined in relation \ref{linear_system}. 
		Our objective is to find a structure for the $\hat{G}$ function that not only guarantees high accuracy but also  avoids a substantial computational cost. Furthermore, this function should also has a robust performance in the presence of noisy data and  mitigate their impact.
		We define $\hat{G}$ function as a nonlinear combination of $\hat{G_i} \,,\,\, i=1,...,m$ convolutions:
		\begin{equation}\label{rhou2}
			\hat{G}(U_j - U^{inc}_j) := F(\hat{G_1}\ast F(\hat{G_2}\ast ... F(\hat{G_m}\ast(U_j - U^{inc}_j) )))
		\end{equation}
		where the function $ F $ is a nonlinear activation function that will be determined later.
		
		A clever technique to obtain functions $\hat{G_i} \,,\,\, i=1,...,m$ is to train them using the available dataset. If we consider the relationship described above as a network, its input would be $ U - U^{inc}$, and its output would be $ \hat{\rho} U $.
		To enable the network to effectively denoise, we add Gaussian noises to the input of this network, which are $ U - U^{inc}$ functions, and we expect it to give us the exact  $\hat{\rho} U$ function as output. It is important to note that during the training of this network, the original $U$ is used as an input. However, when we apply the network in practice, we replace it with an estimated $\overline{U} $. We expect the difference between $U$ and $ \overline{U}$ to be a Gaussian random field. By adding a Gaussian random field to the input, we aim to avoid noise and ensure the network understands the nature of the true wavefield $\hat{\rho} U$. This process reduces the errors resulting from the approximation of $U $ in the system and enhances its stability. The structure of this network is illustrated in  Figure \ref{rhou_Cnn}.

		So far, we have followed a two-step process. Initially, in section \ref{Second}, we obtained an estimate for $U$. Subsequently, we employed the network constructed based on relations \ref{rhou1} and \ref{rhou2} to estimate the values of $ \hat{\rho} U $ (which we have called  $\overline{\rho U} $). 
		Consequently, the density is computed by dividing $\overline{\rho U} $ by $ \overline{U} $. However, it is worth noting that the number of these approximated densities amounts to $4k$, as we have $k$ incident waves, and for each of them, there exist  horizontal, vertical, real, and imaginary components.
		
		Determining which estimate to choose is a crucial decision. Having a structure capable of combining all these estimations into a unique and precise result is of utmost importance. Implementing the Unet network structure is a promising technique, supported by its successful application in previous studies \cite{sanghvi,wei}. We plan to construct a network based on the Unet architecture. This network will take all the estimations we have gathered so far as its input and produce the desired density function as its output. The structure of this network is illustrated in  Figure \ref{Unet}.
		
		\begin{figure}[H]
			\centerline{\includegraphics[scale=0.8]{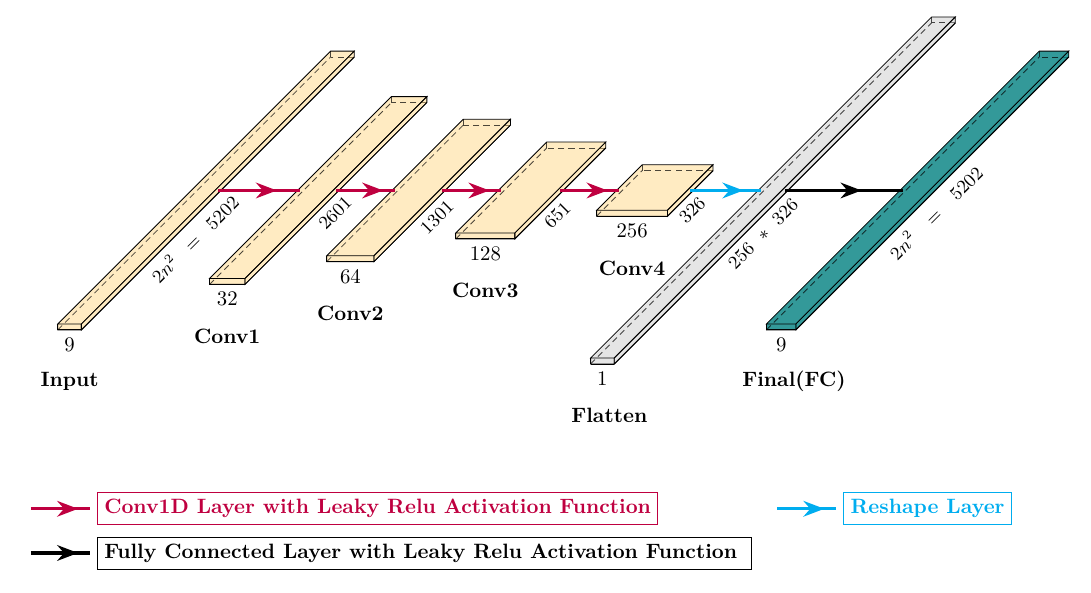}}
			\caption{ The above network is designed to take the real part of displacement fields in the whole area $\Omega$ (for $k=9$ incident waves) from two horizontal and vertical segments, each with a size of $n=51$. This input is directed to achieve an estimation of the real part of field $\hat{\rho} U$ (called $\overline{\rho U} $) in whole area $\Omega$ with a size of $k * 2 n^2$ as an output (because for each incident wave, we have a horizontal and vertical part for displacement field). We train this network with the dataset we made in section \ref{Dataset} and estimations of displacement fields acquired from the beginning of this section. The loss function for this network is defined as $||\hat{\rho} U - \overline{\rho U}||^2_2$, representing the $L_2$ norm between $\overline{\rho U}$ and the original $\hat{\rho} U$. We also train another network with the same structure to estimate the imaginary part of $\hat{\rho} U$.}
			\label{rhou_Cnn}
		\end{figure}
		
		\begin{figure}[H]
			\centerline{\includegraphics[scale=0.58]{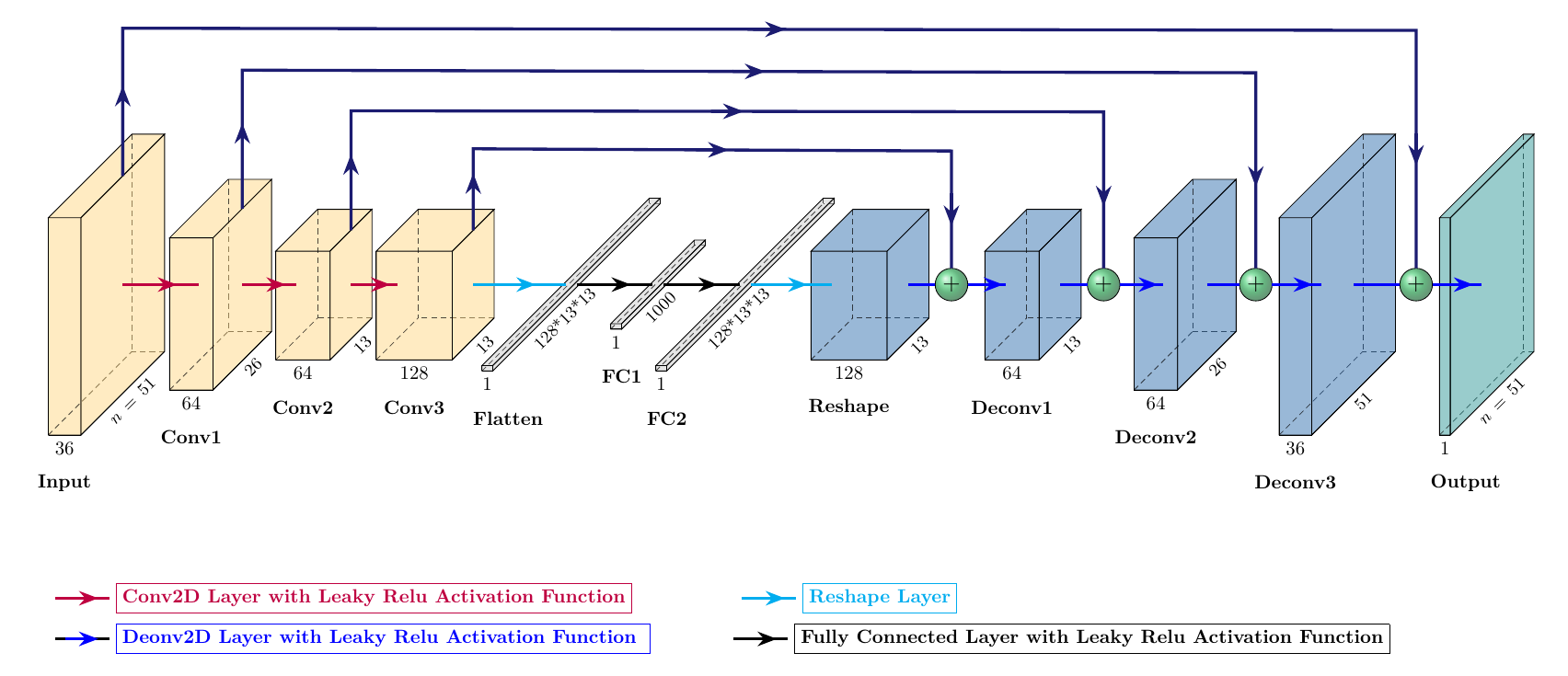}}
			\caption{ The above network is designed to take all $4k$ density estimates obtained in this method (4 is representative of 4 categories of real, imaginary, horizontal, and vertical components of estimation, and k = 9 is related to the number of incident waves) and produce a unique density as output, denoted as $\overline{\rho}$. We train this network with the dataset we made in section {\ref{Dataset}} and estimations of densities that were acquired from the previous network (Figure {\ref{rhou_Cnn}}). The loss function for this network is defined as $||\rho - \overline{\rho}||^2_2$, representing the $L_2$ norm between $\overline{\rho}$ and the original density $\rho$.  }
			\label{Unet}
		\end{figure}
		
		%%%%%%%%%%%%%%%%%%%%%%%%%%%%%%%%%%%%%%%%%%%%%%%%%%%%%%%%%%%%%%%%%%%%%%%%%%%%%%%%%%%%%%%%%%%%%%%%%%%%%%%%%%%%%%%%%%%%%%%%%%%%%%%%%%%%%%%%%%%%%%%%%%%%%%%%%%%%%%%%%%%%%%%%%%%%%%%%%%%%%%%%%%%%%%%%%%%%%%%%%%%%%%%%%%%%%%%%%%%%%%%%%%%%%%%%%%%%%%%%%%%%%%%%%%%%%%%%%%%%%%%%%%%%%%%%%%%%%%%%%%%%%	
		
		\section{Third method (New Variational Autoencoder)}\label{Third}
		%\subsection{Overview from VAE}
		Variational Autoencoders (VAE), a widely used and adaptable category within deep generative models, fall under the subfield of unsupervised deep learning. They can stochastically draw samples from a specified random process to approximate the true underlying distribution. % by leveraging an information-theoretic variational bound
		
		We initiate our process with a dataset $ X = \{  x_i \}_{i=1}^M  $ containing $M$ independent and identically distributed (i.i.d.) samples of a random variable $\boldsymbol{x} \in \mathbb{R}^d$. The objective is to derive a tractable approximation for $p_{\Theta}(\boldsymbol{x})$, enabling the generation of new samples of $\boldsymbol{x}$.  Moreover, we assume that each sample is governed by unobserved latent variables $\boldsymbol{z} \in \mathbb{R}^s$, such that:
		\begin{equation}\label{Pro_Data}
			p_{\Theta}(\boldsymbol{x})  = \int p(\boldsymbol{z}) p_{\Theta}(\boldsymbol{x}| \boldsymbol{z} ) d\boldsymbol{z}
		\end{equation}
		where $\theta $ represents the parameters defining the distribution that we aim to estimate. Typically, the selection of distributions $p_{\Theta}(\boldsymbol{x}| \boldsymbol{z} )$ and $P(\boldsymbol{z})$ is Gaussian.
		
		Preferably, our goal is to minimize the average negative log-likelihood $p_{\Theta}(\boldsymbol{x}) $ over the entire ground-truth dataset $X$. In other words, we seek to solve the optimization problem $\displaystyle\min_{\Theta} \sum_{i=1}^M (-\log(p_{\Theta}(x_i))) $. Unfortunately, the required marginalization over $\boldsymbol{z}$ is generally intractable. The challenge arises from the approximation of $p_{\Theta}(x_i)$, where we sample a large number of $p_{\Theta}(x_i | z_j)$ for $\boldsymbol{z}$ values $\{z_1,...,z_j,..., z_t\}$. %in estimation of $ p_{\Theta}(x_i) \simeq {\sum_{j=1}^t p_{\Theta}(x_i | z_j)/t}  $. 
		In high-dimensional spaces, the necessity for a large $t$ arises before obtaining an accurate estimate of $p_{\Theta}(x_i)$. Furthermore, in practice, for most $z_i$, $p_{\Theta}(x_i | z_j) $ is likely to be nearly zero, contributing insignificantly to our estimate of $p_{\Theta}(x_i)$. The fundamental concept behind the VAE is to strategically sample values of $z_j$ that are most likely to have produced $x_i$, then compute $p_{\Theta}(x_i)$ only from those values. As a result, a new function $ q_{\Phi}(\boldsymbol{z}| \boldsymbol{x} ) $ must be introduced which can take a value $x_i$ and provide a distribution over $\boldsymbol{z}$ values that are likely to produce $x_i$. As such, this makes it relatively easy for us to compute $\mathbb{E}_{q_{\Phi}(\boldsymbol{z}|\boldsymbol{x})}  p(\boldsymbol{x}|\boldsymbol{z})$.
		
		Hence, aside from minimizing the negative log-likelihood $p_{\Theta}(\boldsymbol{x}) $, we must adjust the parameters of $\Phi $ to minimize the difference between $q_{\Phi}(\boldsymbol{z}|\boldsymbol{x}) $ and the true posterior $p_{\Theta}(\boldsymbol{z}|\boldsymbol{x}) $. For this purpose, the ELBO function is defined as follows:
		
		\begin{equation}\label{VAE_Loss1}
			\mathcal{L}(\Phi , \Theta) = \dfrac{1}{M} \sum_{i=1}^M   (-\log(p_{\Theta}(x_i)))
			+   \dfrac{1}{M} \sum_{i=1}^M \mathcal{D}_{KL}\Big(  q_{\Phi}(\boldsymbol{z}|{x_i}) ||p_{\Theta}(\boldsymbol{z}|{x_i}) ) \Big)
		\end{equation}
		Here $\mathcal{D}_{KL}(p,q) :=\mathbb{E}_{\boldsymbol{x}\sim p}\log\Big(\dfrac{p(\boldsymbol{x})}{q(\boldsymbol{x})}\Big) $, always a non-negative quantity, denotes the Kullback-Leibler (KL) divergence between two distributions $p$ and $q$. For optimization convenience, the equation \ref{VAE_Loss1} can be expressed as:
		
		\begin{equation}\label{VAE_Loss2}
			\mathcal{L}(\Phi , \Theta) = -\dfrac{1}{M} \sum_{i=1}^M \int  q_{\Phi}(\boldsymbol{z}|{x_i} ) 
			\log \Big(p_{\Theta}({x_i}| \boldsymbol{z})\Big)d\boldsymbol{z}  +   \dfrac{1}{M} \sum_{i=1}^M \mathcal{D}_{KL}\Big(  q_{\Phi}(\boldsymbol{z}|{x_i}) || p(\boldsymbol{z} ) \Big)
		\end{equation}
		In these expressions, $q_{\Phi}(\boldsymbol{z}|{x_i}) $ can be seen as an encoder model defining a conditional distribution over the latent variable $\boldsymbol{z}$, while $p_{\Theta}(x_i|\boldsymbol{z}) $ can be interpreted as a decoder model.

		The VAE, as formulated in Equation \ref{VAE_Loss2}, consists of two components: a data-fitting loss inspired by the structure of an Autoencoder (AE) and a regularization factor based on KL-divergence. The first one encourages assigning high probability to latent variable $\boldsymbol{z}$ that leads to accurate reconstructions of each $x_i$. Notably, when $q_{\Phi}(\boldsymbol{z}|\boldsymbol{x}) $ is a Dirac delta function, this term precisely mirrors a deterministic AE with a data reconstruction loss defined by $-\log(  p_{\Theta}(\boldsymbol{x}| \boldsymbol{z})) $. Additionally, the KL regularizer $ \mathcal{D}_{KL}\Big(  q_{\Phi}(\boldsymbol{z}|x_i) || p(\boldsymbol{z} ) \Big)  $ guides the encoder distribution towards the prior.

		For continuous data, the common assumption is as follows:
		\begin{equation}\label{Gaussian}
			p(\boldsymbol{z}) = \mathcal{N}(\boldsymbol{z} | 0 , I) 
			\quad , \quad
			q_{\Phi}(\boldsymbol{z}|\boldsymbol{x}) = \mathcal{N}(\boldsymbol{z} | \mu_{\Phi}(\boldsymbol{x}) , \mathcal{V}_{\Phi}(\boldsymbol{x})) 
			\quad , \quad
			p_{\Theta}(\boldsymbol{x} | \boldsymbol{z}) = \mathcal{N}(\boldsymbol{x} | \mu_{\Theta}(\boldsymbol{z}) , \alpha I)
		\end{equation}
		In this formulation, $\alpha > 0$ represents a scalar variance parameter. The Gaussian moments, $ \mu_{\Phi}(\boldsymbol{x}) \in \mathbb{R}^s,  \,\mu_{\Theta}(\boldsymbol{z}) \in \mathbb{R}^d $, and $ \mathcal{V}_{\Phi}(\boldsymbol{x}) = diag(v_1^2(\boldsymbol{x}),...,v_s^2(\boldsymbol{x}))  $ are computed through feedforward neural network layers.	 The encoder network, parameterized by $\Phi $, takes $\boldsymbol{x}$ as input and produces $ \mu_{\Phi} $ and $ \mathcal{V}_{\Phi}$. Similarly, the decoder network, parameterized by $\Theta $, transforms a latent variable $\boldsymbol{z}$ into $  \mu_{\Theta} $.	From the relationship given in Equation \ref{Gaussian}, it can be deduced:
		\begin{equation}\label{KL}
			2 \mathcal{D}_{KL}\Big(  q_{\Phi}(\mathbf{z}|x_i)  || p(\mathbf{z} ) \Big) = 
			trace( \mathcal{V}_{\Phi}(x_i)) -\log(|\mathcal{V}_{\Phi}(x_i)|)  + ||\mu_{\Phi}(x_i)||_2^2 
		\end{equation}
		We removed constants that are not relevant. Furthermore, $|.|$ denotes the matrix's determinant. Given these assumptions, the generic VAE objective from \ref{VAE_Loss2} can be refined to:
		\begin{equation}\label{VAE_Loss}
			\mathcal{L}(\Phi , \Theta) = \dfrac{1}{2M} \sum_{i=1}^M \bigg\{     \mathbb{E}_{q_{\Phi}(\boldsymbol{z}|x_i)}  \Big(\dfrac{||x_i - \mu_{\Theta}(\boldsymbol{z}) ||_2^2}{\alpha} \big)         +
			trace( \mathcal{V}_{\Phi}(x_i)) -\log(|\mathcal{V}_{\Phi}(x_i)|)  + ||\mu_{\Phi}(x_i)||_2^2
			\bigg\}
		\end{equation}
		This formulation can be effectively minimized using stochastic optimization algorithms such as SGD and Adam. 

		Probabilistic methods in inversion problems aim to consider a random variable as a solution instead of an unknown value. Consequently, the solution to the probabilistic inverse problem is a probability distribution instead of a single estimated value. Although the goal of generative modeling, such as VAE, fundamentally differs from scientific inverse problems, we claim that VAEs are also well suited for it. To estimate the inverse solution using VAE, some studies have been conducted.  particularly, in one of the most interesting studies, with a structural modification in VAE, \cite{goh} reveals the latent variable $\boldsymbol{z}  $ as the answer to the inverse problem.  However, it's worth noting that this method may face challenges related to its relatively lower accuracy.

		\subsection{New-VAE structue}\label{new-vae}
		With modifications to the conventional VAE structure, we aim to provide the VAE network with this capability to not only generate samples but also solve the inverse problem stochastically. In fact, in our problem, we want to introduce a random variable $\boldsymbol{\rho} $ as a solution to the inverse problem using the surface displacement field, which, for simplicity, we call it $ \boldsymbol{u} $.

		By integrating two VAE networks and defining their interaction, we aim to derive a stochastic solution for the inverse problem. Consider one VAE1 with latent variable $\boldsymbol{z}_1$ for variable $\boldsymbol{\rho} $ and another VAE2 with latent variable $ \boldsymbol{z}_2$ for variable $\boldsymbol{u}  $. The connection between these $ \boldsymbol{z}_1$ and $ \boldsymbol{z}_2$ can be the key to connecting the density $ \boldsymbol{\rho} $  to the field $\boldsymbol{u}  $  or vice versa. However, this connection can be complicated. To make this process easier, instead of $ \boldsymbol{z}_1$ and $ \boldsymbol{z}_2$, we use one hidden variable $ \boldsymbol{z}$ so that this variable is related to both $\boldsymbol{\rho}$ and $\boldsymbol{u}  $.
		The outline of this procedure and the relation between variables is depicted in Figure \ref{VAE_New}.
		
		\begin{figure}[H]
			\centerline{\includegraphics[scale=0.65]{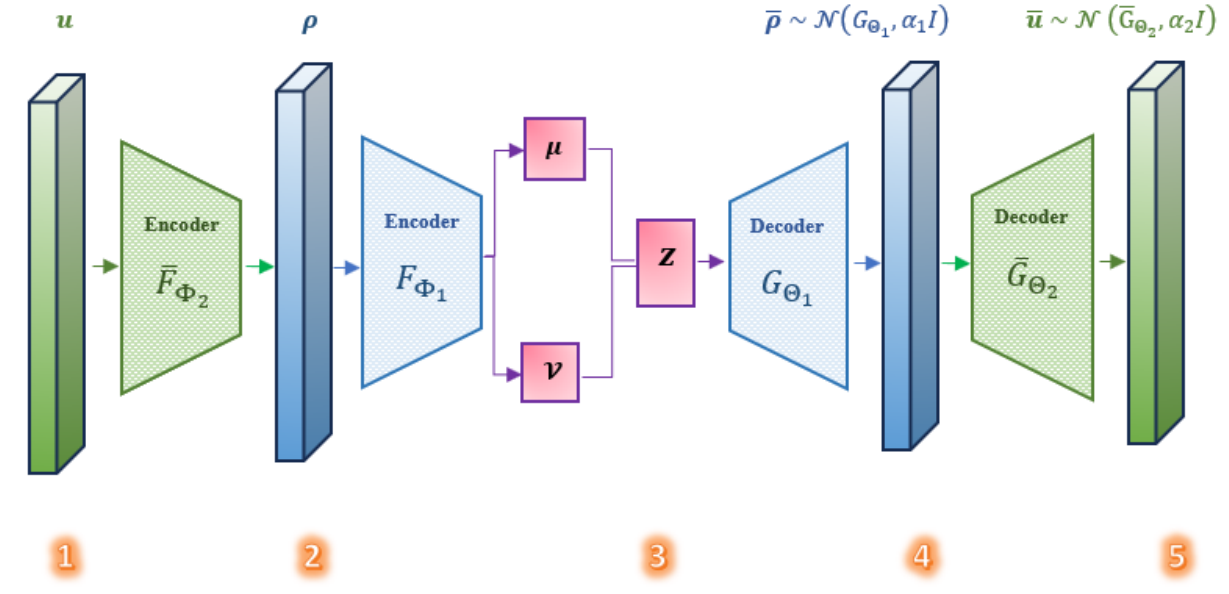}}
			\caption{Structure and relation between variables in New-VAE }
			\label{VAE_New}
		\end{figure}
		
		Our network consists of three variables $\boldsymbol{\rho}$, $\boldsymbol{u}$, and $\boldsymbol{z}$. The inner core of the network (Figure \ref{VAE_New train1}) is a VAE that shows the relationship between variables $\boldsymbol{\rho}$ and $\boldsymbol{z}$. Its encoder and decoder are denoted by $ q_{\Phi_1}(\boldsymbol{z}|\boldsymbol{\rho}) =  \mathcal{N}(\boldsymbol{z} | \mu_{\Phi_1}(\boldsymbol{\rho}) , \mathcal{V}_{\Phi_1}(\boldsymbol{\rho})) $ and  $p_{\Theta_1}(\boldsymbol{\rho}|\boldsymbol{z}) = \mathcal{N}(\boldsymbol{\rho} | G_{\Theta_1}(\boldsymbol{z}) , \alpha_1 I) $, respectively.
		On the other hand, the outer part of the network (figure \ref{VAE_New train2}), is a VAE that shows the relationship between variables $\boldsymbol{u}$ and $\boldsymbol{z}$. We show its encoder with $ q_{\Phi_1,\Phi_2}(\boldsymbol{z}|\boldsymbol{u}) =  \mathcal{N}(\boldsymbol{z} | \mu_{\Phi_1,\Phi_2}(\boldsymbol{u}) , \mathcal{V}_{\Phi_1,\Phi_2}(\boldsymbol{u})) $  and its decoder with $p_{\Theta_1,\Theta_2}(\boldsymbol{u}|\boldsymbol{z}) = \mathcal{N}(\boldsymbol{u}| \bar{G}_{\Theta_2}(\bar{\boldsymbol{\rho}}) , \alpha_2 I) $ where $\bar{\boldsymbol{\rho}} = \mathcal{N}(\boldsymbol{\bar{\boldsymbol{\rho}}} | G_{\Theta_1}(\boldsymbol{z}) , \alpha_1 I)$.
		
		In conventional VAE, we looked for parameters $\Phi $ and $ \Theta $  such that the function $ -\log(p_{\Theta}(\boldsymbol{x} )) + \mathcal{D}_{KL}\Big(  q_{\Phi}(\boldsymbol{z} |\boldsymbol{x} ) || p_{\Theta}(\boldsymbol{z} |\boldsymbol{x}  ) \Big) $  is minimized, and thus, the posterior can be approximated. 
		As a result, to correctly find the relationship between the latent variables $ \boldsymbol{z}$, $\boldsymbol{\rho}$, and $\boldsymbol{u} $ in the New-VAE method, we must find the parameters of $\Phi_1, \Phi_2, \Theta_1$, and $ \Theta_2 $ such that the following function is minimized:
		
		\begin{equation} \label{VAE_new_kl}
			\begin{split}
				\mathcal{L}(\Phi_1 , \Phi_2 , \Theta_1 , \Theta_2) 
				&=  \underbrace{ \overbrace{\dfrac{1}{M} \sum_{i=1}^M   (-\log(p_{\Theta_1}(\rho_i)))}^{\mathrm{Reconstruction\,\,\,  Loss_1}}
					+ \overbrace{\dfrac{1}{M} \sum_{i=1}^M\mathcal{D}_{KL}\Big(  q_{\Phi_1}(\boldsymbol{z}|\rho_i)  || p_{\Theta_1}(\boldsymbol{z} | \rho_i  ) \Big)}^{\mathcal{D}_{{KL}_1}} }_{\mathbf{Loss_1}}  \\
				& + \underbrace{ \overbrace{\dfrac{1}{M} \sum_{i=1}^M   (-\log(p_{\Theta_1 , \Theta_2}(u_i)))}^{\mathrm{Reconstruction\,\,\,  Loss_2}} + \overbrace{\dfrac{1}{M} \sum_{i=1}^M\mathcal{D}_{KL}\Big(  q_{\Phi_1 ,  \Phi_2 }(\boldsymbol{z} |u_i)  || p_{\Theta_1 , \Theta_2}(\boldsymbol{z} | u_i ) \Big)}^{\mathcal{D}_{{KL}_2}}}_{\mathbf{Loss_2}}                   \\
				&  = -\underbrace{\dfrac{1}{M} \sum_{i=1}^M \int  q_{\Phi_1}(\boldsymbol{z} |\rho_i ) 
					\log \Big(p_{\Theta_1}({\rho}_i| \boldsymbol{z} )\Big)d\boldsymbol{z}   +   \dfrac{1}{M} \sum_{i=1}^M \mathcal{D}_{KL}\Big(  q_{\Phi_1}(\boldsymbol{z}|\rho_i) || p(\boldsymbol{z}  ) \Big)}_{\mathbf{Loss_1}} \,                  \\
				& - \underbrace{\dfrac{1}{M} \sum_{i=1}^M \int  q_{\Phi_1 , \Phi_2}(\boldsymbol{z} |u_i ) 
					\log \Big(p_{\Theta_1 , \Theta_2}({u}_i|\boldsymbol{z} )\Big)d\boldsymbol{z}   +   \dfrac{1}{M} \sum_{i=1}^M \mathcal{D}_{KL}\Big(  q_{\Phi_1 , \Phi_2}(\boldsymbol{z} |u_i) || p(\boldsymbol{z}  ) \Big)  }_{\mathbf{Loss_2}}                                            \\	 
				& = \underbrace{\dfrac{1}{2M} \sum_{i=1}^M \bigg\{     \mathbb{E}_{q_{\Phi_1}(\boldsymbol{z} |\rho_i)}  \Big(\dfrac{||\rho_i - \mu_{\Theta_1}(z) ||_2^2}{\alpha_1} \Big)         +
					trace( \mathcal{V}_{\Phi_1}(\rho_i)) -\log(|\mathcal{V}_{\Phi_1}(\rho_i)|)  + ||\mu_{\Phi_1}(\rho_i)||_2^2
					\bigg\}}_{\mathbf{Loss_1}}
				\\
				& + \underbrace{\dfrac{1}{2M} \sum_{i=1}^M \bigg\{     \mathbb{E}_{q_{\Phi_1,\Phi_2}(\boldsymbol{z} |u_i)}  \Big(\dfrac{||u_i - \mu_{\Theta_1,\Theta_2}(z) ||_2^2}{\alpha_2} \Big)         +
					trace( \mathcal{V}_{\Phi_1,\Phi_2}(u_i)) -\log(|\mathcal{V}_{\Phi_1,\Phi_2}(u_i)|)  + ||\mu_{\Phi_1,\Phi_2}(u_i)||_2^2
					\bigg\}}_{\mathbf{Loss_2}}	
			\end{split}
		\end{equation}

		To intelligently train this network and determine the optimal parameters, we follow the sequential optimization processes. Initially, we minimize $\mathrm{Loss_1}$ using multiple iterations of one optimization  algorithm as depicted in Figure \ref{VAE_New train1}, resulting in an initial estimate for the parameters $\Phi_1$ and $\Theta_1$. Subsequently, we proceed to minimize $\mathrm{Loss_2}$ through several iterations of the one optimization  algorithm, leading to the optimal parameters $\Phi_1$, $\Phi_2$, $\Theta_1$, and $\Theta_2$ as illustrated in Figure \ref{VAE_New train2}. This process iteratively minimizes these functions by using the optimized values from each step as the initial values for the next step, which ultimately converges to an equilibrium point.

		Some of the advantages and capabilities provided by New-VAE include the following:
		\begin{itemize}
			\item (Inverse aspect) New-VAE can to solve the inverse problem. To achieve this, we input the displacement field $\boldsymbol{u}$ in step 1, as  depicted in Figure \ref{VAE_New}, and through feedforward processes, we obtain a deterministic solution in step 2. Now, with an additional feedforward  obtained from the previous step, we reach a stochastic solution in step 4. In seismic wave tasks, where there is uncertainty in parameters and approximation of quantities, stochastic solutions are often preferable, allowing the expert to make informed choices. It is important to note that we have the option to learn a covariance matrix for variable $\overline{\boldsymbol{\rho}}$ in step 4 of Figure \ref{VAE_New}, just as we did for the variable $\boldsymbol{z}$. However, for the scope of this paper, we chose not to pursue this extension.
			
			\item (Generator) By determining the  distribution of latent variable $\boldsymbol{z}$,   we will somewhat find the distribution $\boldsymbol{\rho} $, which enables us to generate samples. In section \ref{Dataset}, we will explain that the construction of the random field $\boldsymbol{\rho}$ and consequently, the displacement field $\boldsymbol{u}$ related to it requires solving a large linear equation. This procedure has a high calculation cost for both. Having an optimized New-VAE network (Figure \ref{VAE_New}), with a feedforward from 3 to 4 and 5, we can generate density and  the corresponding displacement field, respectively.  Utilizing this process enables us to augment the dataset with significantly reduced computational costs.
			
			\item (Forward aspect) Another capability of this network is to solve the direct problem. Specifically, by inputting the density $\boldsymbol{\rho}$ at step 4 in Figure \ref{VAE_New} and utilizing a feedforward process, we can obtain an estimate for $\boldsymbol{u}$ at step 5. In problems with the structure $F(\boldsymbol{\rho})=\boldsymbol{u} $ where $F$ represents a large linear matrix or has an unconventional nonlinear structure, since the CNN networks reduce the number of parameters, using this network instead of directly obtaining the solution can be helpful in terms of computational cost.   In our Elastic problem, to build the dataset, once the subsurface density  is known, the displacement field can be obtained by solving a large linear system over the whole region. In  solving the inverse problem by deep learning, we often  use the set of subsurface densities and displacement fields on the surface for training. Therefore, when we create a dataset, it is not usually necessary to have the displacement field over the whole region. As a result, this network can be used to effectively produce a large-scale dataset, as illustrated in Figure \ref{VAE_New} at steps 4 to 5.
			
		\end{itemize}
		
		\begin{figure}[H]
			\centering
			\begin{subfigure}[b]{0.6\textwidth}
				\includegraphics[width=\linewidth]{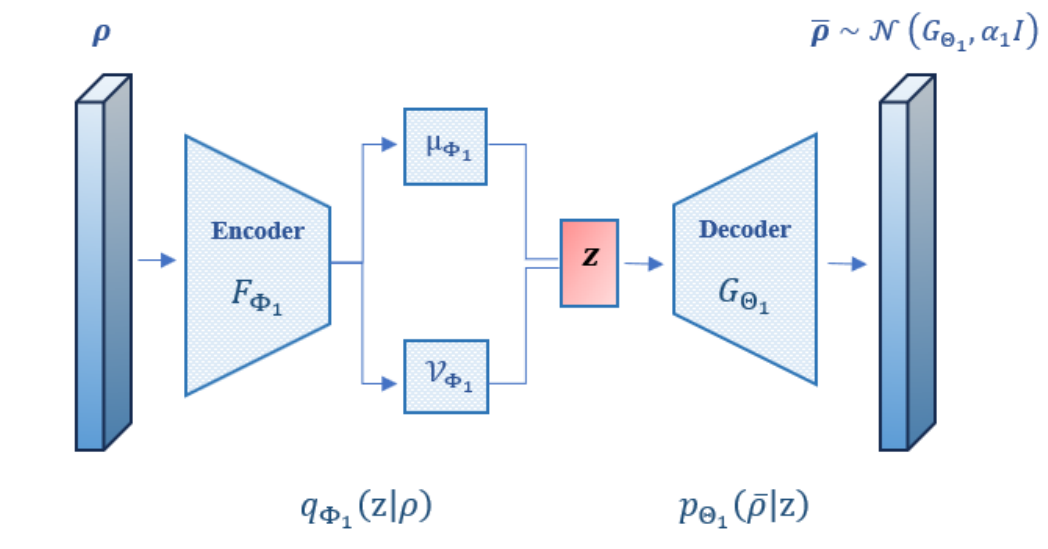}
				\caption{The traditional VAE network, as described at the beginning of this section, obtains the relation between variables $\rho $ and $z $ by minimizing the function $\mathrm{Loss_1} $ over  $\Phi_1, \Theta_1 $.} 
				\label{VAE_New train1}
			\end{subfigure}%
			
			\begin{subfigure}[b]{0.8\textwidth}
				\includegraphics[width=\linewidth]{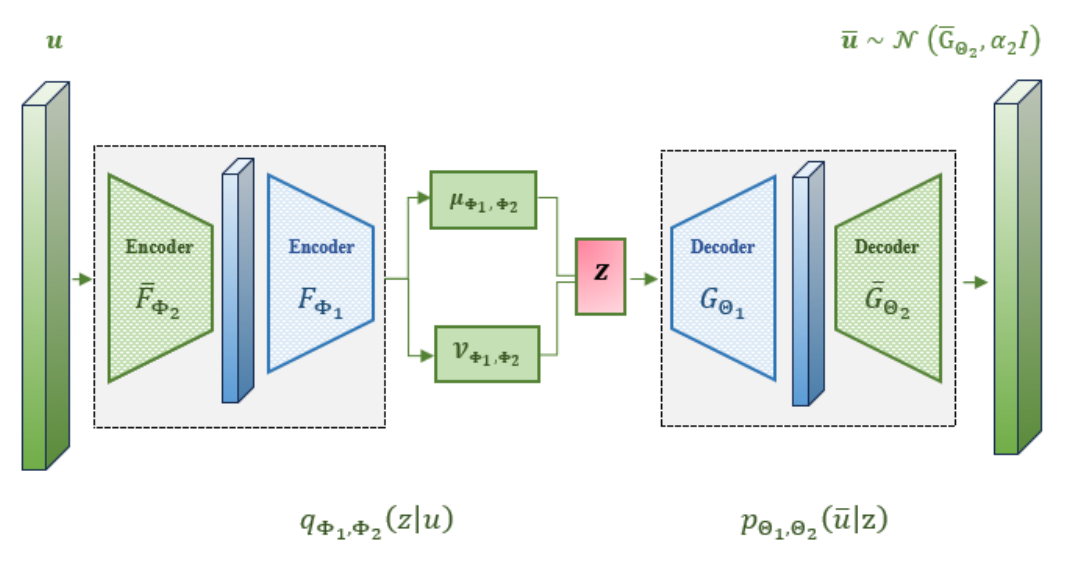}
				\caption{The VAE for finding the relation between variables $u$ and $z$. The blue part of this VAE is initialized with the optimized value from step (a). The optimal values of $ \Phi_1, \Phi_2, \Theta_1 $ and $ \Theta_2 $ are obtained by minimizing the function $\mathrm{Loss_2}$.}
				\label{VAE_New train2}
			\end{subfigure}%
			
			\caption{ New-VAE training steps. Each of steps (a) and (b) is optimized, respectively, using several iterations of an optimization algorithm.  The optimal value of each step is used as the initial value in the next step.}
			\label{VAE_New train}
		\end{figure}

		The structure of networks \ref{VAE_New train1} and  \ref{VAE_New train2} is illustrated in Figures \ref{VAE_structure1} and \ref{VAE_structure2},  respectively.
		
		The next consideration after understanding how New-VAE operates is how we can incorporate the physics of the problem into it. According to Figure \ref{VAE_New}, which shows the general structure of the third method, we  notice that it is possible to replace the process from steps 1 to 2 with a deterministic method that incorporates the physics of the problem.
		For example, in this paper, we utilized the Linear-to-Nonlinear method to obtain the final results. By doing this, $\bar{F}(\Phi_2) $ has been determined before the learning process.

		%%%%%%%%%%%%%%%%%%%%%%%%%%%%%%%%%%%%%%%%%%%%%%%%%%%%%%%%%%%%%%%%%%%%%%%%%%%%%%%%%%%%%%%%%%%%%%%%%%%%%%%%%%%%%%%%%%%%%%%%%%%%%%%%%%%%%%%%%%%%%%%%%%%%%%%%%%%%%%%%%%%%%%%%%%%%%%%%%%%%%%%%%%%%%%%%%%%%%%%%%%%%%%%%%%%%%%%%%%%%%%%%%%%%%%%%%%%%%%%%%%%%%%%%%%%%%%%%%%%%%%%%%%%%%%%%%%%%%%%%%%%%%%%%%%%%%%%%%%%%%%%%%%%%%%%%	

		\begin{figure}[H]
			\centering
			\begin{subfigure}[b]{1\textwidth}
				\includegraphics[width=\linewidth]{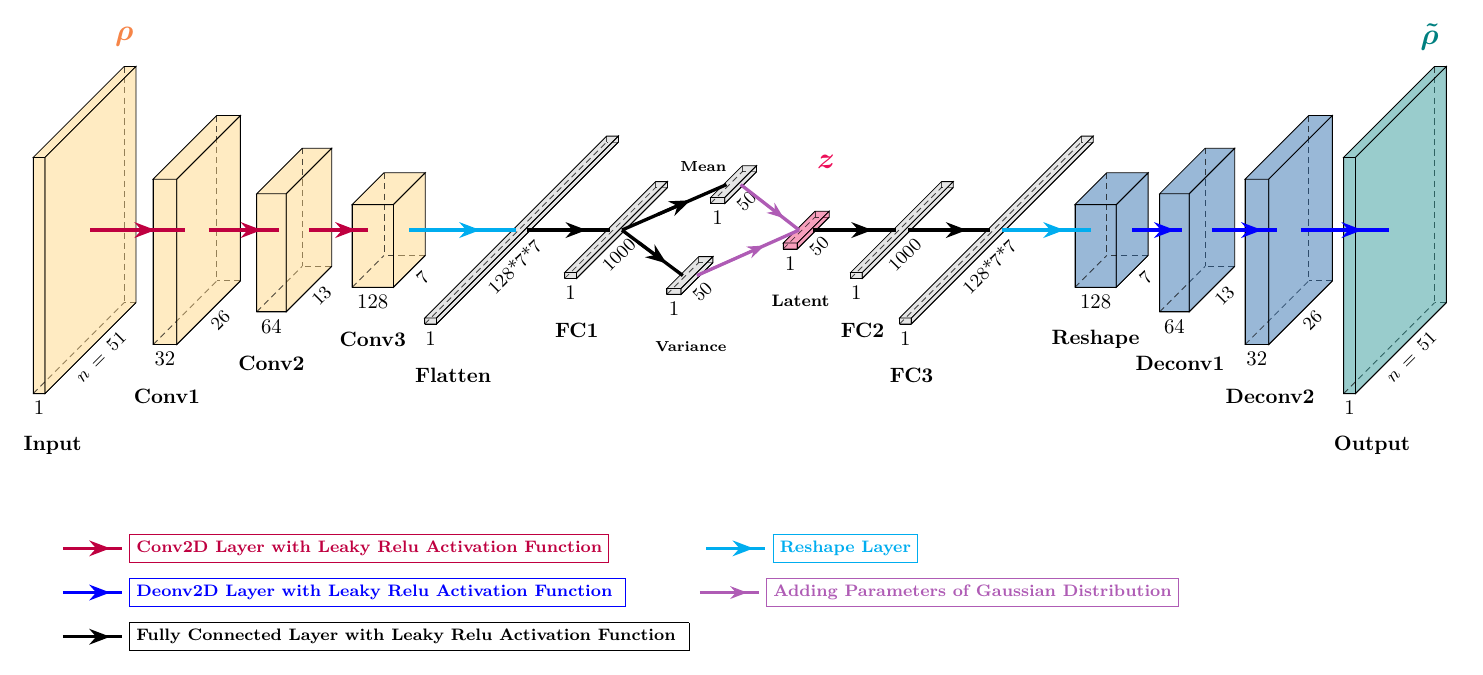}
				\caption{The VAE network structure related to Figure {\ref{VAE_New train1}} is shown, which shows the relation between variables $\rho$ and $z$.}
				\label{VAE_structure1}
			\end{subfigure}%
			\vspace{1 cm}
			\begin{subfigure}[]{1\textwidth}
				\includegraphics[width=\linewidth]{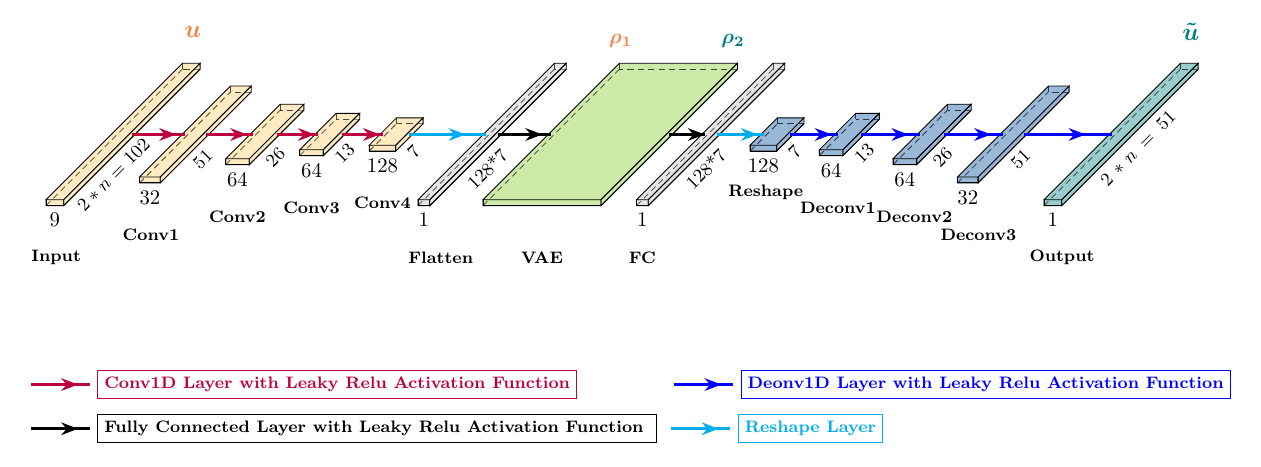}
				\caption{The VAE network structure related to Figure \ref{VAE_New train2} is shown, which shows the relation between variables $u$ and $z$.}
				\label{VAE_structure2}
			\end{subfigure}%
			
			\caption{The details of the networks are illustrated in Figure \ref{VAE_New train}. Please note that in Figure \ref{VAE_structure2}, the middle green square structure has the same structure as Figure \ref{VAE_structure1}. }
			\label{VAE_structre}
		\end{figure}

		%%%%%%%%%%%%%%%%%%%%%%%%%%%%%%%%%%%%%%%%%%%%%%%%%%%%%%%%
		%%%%%%%%%%%%%%%%%%%%%%%%%%%%%%%%%%%%%%%%%%%%%%%%%%%%%%%%
		%%%%%%%%%%%%%%%%%%%%%%%%%%%%%%%%%%%%%%%%%%%%%%%%%%%%%%%%
		%%%%%%%%%%%%%%%%%%%%%%%%%%%%%%%%%%%%%%%%%%%%%%%%%%%%%%%%
		\section{Fourth method ( Physics-Based )}\label{Fourth}
		
		The physics-based method involves minimizing the distance between the data predicted by the forward model and the recorded data over the parameter space. We formulate it as the minimization problem:
		\begin{equation}
			\begin{split}     	
				&\min _\rho \mathcal{J} \big[ U(\rho) \big] =\frac{1}{2} \sum_{j=1}^{k} \left\|RU_j-d_j^{obs}\right\|^2 ,
				\\
				\text{Subject to} \hspace{0.5 cm} &(U_j - U_j^{inc})_{2n^2 \times 1} = A_{2n^2 \times 2n^2} \,(\hat{\rho} U_j)_{2n^2 \times 1} , \hspace{0.5 cm} j=1,...,k
			\end{split}
			\label{Opt1}
		\end{equation} 
		Where $A$, $U_j$, $U^{\text{inc}}_j$, and $\hat{\rho}$ are defined as in relation {\ref{linear_system}}. The term $d_j^{\text{obs}}$ represents the displacement field observed on the surface. Additionally, $R$ is the mapping that projects the displacement fields $U_j$ in the domain $\Omega$ to their  surface locations.
		In this paper, we address the optimization problem {\ref{Opt1}} using two approaches: the Gradient Descent and the Truncated Newton method (with Gauss Approximation). We utilize the adjoint state technique to achieve this, as described in \mbox{\cite{metivier}}.

		We change the variables to make the complex-valued optimization problem  {\ref{Opt1}} to  a real-valued problem, which simplifies the problem.
		\begin{equation}
			U_j = 		
			\begin{bmatrix}
				\Re(U_j)            \\ 
				\Im(U_j)      
			\end{bmatrix}_{4n^2 \times 1}    ,\,\,\,
			U_j^{inc} = 		
			\begin{bmatrix}
				\Re(U_j^{inc})            \\ 
				\Im(U_j^{inc})    
			\end{bmatrix}_{4n^2 \times 1}   , \,\,\,
			A =    
			\begin{bmatrix}
				\Re(A) \,\,\, \, &-\Im(A)            \\ 
				\Im(A) \,\,\, \, &\Re(A)     
			\end{bmatrix}_{4n^2  \times 4n^2}  , \,\,\,
			\rho =   
			\begin{bmatrix}
				\hat{\rho} , \hat{\rho} 
			\end{bmatrix}_{1 \times 4n^2} 
			\label{New-def}
		\end{equation}
		Now, with the new definitions, optimization problem {\ref{Opt1}} is defined as follows:  
		
		\begin{subnumcases}{}
			\min _\rho\mathcal{J} \big[ U(\rho) \big] =\frac{1}{2} \sum_{j=1}^{k} \left\|RU_j-d_j^{obs}\right\|^2 , \hspace{1.5 cm} \label{Opt2}
			\\
			\text{Subject to} \hspace{0.5 cm}  \underbrace{ ( A.\rho - I )}_{A_{\rho}} U_j = -U_j^{inc} \hspace{0.5 cm} j=1,...,k
			\label{opt2}
		\end{subnumcases}    
		where "." represents the elementwise multiplication of the vector $\rho$ in the rows of the matrix $A$.
		
		\subsection{Gradient computation}  
		To compute the $ \nabla \mathcal{J}(\rho)$ in relation  {\ref{Opt2}} and {\ref{opt2}}, it is necessary to calculate $\frac{\partial U_j}{\partial\rho}$, which is computationally challenging. Therefore, we use the adjoint state method to obtain the gradient. For this purpose, we define a Lagrangian function. The Lagrangian function associated with the problem {\ref{Opt2}} and {\ref{opt2}} is given by: 
		\begin{equation}
			\mathcal{L}\big[\rho, U, \lambda\big] = \frac{1}{2} \sum_{j=1}^{k} \left\|RU_j-d_j^{inc}\right\|^2  + 
			\sum_{j=1}^{k} < A_{\rho}U_j+U_j^{inc}, \lambda_j >
		\end{equation}	
		where $<.\, , . >$ denotes the inner product and $\lambda_j $ represents the adjoint	vector.
		
		Note that $\mathcal{L}\big[\rho, U(\rho), \lambda\big] = \mathcal{J} \big[ U(\rho)\big]$  independent of $\lambda$ (adjoint vector), where $U=U(\rho)$ is the forward problem solution. Differentiating this expression will yield the desired gradient.

		\begin{equation}
			\nabla \mathcal{J}(\rho)= \dfrac{d}{d\rho}\mathcal{J} \big[ U(\rho)\big] = \sum_{j=1}^{k}\dfrac{\
				\partial \mathcal{L}}{\partial U_j}  \dfrac{\partial U_j}{\partial\rho}	+  \frac{\partial \mathcal{L} }{\partial\rho}
			\label{Gradient}
		\end{equation}	
		The Partial derivatives of $\mathcal{L}$ are computed as:    
		\begin{subnumcases}{}
			\dfrac{\partial \mathcal{L} }{\partial\rho}  = \sum_{j=1}^{k}  (A.U_j)^T \lambda_j  \label{Lag1} \\
			\dfrac{\partial \mathcal{L} }{\partial U_j}  =  R^T( RU_j - d_j^{obs})  + A_{\rho}^T \lambda_j \qquad j=1,...,k \label{Lag2} \\
			\dfrac{\partial \mathcal{L} }{\partial \lambda_j} =  A_{\rho}U_j+U_j^{inc} \qquad j=1,...,k \label{Lag3}
		\end{subnumcases}
		
		To compute the $ \nabla \mathcal{J}(\rho)$ as specified in relation {\ref{Gradient}}, we set $\lambda_j$ so that $\frac{\partial \mathcal{L}}{\partial U_j}$ equal to zero. The gradient can then be calculated using $ \frac{\partial \mathcal{L} }{\partial\rho}$. So, this process involves three steps. First, we solve the direct problem {\ref{opt2}} to obtain $U_j$ at point $\rho$. In the second step, with $U_j$ and $\rho$ known, by setting equation {\ref{Lag2}} equals zero, we obtain the values of $\lambda_j$. In the third step, the $ \nabla \mathcal{J}(\rho)$ is computed from equation {\ref{Lag1}}. Therefore, the Gradient Descent algorithm for problem {\ref{Opt2}} and {\ref{opt2}} can be implemented as described in Algorithm {\ref{alg5}}.

		\begin{algorithm}
			\caption{( Gradient Descent Algorithm )}
			\label{alg5}
			\hspace*{\algorithmicindent} \textbf{Input : } $\rho_0 , \epsilon_1 $ \vspace{0.1 cm} \\ 
			\hspace*{\algorithmicindent} \textbf{Output :} $\arg \displaystyle\min_{\rho} \mathcal{J} \big[ U(\rho) \big]$
			\begin{algorithmic}[1]
				\While{$\epsilon_1 \leq \mathcal{J} \big[ U(\rho) \big]  $}
				\vspace{0.1 cm}
				\State  Solve $A_{\rho}(U_j)=-U_j^{inc} \qquad j=1,...,k $ 	
				\vspace{0.1 cm}			
				\State Solve $   A_{\rho}^T (\lambda_j)  = -R^T( RU_j - d_j^{obs}) \qquad j=1,...,k$
				\vspace{0.1 cm}
				\State Compute $ \nabla \mathcal{J}(\rho) = \sum_{j=1}^{K} (A.U_j)^T \lambda_j$
				\vspace{0.1 cm}
				\EndWhile 
				\State Find $\gamma$ such that it satisfies the strong Wolfe conditions
				\vspace{0.1 cm}
				\State $\rho = \rho - \gamma \nabla \mathcal{J}(\rho)$ \\
				\vspace{0.1 cm}
				\Return  $\rho$
				
			\end{algorithmic}
		\end{algorithm}
		
		\begin{remark}	
			Note that the gradient vector $ \nabla \mathcal{J}(\rho)$ has a length of $4n^2$, which results from its definition in formula {\ref{New-def}}. To ensure it is well-defined and correct, after computation of $ \nabla \mathcal{J}(\rho)$  in step 4 of the Algorithm {\ref{alg5}}, it should be altered in the form of $ \nabla \mathcal{J}(\rho) = \nabla \mathcal{J}(\rho)[0:n^2] + \nabla \mathcal{J}(\rho)[n^2:2n^2] + \nabla \mathcal{J}(\rho)[2n^2:3n^2]+ \nabla \mathcal{J}(\rho)[3n^2:4n^2]$ . To simplify the notation we will not mention this point in the entire of this section.
		\end{remark}

		\subsection{The Gauss-Newton method for approximation Hessian} 
		Nonlinear minimization problems are typically solved using Newton-based methods. The Newton method involves building a sequence $\rho_t$ starting from an initial guess $\rho_0$, using the recurrence relation: 
		\begin{equation}
			\rho_{t+1} = \rho_t + \gamma_t \Delta \rho
			\label{sequence}
		\end{equation}		
		where $\gamma_t$ is a scalar parameter associated with a globalization method (linesearch) and the increment $ \Delta \rho$ satisfies:
		\begin{equation}
			H(\rho_t) \Delta \rho = -\nabla \mathcal{J}(\rho_t)
			\label{Hessi_Vector}
		\end{equation}	
		
		From the definition of $\mathcal{J} $, the Hessian operator $H(\rho)$ can be written as:
		\begin{equation}	
			H(\rho) \quad=\quad  \underbrace{\sum_{j=1}^{K} \dfrac{\partial U_j}{\partial \rho}^T R^T R \dfrac{\partial U_j}{\partial \rho}}_{H_1(\rho)} \quad+\quad
			\underbrace{\sum_{j=1}^{K} R^T(RU_j - d_j^{obs}) \dfrac{\partial^2 U_j}{\partial^2 \rho}}_{H_2(\rho)}
			\label{Hessi}
		\end{equation}	
		The matrix $H_1(\rho)$ is known as the Gauss-Newton approximation of the
		Hessian operator ,which is obtained by neglecting the $H_2(\rho)$ matrix.
		
		\subsection{Truncated Newton Method (with Gauss Approximation)}
		Unlike quasi-Newton algorithms, this method does not rely on an approximation of the inverse Hessian operator. Instead, the descent direction $\Delta \rho$ is computed as the solution to equation {\ref{Hessi_Vector}} using an iterative algorithm such as the conjugate gradient algorithm. It is important to note that in this paper, instead of considering equation {\ref{Hessi_Vector}}, we use its Gauss-Newton approximation and implement the Truncated Newton method on it as follows:
		\begin{equation}
			H_1(\rho_t) \Delta \rho = -\nabla \mathcal{J}(\rho_t)
			\label{Hessi_Vector2}
		\end{equation}	
		For this purpose, we consider the constrained minimization problem for an arbitrary $\{w_j \in \mathbb{R}^{4n^2} \,\, , \,\, j=1,...,k\}$:
		\begin{equation}
			\begin{split}      	
				& \min _\rho\mathcal{Q} \big[ U(\rho) \big] = \sum_{j=1}^{k} < U_j , w_j> , \hspace{1.5 cm}
				\\
				\text{Subject to} \hspace{0.5 cm} & A_{\rho} U_j = -U_j^{inc} \hspace{0.5 cm} j=1,...,k
			\end{split}
			\label{Opt3}
		\end{equation}	 
		The Lagrangian function associated with the problem {\ref{Opt3}} is:
		\begin{equation}
			\mathcal{L}_w\big[\rho, U, \mu\big] = \sum_{j=1}^{k} < U_j , w_j>  + 
			\sum_{j=1}^{k} < A_{\rho}U_j+U_j^{inc}, \mu_j >
		\end{equation}	
		
		Note that $\mathcal{L}_w\big[\rho, U(\rho), \mu\big] = \mathcal{Q} \big[ U(\rho)\big]  $  independent of $\mu$ (adjoint vector), where $U=U(\rho)$ is the forward problem solution. Differentiating this expression will yield the desired gradient:
		
		\begin{equation}
			\nabla \mathcal{Q}(\rho)= \dfrac{d}{d\rho}\mathcal{Q} \big[ U(\rho)\big] = \sum_{j=1}^{k}\dfrac{\
				\partial \mathcal{L}_w}{\partial U_j}  \dfrac{\partial U_j}{\partial\rho}	+  \frac{\partial \mathcal{L}_w }{\partial\rho}
			\label{}
		\end{equation}	
		The Partial derivatives  of $\mathcal{L}_w$ are computed as:     
		
		\begin{subnumcases}{}
			\dfrac{\partial \mathcal{L}_w }{\partial\rho}  = \sum_{j=1}^{k} (A.U_j)^T \mu_j  \label{Lag4} \\
			\dfrac{\partial \mathcal{L}_w }{\partial U_j}  =   A_{\rho}^T \mu_j  + w_j \qquad j=1,...,k \label{Lag5} \\
			\dfrac{\partial \mathcal{L}_w }{\partial \mu_j} =  A_{\rho}U_j+U_j^{inc} \qquad j=1,...,k  \label{Lag6}
		\end{subnumcases}

		Note that according to {\ref{Opt3}}, the relation $\nabla \mathcal{Q}(\rho) = \sum_{j=1}^{k} {\dfrac{\partial U_j}{\partial \rho}}^T w_j $ is resulted. By replacing $w_j$ with $ R^T R \dfrac{\partial U_j}{\partial \rho} v $:
		
		\begin{equation}
			\nabla \mathcal{Q}(\rho) = \sum_{j=1}^{k} {\dfrac{\partial U_j}{\partial \rho}}^T w_j = \sum_{j=1}^{k}\dfrac{\partial U_j}{\partial \rho}^T R^T R \dfrac{\partial U_j}{\partial \rho} v = H_1(\rho)v
			\label{}
		\end{equation}	       
		Therefore, it is sufficient to compute the $\dfrac{\partial U_j}{\partial \rho} v $ to obtain the $H_1(\rho)v$. To calculate $ \dfrac{\partial U_j}{\partial \rho}v$, we take the Gateaux derivative in the direction of $v$ from equation {\ref{opt2}}:
		
		\begin{equation}
			A_{\rho}\Big(\dfrac{\partial U_j}{\partial \rho} v\Big) = -(A.U_j)v \qquad j=1,...,k
			\label{Jacob_Vector}
		\end{equation}

		As a result, by solving equation {\ref{Jacob_Vector}}, we obtain $w_j$. Then, by setting $\mu_j$ such that $\frac{\partial \mathcal{L}_w}{\partial U_j} = 0$ ,  $\nabla \mathcal{Q}(\rho)$ can be calculated by $ \frac{\partial \mathcal{L}_w }{\partial\rho}$. With $H_1(\rho)v $ we can now obtain the desired path $\Delta(\rho) $ by solving equation  $H_1(\rho)v = -\nabla \mathcal{J}(\rho) $ using the conjugate gradient method. The general algorithm for this method is outlined in Algorithm {\ref{alg4}}.

		\begin{algorithm}[H]
			\caption{( Truncated Newton Algorithm with Gauss Approximation )}
			\label{alg4}
			\hspace*{\algorithmicindent} \textbf{Input : } $\rho_0 , \epsilon_1 , \epsilon_2$ \vspace{0.1 cm} \\ 
			\hspace*{\algorithmicindent} \textbf{Output :} $\arg \displaystyle\min_{\rho} \mathcal{J} \big[ U(\rho) \big]$
			\begin{algorithmic}[1]
				\While{$\epsilon_1 \leq \mathcal{J} \big[ U(\rho) \big]  $}
				\vspace{0.1 cm}
				\State  Solve $A_{\rho}(U_j)=-U_j^{inc} \qquad j=1,...,k $ 	
				\vspace{0.1 cm}			
				\State Solve $   A_{\rho}^T (\lambda_j)  = -R^T( RU_j - d_j^{obs}) \qquad j=1,...,k$
				\vspace{0.1 cm}
				\State Compute $ \nabla \mathcal{J}(\rho) = \sum_{j=1}^{k} (A.U_j)^T \lambda_j$
				\vspace{0.1 cm}
				\State $\Delta\rho =0 \,\,\,,\,\,\, r_0=\nabla \mathcal{J}(\rho) \,\,\,,\,\,\, v_0=-\nabla \mathcal{J}(\rho)\,\,\,,\,\,\, i=0$
				\vspace{0.1 cm}
				\While{$\epsilon_2 \leq ||r_i||^2  $}
				\vspace{0.1 cm}
				\State Solve $ A_{\rho}\Big(\frac{\partial U_j}{\partial \rho} v_i\Big) = -(A.U_j)v_i \qquad j=1,...,k$
				\vspace{0.1 cm}
				\State  Solve $A_{\rho}^T (\mu_j)  =  - R^T R \frac{\partial U_j}{\partial \rho} v_i \qquad j=1,...,k$
				\vspace{0.1 cm}
				\State Compute $H_1(\rho)v_{i} = \sum_{j=1}^{k} (A.U_j)^T\mu_j  $
				\vspace{0.1 cm}
				\State $\alpha_i = <r_{i},r_{i}> /<H_1(\rho)v_{i} , v_{i}> $
				\vspace{0.1 cm}
				\State $\Delta \rho = \Delta \rho + \alpha_i v_i$
				\vspace{0.1 cm}
				\State $r_{i+1} = r_{i} + \alpha_i H_1(\rho)v_i $
				\vspace{0.1 cm}
				\State $v_{i+1} = -r_{i+1} + (<r_{i+1},r_{i+1}>/ <r_{i},r_{i}>)v_{i} $
				\vspace{0.1 cm}
				\State $i = i+1$
				\EndWhile 
				\vspace{0.1 cm}
				\State Find $\gamma$ such that it satisfies the strong Wolfe conditions
				\vspace{0.1 cm}
				\State $\rho = \rho + \gamma \Delta \rho$
				\vspace{0.1 cm}
				\EndWhile \\
				\Return  $\rho$
				
			\end{algorithmic}
		\end{algorithm}

		\section{Dataset}\label{Dataset}
		\subsection{Generating random field}
		In a two-dimensional domain $\Omega \subset  \mathbb{R}^2 $, a random field $\big\{ \rho(\boldsymbol{x}) : \boldsymbol{x} \in \Omega \big\}$ consists of a collection of real-valued random variables defined on a probability space $(\mathcal{S} , \mathcal{F} , \mathcal{P}) $. In other words, for every $ \boldsymbol{x} \in \Omega , \, \rho(\boldsymbol{x}) : \mathcal{S} \rightarrow \mathbb{R}  $ represents a random variable. A particularly significant category within random fields comprises stationary random fields, characterized by constant mean values, finite variances, and covariance between random variable $\rho(\boldsymbol{x})$ and $\rho(\boldsymbol{y}) $, only dependent on the spatial displacement vector $  \boldsymbol{x} -  \boldsymbol{y} $.
		Our objective is to construct stationary random fields in such a way that the vector of random variables $ \boldsymbol{\rho} = [\rho(\boldsymbol{x}_1),\rho(\boldsymbol{x}_2), . . . ,\rho(\boldsymbol{x}_N)]^{\intercal} $ follows a multivariate Gaussian distribution for all $\boldsymbol{x}_1, . . . , \boldsymbol{x}_N \in \Omega$. In other words, $\boldsymbol{\rho} \sim \mathcal{N}(\mu,\Sigma)$, where the mean vector  $\mu $ and the covariance matrix $\Sigma $ are defined as $\mu_i = \mu(\boldsymbol{x}_i) $ and $\Sigma_{ij} = \Sigma(\boldsymbol{x}_i, \boldsymbol{x}_j), i, j = 1, . . . ,N$. By using Algorithm \ref{alg1}, we can construct Gaussian random fields with a mean vector $\mu$ and a covariance matrix $\Sigma$.

		\begin{algorithm}
			\caption{(Gaussian Random Field  Generator)}\label{alg1}
			\hspace*{\algorithmicindent} \textbf{Input : } Mean vector $\boldsymbol{\mu} =(\mu_1 , ... ,\mu_n)^{\intercal} $ and Covariance matrix $\Sigma $ \\
			\hspace*{\algorithmicindent} \textbf{Output :} $\boldsymbol{\rho}$ \qquad (Gaussian Random Field)
			\begin{algorithmic}[1]
				\State Find a square root $ H $ of $\Sigma $ , so that $\Sigma = HH^{\intercal} $
				\State Generate $Z_1, ... ,Z_n  \stackrel{iid}{\sim} \mathcal{N}(0,1) $ .
				\State  Let 
				$\boldsymbol{Z} = (Z_1, ... ,Z_n)^{\intercal} $ 
				\State $ \boldsymbol{\rho} = \boldsymbol{\mu} + H\boldsymbol{Z}$ \\
				\Return $\boldsymbol{\rho}$ 
				
			\end{algorithmic}
		\end{algorithm}

		To analyze Algorithm \ref{alg1} in detail, it should be noted that since the covariance matrix is symmetric and positive semi-definite, it is always possible to find a real-valued lower triangular matrix $ H$ using Cholesky's square-root method such that  $\Sigma = HH^{\intercal} $. The advantage of algorithm \ref{alg1} is that it can efficiently generate numerous realizations in real-time when provided with a covariance matrix and its Cholesky decomposition. However, it should be noted that Cholesky decomposition requires $\mathcal{O}(N^3) $ floating-point operations, which can be a limiting factor.

		In this study, to create stationary random fields, we focus on the case where $\mu =0 $ and the covariance is as follows:
		\begin{equation}\label{cov}
			\Sigma(\boldsymbol{x},\boldsymbol{y})  =  \exp \bigg(  -(\dfrac{x_1-y_1}{a_1})^2 - (\dfrac{x_2-y_2}{a_2})^2  \bigg) \qquad \boldsymbol{x},\boldsymbol{y} \in \Omega 
		\end{equation}
		
		Although the matrices obtained from equation \ref{cov} are symmetric, since they are dependent on their parameters $[a_1,a_2]$, these matrices are not necessarily semi-positive definite, and it is impossible to decompose them using  Cholsky.  So, they cannot be obtained from algorithm \ref{alg1}. As a result, by adding a constant of identity matrix to these matrices, we make them positive definite. 
		There are alternative methods to create random stationary fields. By changing the covariance structure and with special tools of linear algebra \cite{kroese}, these fields can be optimally produced with a much lower computational cost. Since it is not the main topic of our paper, we do not employ these strategies.

		To produce random fields for the dataset, we set domain $\Omega = [-1,1]\times [-1,1] $ and divide $\Omega $ to $N = n\times n$ uniform rectangular elements. Using observation and experimentation, we create 28 pairs of $(a_1,a_2) $ values in such a way that they cover nearly all situations of earth density. For each pair of $(a_1,a_2) $ we have a covariance matrix as \ref{cov}, and construct 1000 realizations from it. In Figure \ref{random_field}, realizations obtained from specific pairs of $(a_1,a_2) $ are observable.
		
		\begin{figure}[H]
			\centering
			\begin{subfigure}[b]{0.2\textwidth}
				\includegraphics[width=\linewidth]{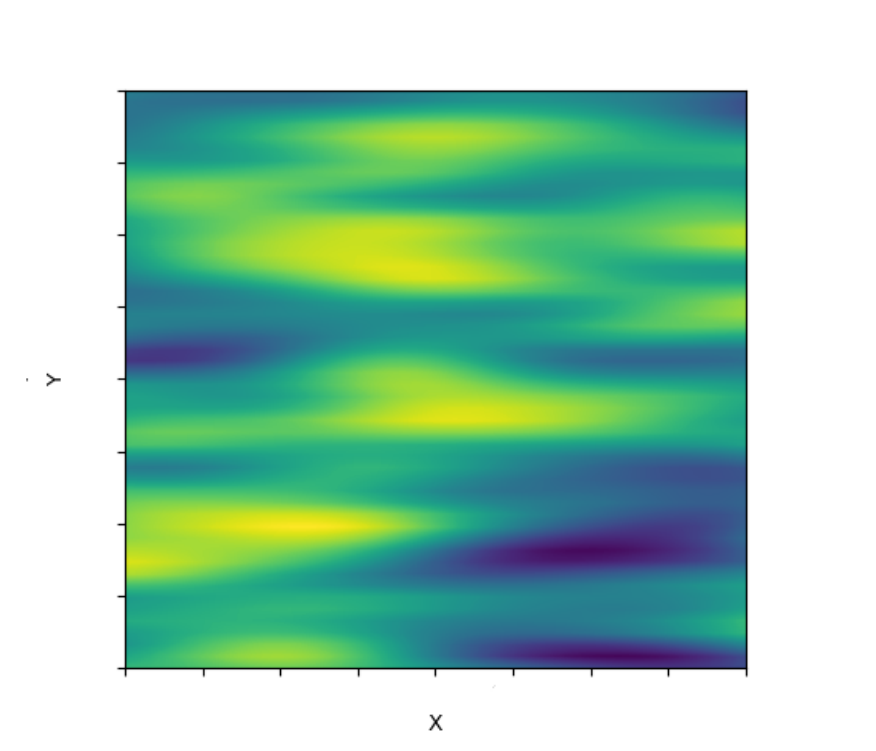}
				\caption{[2,3]}
			\end{subfigure}%
			\begin{subfigure}[b]{0.2\textwidth}
				\includegraphics[width=\linewidth]{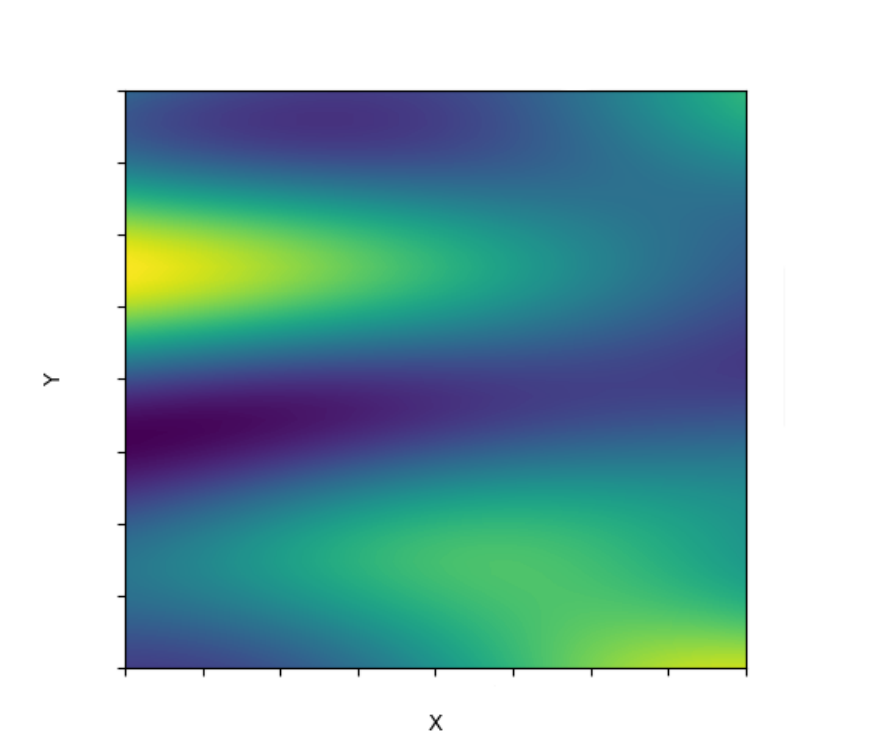}
				\caption{[4,6]}
			\end{subfigure}%
			\begin{subfigure}[b]{0.2\textwidth}
				\includegraphics[width=\linewidth]{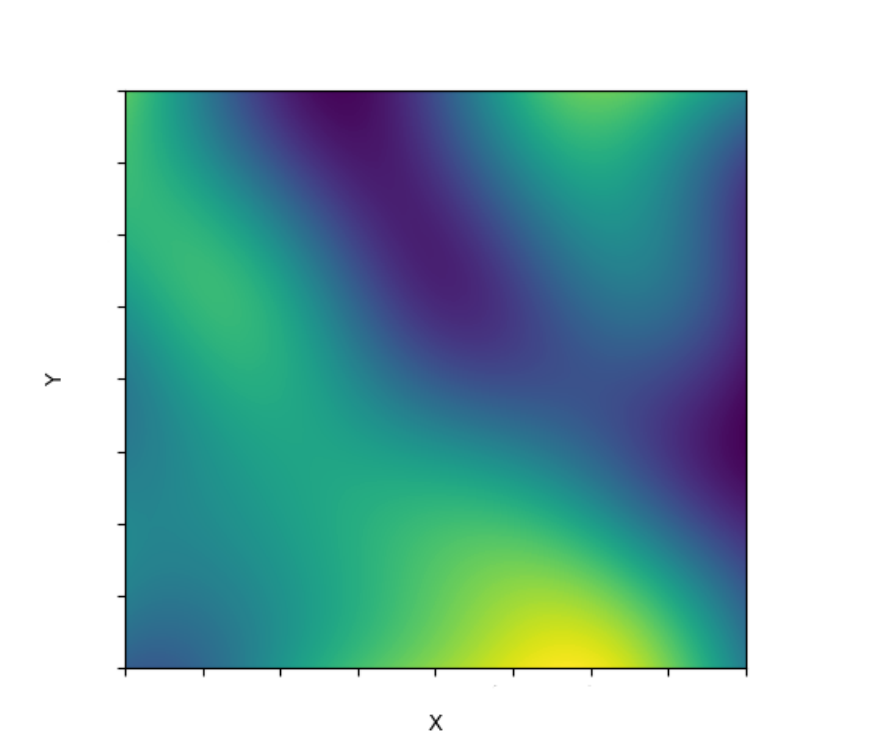}
				\caption{[7,12]}
			\end{subfigure}%
			\begin{subfigure}[b]{0.2\textwidth}
				\includegraphics[width=\linewidth]{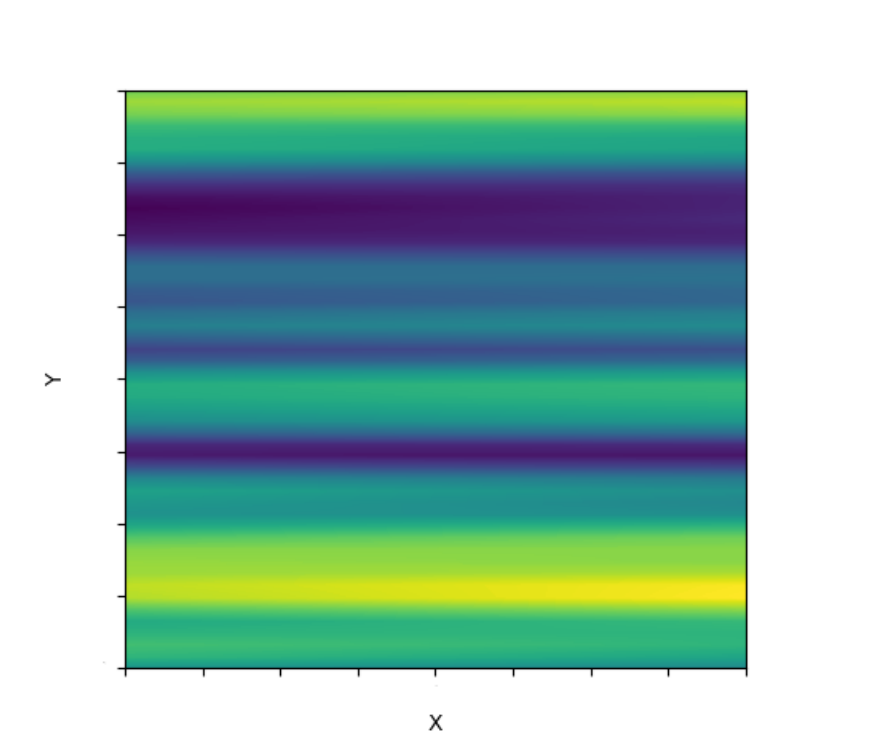}
				\caption{[20,3]}
			\end{subfigure}
			\begin{subfigure}[b]{0.2\textwidth}
				\includegraphics[width=\linewidth]{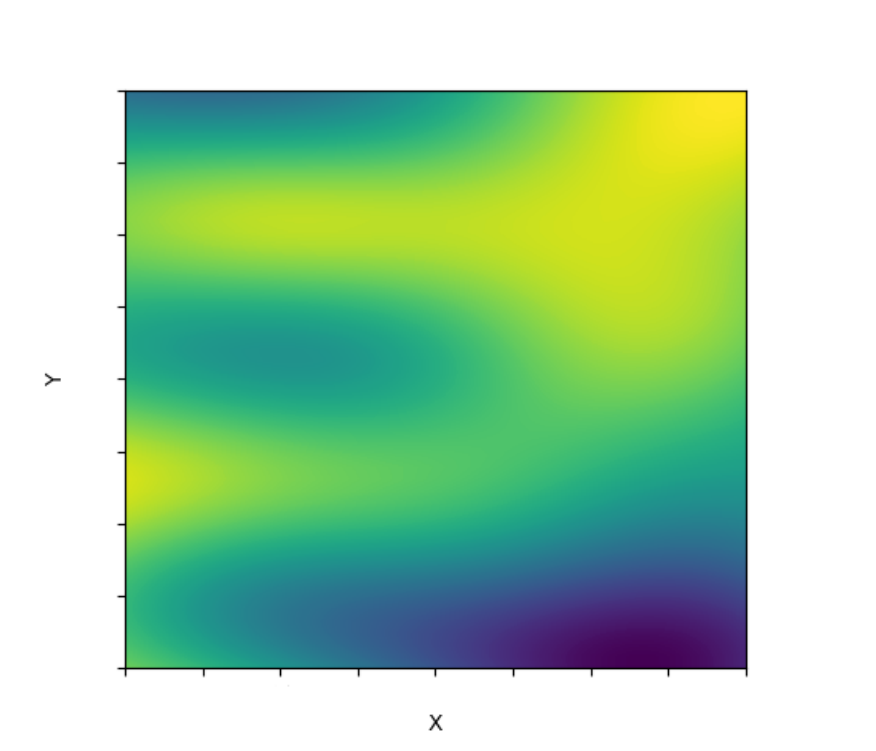}
				\caption{[5,9]}
			\end{subfigure}%
			\begin{subfigure}[b]{0.2\textwidth}
				\includegraphics[width=\linewidth]{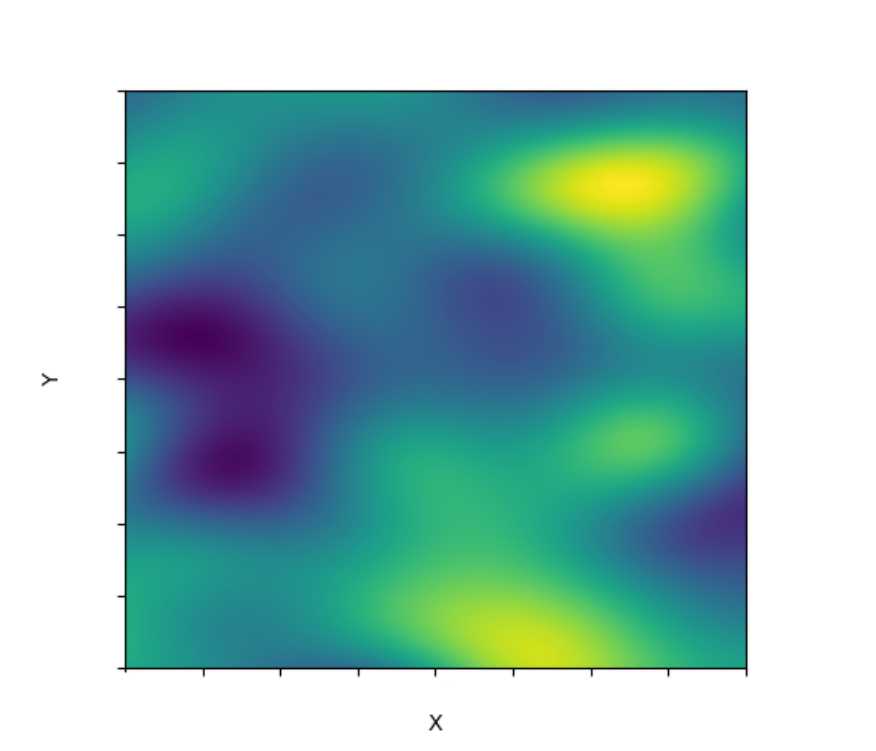}
				\caption{[20,9]}
			\end{subfigure}%
			\begin{subfigure}[b]{0.2\textwidth}
				\includegraphics[width=\linewidth]{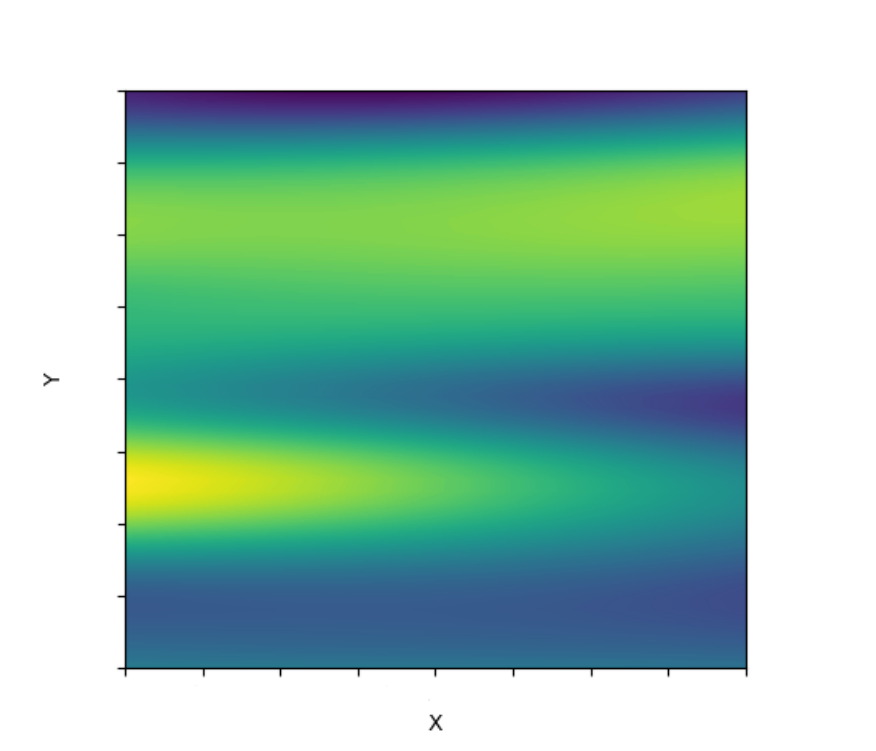}
				\caption{[26,1]}
			\end{subfigure}%
			\begin{subfigure}[b]{0.2\textwidth}
				\includegraphics[width=\linewidth]{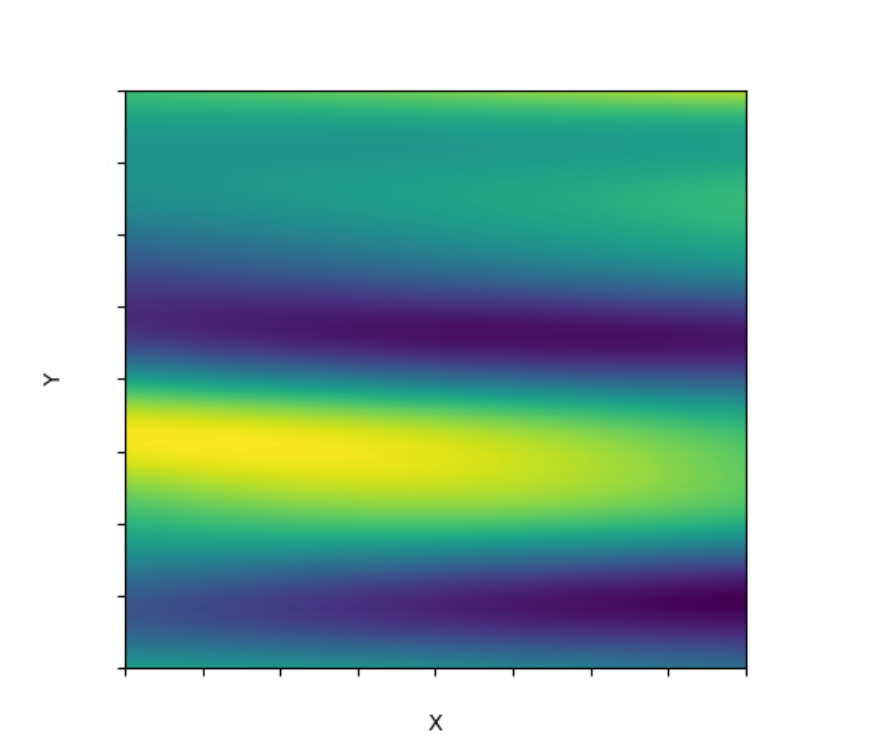}
				\caption{[7,8]}
			\end{subfigure}
			\caption{Random fields with some $[a_1,a_2] $ }
			\label{random_field}
		\end{figure}
		
		\subsubsection{Fault model}
		In geology, a fault is a crack or break in rock where there has been a noticeable shift due to movements in the earth's crust.
		In order to make our model closer to reality, we  add faults in our random fields.
		To achieve this, we first build fields that are twice as large as the prior random fields, which means we produce $2n\times 2n$ fields. Then, we select a random point $x_0 $ from the central $n\times n $ square  field and one slope m randomly.
		Using parameters $m$ and $x_0$, we can define $l_1$ as a line. Additionally, let us consider $ l_2$ as another horizontal line passing through point $x_0$. These two lines partition the field into four distinct sections. Now, randomly displace one of these quadrants along $l_1$ by a random distance, $h$.

		We can repeat this process multiple times to create faults at different angles and positions throughout our random fields. Finally, we choose this new central square  as the final field. An example of this procedure is depicted in Figure \ref{fracture1} and some random fields with faults in Figure \ref{fracture2}.

		\begin{figure}[H]
			\begin{subfigure}[b]{0.25\textwidth}
				\includegraphics[width=\linewidth]{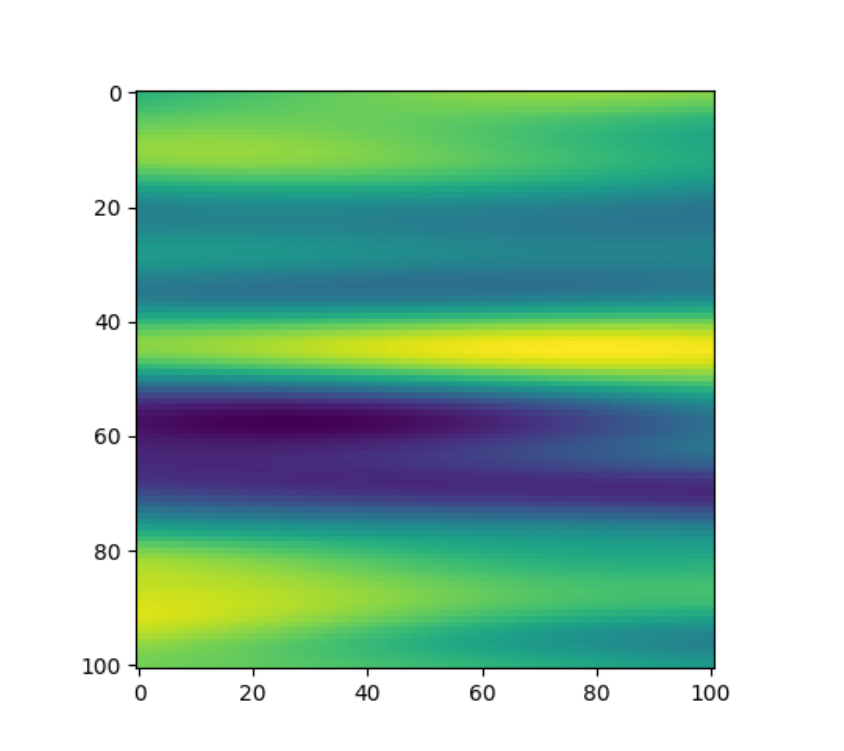}
				\caption{Initial $2n\times 2n $ random field}
			\end{subfigure}%
			\begin{subfigure}[b]{0.25\textwidth}
				\includegraphics[width=\linewidth]{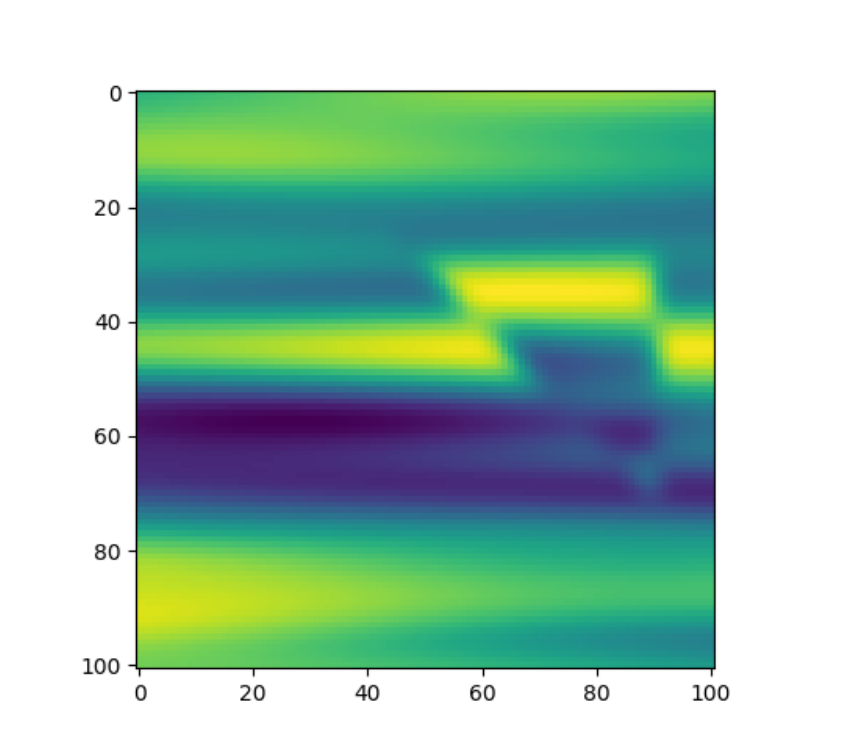}
				\caption{First fault}
			\end{subfigure}%
			\begin{subfigure}[b]{0.25\textwidth}
				\includegraphics[width=\linewidth]{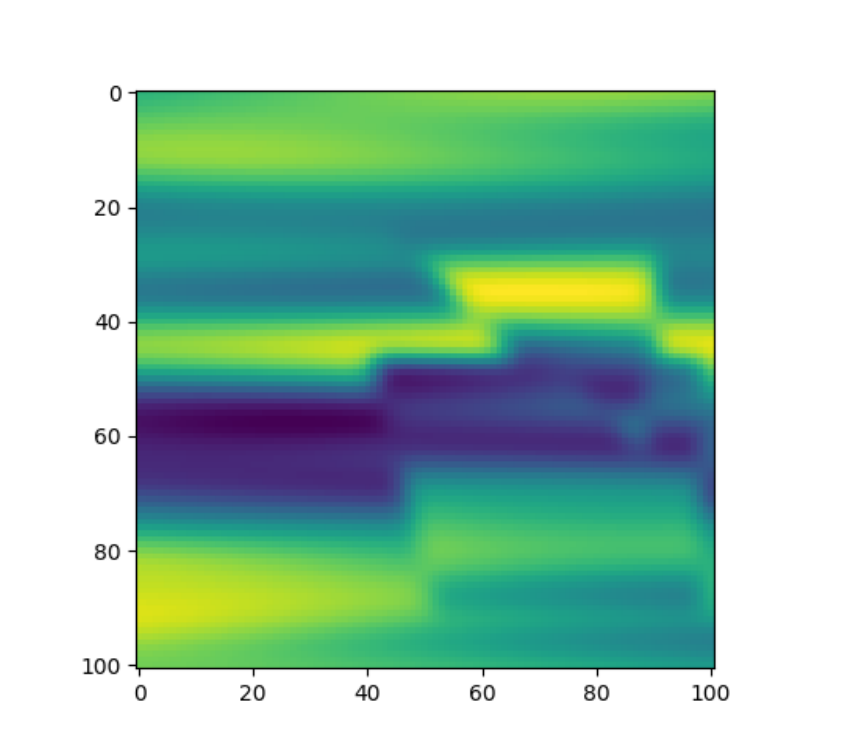}
				\caption{Second fault}
			\end{subfigure}%
			\begin{subfigure}[b]{0.25\textwidth}
				\includegraphics[width=\linewidth]{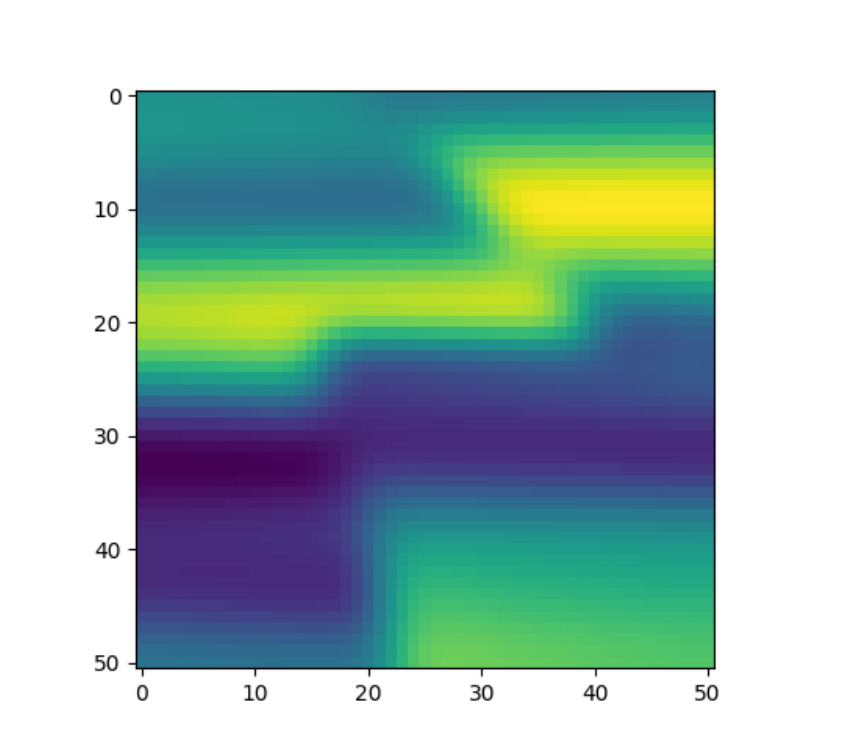}
				\caption{Final $n\times n $ random field }
			\end{subfigure}
			
			\caption{The process of creating faults in the random field }
			\label{fracture1}
		\end{figure}

		\begin{figure}[H]
			\centering
			\begin{subfigure}[b]{0.25\textwidth}
				\includegraphics[width=\linewidth]{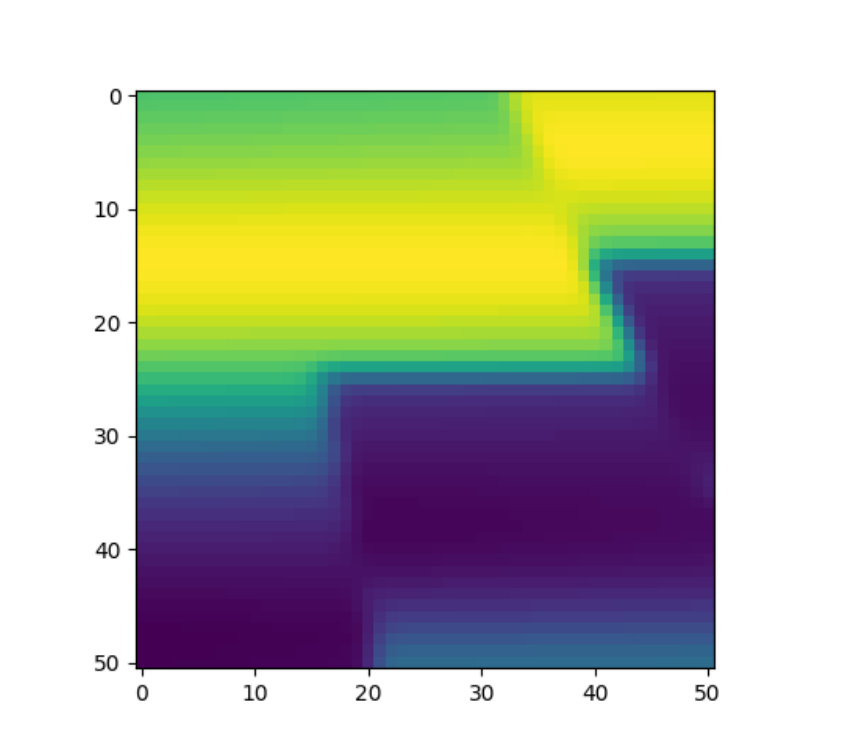}
			\end{subfigure}%
			\begin{subfigure}[b]{0.25\textwidth}
				\includegraphics[width=\linewidth]{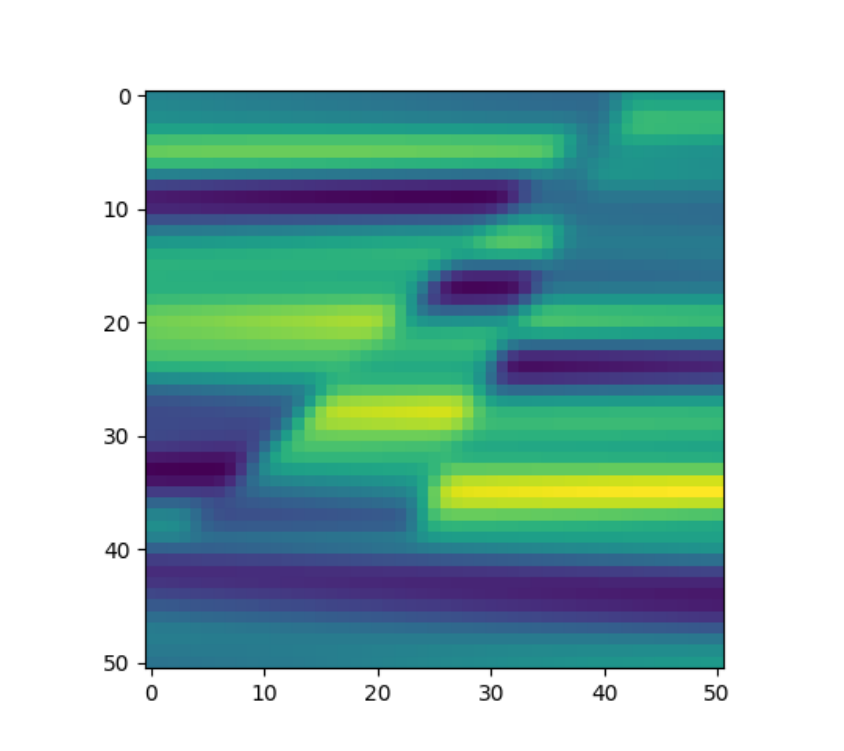}
			\end{subfigure}%
			\begin{subfigure}[b]{0.25\textwidth}
				\includegraphics[width=\linewidth]{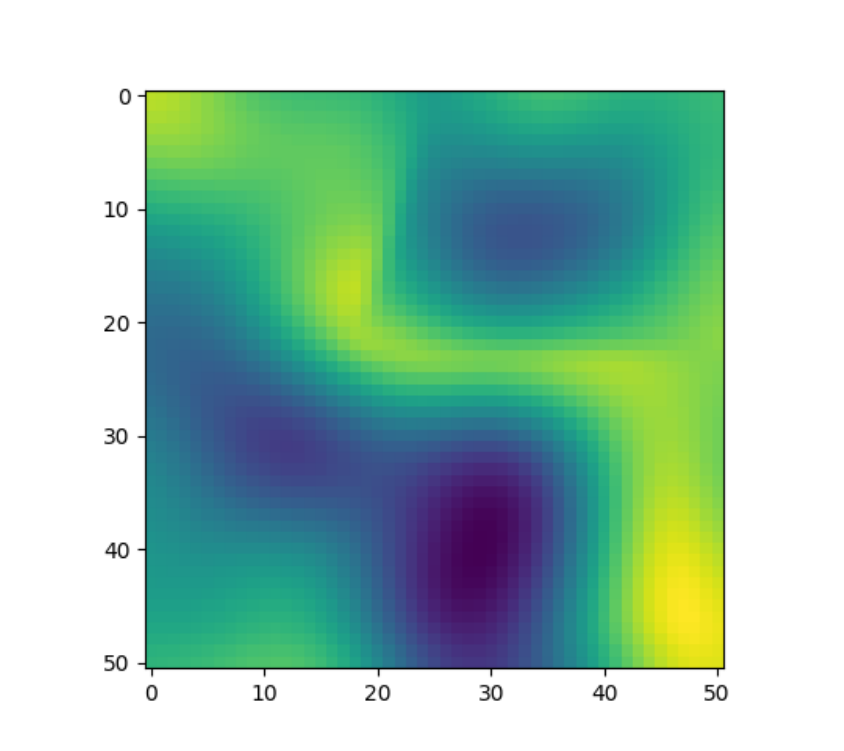}
			\end{subfigure}%
			\begin{subfigure}[b]{0.25\textwidth}
				\includegraphics[width=\linewidth]{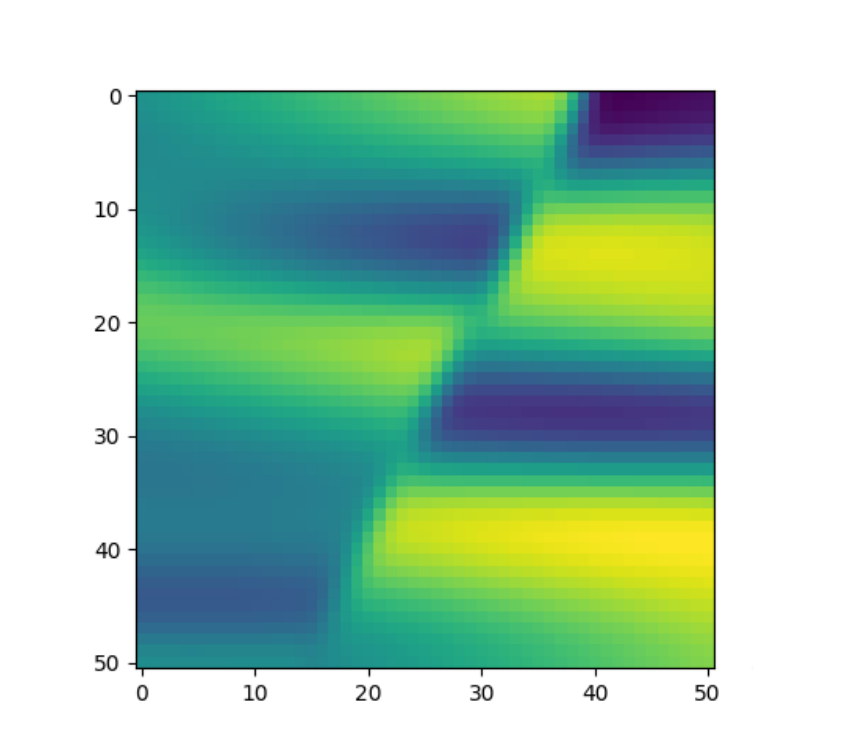}
			\end{subfigure}%
			
			\caption{Some faults in random fields }
			\label{fracture2}
		\end{figure}

		%%%%%%%%%%%%%%%%%%%%%%%%%%%%%%%%%%%%%%%%%%%%%%%%%%%%%%%%%%%%%%%%%%%%%%%%%%%%%%%%%%%%%%%%%%%%%%%%%%%%%%%%%%%%%%%%%%%%%%%%%%%%%%%%%%%%%%%%%%%%%%%%%%%%%%%%%%%%%%

		\subsection{Displacement field}
		We have generated random fields up to now. Our next objective is to derive the displacement field using these random fields and other relevant physical parameters.
		
		First, we determine the physical parameters of the problem in such a way that  they are close to the reality.
		In section \ref{Function setting}, we have assumed that the density of the homogeneous region equals 1. For instance, let us take this region as water, which has an approximate density of $ 1 gr /{cm}^3 $.
		Then, we adjust the range of densities to be within two reasonable values that we believe the density of the earth should be. So, we shift the density of random fields to  $[ 2 , 12] gr /{cm}^3$.
		
		The Lame parameters  $\lambda $ and $\mu$ depend on the density, but in this paper, we assume that the two parameters are equal constants and their values are  $10^{3} \, Gpa$. The circular frequency is equal to $ \omega=10$, and incident waves are made according to the formula \ref{incident}.

		We have a displacement field for each subsurface density and incident wave. To improve our ability to solve the inverse problem, we utilize $k$ directions $\theta$, resulting in $k$ incident waves for each density.
		We select $k=9$ directions from the distribution below, and from these selections, we obtain the corresponding displacement field $\boldsymbol{u}$.
		$$ \theta  \stackrel{iid}{\sim} Unif(\dfrac{-\pi}{2} , \dfrac{\pi}{2} ) $$
		
		Once we have defined these parameters, we can use the Lippmann-Schwinger integral equation  \ref{Lippman} to derive the direct solution to the elastic problem.

		To get $\boldsymbol{u}=(u_1(x),u_2(x)) $ by discretizing integral equation \ref{Lippman}, we get the following linear equation:
		\begin{equation}
			U - U^{inc} = A U 
		\end{equation}
		where $U = [u_1(x_1),...,u_1(x_N),\, u_2(x_1),...,u_2(x_N)]^{\intercal} $ , 
		$U^{inc} = [u^{inc}_1(x_1),...,u^{inc}_1(x_N),\, u^{inc}_2(x_1),...,u^{inc}_2(x_N)]^{\intercal} $ and the arrays of the matrix A are:
		\begin{align}
			A(i,j) &=
			\begin{cases}
				-\omega^2 \big(1-\rho(j)\big) \int_{D_{j}} G_{1,1}(x_i,x_j,\omega)dx_j     
				& \text{if } i \leq N \quad\text{and}\quad j \leq N \\
				-\omega^2 \big(1-\rho(j - N)\big)  \int_{D_{j}}G_{1,2}(x_i,x_j,\omega)dx_j   
				& \text{if } i \leq N \quad\text{and}\quad j  >   N \\
				-\omega^2 \big(1-\rho(j)\big)  \int_{D_{j}} G_{2,1}(x_i,x_j,\omega)dx_j              
				& \text{if } i > N    \quad\text{and}\quad j \leq N \\
				-\omega^2 \big(1-\rho(j - N)\big)  \int_{D_{j}} G_{2,2}(x_i,x_j,\omega)dx_j    
				& \text{if } i > N    \quad\text{and}\quad j  >   N \\
			\end{cases}
		\end{align}
		where $G_{i,j} $ are arrays of the matrix $\boldsymbol{G} $ and $D_j $ is the area related to variable $x_j $.
		
		To obtain $U$, we simply need to solve equation $ (I - A)_{2N\times 2N}U_{2N\times 1} =  U^{inc}_{2N\times 1} $. To expedite this process, considering that we have $k=9$ incident waves for each subsurface density, we solve this equation in batches. In our experiment, each batch consists of $k=9$ equations, and we have leveraged GPU and the TensorFlow library for batch-solving these equations.
		
		%%%%%%%%%%%%%%%%%%%%%%%%%%%%%%%%%%%%%%%%%%%%%%%%%%%%%%%%%%%%%%%%%%%%%%%%%%%%%%%%%%%%%%%%%%%%%%%%%%%%%%%%%%%%%%%%%%%%%%%%%%%%%%%%%%%%%%%%%%%%%%
		\section{Computational cost}\label{Complexity}
		In this section, we assess the computational cost of the presented methods. For simplicity, we only focus on the primary operations within each method.
		Under this simplifying assumption, fully connected layers dominate convolution layers, activation functions, concatenation, and summation during  forward and backward propagation processes. Furthermore, we denote the epochs in the networks as $t$, the size of the training set as $m$, the count of incident waves as $k$, and the features of each sample as $N=n^2$.

		\begin{enumerate}
			\item 
			In the first method, \textbf{ Direct Deep Learning Inversion}, the computational cost would be $\mathcal{O} (tm N\sqrt{N}k)$,  according to Figure \ref{Dens_Cnn}.
			\item 
			In the second approach, we initially train the displacement field for the real and imaginary components, as illustrated in Figure \ref{Disp_Cnn}. The computational cost of this training process is $\mathcal{O}(tmN\sqrt{N}k^2)$.
			
			\begin{itemize}
				\item 
				In the \textbf{Least Squares} technique within this method,
				the final system is $B_{(2kN\times N)}\rho = b$, and since the computational cost of the least squares method  for this system is $\mathcal{O}(k^2N^3)$, the total computational cost is $\mathcal{O}(k^2N^3 + tm N\sqrt{N}k^2) $.

				\item
				In the  \textbf{Inverse Convolution} technique within this method, we incur a computational cost of $\mathcal{O}(N^2)$ to train the matrix $\tilde{G} $ based on algorithm \ref{alg2} and $\mathcal{O}(tmkNd_1)$ 
				for the U-net as shown in Figure \ref{Unet} (where in this paper, $d_1=1000$). Consequently, the total computational cost would be $\mathcal{O}(N^2 + tmkNd_1 + tm N\sqrt{N}k^2 )$.		
				\item
				In the  \textbf{Linear-to-Nonlinear} technique within this method, we utilize two networks. The first network, as depicted in Figure \ref{rhou_Cnn}, incurs a computational cost of $\mathcal{O}(tmk^2 N^2)$. On the other hand, the second network, Unet, whose structure is depicted in Figure \ref{Unet}, requires $\mathcal{O}(tmkNd_1)$.
				Consequently, the total computational cost is $\mathcal{O}(tmk^2 N^2 + tmkNd_1 + tm N\sqrt{N}k^2) = \mathcal{O}(tmk^2 N^2)$.
				
			\end{itemize}
			
			\item 	
			The third method, the \textbf{New-VAE} technique, involves an internal network, depicted in Figure \ref{VAE_structure1}, with a computational cost of $\mathcal{O}(tmNd_2)$ (where in this paper, $d_2=500$), and an external network with a cost of $\mathcal{O}(tmkN\sqrt{N})$ as shown in Figure \ref{VAE_structure2}. Consequently, the total computational cost is $\mathcal{O}(tmNd_2 + tmkN\sqrt{N}) = \mathcal{O}(tmkN\sqrt{N})$.
			
			\item 
			In the fourth approach, physics-based techniques, we should solve two linear systems (the direct problem) for the main computation of gradient. This process has a computational cost of $\mathcal{O}(2kN^3)$.
			\begin{itemize}
				\item 
				In the \textbf{Gradient Descent} technique, If we denote the number of iterations by $t_1$, the computational cost is $\mathcal{O}(2t_1kN^3)$.
				\item
				In the \textbf{Truncated Newton (Gauss Approximation)} technique, let $t_2$ represent the number of iterations of the inner loop in Algorithm {\ref{alg4}}. Since two linear systems are solved in each iteration of this loop, the computational cost is $\mathcal{O}(2t_1kN^3 + 2t_1t_2kN^3)$.
			\end{itemize}	
		\end{enumerate}	 
		
		It is important to mention that the computations are processed simultaneously across numerous cores, substantially decreasing computing costs. The training procedure for all networks and the creation of the dataset are conducted using a single Nvidia GTX Titan X GPU.

		%%%%%%%%%%%%%%%%%%%%%%%%%%%%%%%%%%%%%%%%%%%%%%%%%%%%%%%%%%%%%%%%%%%%%%%%%%%%%%%%%%%%%%%%%%%%%%%%%%%%%%%%%%%%%%%%%%%%%%%%%%%%%%%%%%%%%%%%%%%%%%
		\section{Results}\label{Results}	
		By dividing the dataset into training and test, each comprising 27000 and 1000 samples, respectively, the numerical and visual results of the proposed methods are given for both sets. Table \ref{table1} presents the accuracy of data-driven methods to estimate the density function with the performance of auxiliary networks.  Table {\ref{table3}} also shows the accuracy of the physics-based method with two different initial guesses: the first guess is a constant function with a value equal to the mean of the original density, and the second initial guess is nearly close to the grand truth. In Table {\ref{table1}} and {\ref{table3}}, $\Omega$ represents the domain containing the points with the given coordinates, $\partial \Omega$ refers to the surface location and $K$ denotes a set of scattered waves from each incident wave. It is important to note that since the values of $\boldsymbol{u}$ are complex numbers, their real and imaginary parts were trained separately, and the accuracy of the absolute value of them are presented in Table \ref{table1} and \ref{table3}. Moreover, the parameter of $L_2$ regularization in the Least Squares method equals 0.02. Table \ref{table2} exhibits the values of loss functions derived from relation \ref{VAE_new_kl} in the third method when the parameters have been trained. Notably, the relative $L_2$ error formulas in Table \ref{table1}, Table \ref{table3} and Table \ref{table2} actually represent the average of relative $L_2$ errors calculated over the entire training and test datasets. 
		Additionally, the value of the loss function mentioned in Algorithm \ref{alg2} equals 15.

		\begin{table}[H]
			\caption{Relative $L_2$ error of different methods }
			\label{table1}
			\centering
			\setlength{\tabcolsep}{8pt} % Default value: 6pt
			\renewcommand{\arraystretch}{1.5} % Default value: 1
			\begin{tabular}{llcccccc}
				\hline 
				\hline
				\multicolumn{2}{l}{\multirow{2}{*}{}}    & \multicolumn{2}{c}{$\boldsymbol{ \dfrac{ ||\rho u- \overline{\rho u}||_{L^2(\Omega\times K)}}{|| \rho u||_{L^2(\Omega\times K)}}}$}    & \multicolumn{2}{c}{$\boldsymbol{ \dfrac{ || u- \overline{u}||_{L^2(\Omega\times K)}}{||  u||_{L^2(\Omega\times K)}}}$}                                      & \multicolumn{2}{c}{$\boldsymbol{ \dfrac{ ||\rho - \overline{\rho }||_{L^2(\Omega)}}{|| \rho ||_{L^2(\Omega)}}}$}    \\ \cmidrule(lr){3-4} \cmidrule(lr){5-6}  \cmidrule(lr){7-8}

				\multicolumn{2}{c}{}                     & \multicolumn{1}{c}{Training} & Test & \multicolumn{1}{c}{Training}                      &          Test             & \multicolumn{1}{c}{Training} &  Test 
				\\

				\multicolumn{2}{l}{\textbf{First Method (Direct)}}           & \multicolumn{1}{c}{-} & - & \multicolumn{1}{c}{-}    &    -   & \multicolumn{1}{c}{0.035} & 0.036  \\ 
				\cline{1-2}

				\multicolumn{1}{l}{\multirow{3}{*}{\textbf{Second Method}}} & \textbf{Least Square} & \multicolumn{1}{c}{-} & - & \multicolumn{1}{c}{0.046} & 0.049 & \multicolumn{1}{c}{0.097} & 0.099  
				\\ \cline{2-2}
				
				\multicolumn{1}{l}{-}          &  \textbf{Inverse Convolution}& \multicolumn{1}{c}{-} & - & \multicolumn{1}{c}{0.046}            &        0.049          & \multicolumn{1}{c}{0.028} & 0.048  
				\\   \cline{2-2}
				
				\multicolumn{1}{c}{}                 & \textbf{Linear-to-Nonlinear} & \multicolumn{1}{c}{0.053} & 0.058 & \multicolumn{1}{c}{0.046}                  &        0.049           & \multicolumn{1}{c}{0.023} &  0.028  
				\\ \cline{1-2}

				\multicolumn{2}{l}{\textbf{Third Method (New-VAE)}}      & \multicolumn{1}{c}{-} & - & \multicolumn{1}{c}{-}              &       -       & \multicolumn{1}{c}{0.026}  & 0.030  \\ \hline \hline
			\end{tabular}
		\end{table}
		%%%%%%%%%%%%%%%%%%%%%%%%%%%%%%%%%%%%%%%%%%%%%%%%%%%%%%%%%%%%%%%%%%%%%%%%%%%%%%%%%%%%%%%%%%%%%%
		%%%%%%%%%%%%%%%%%%%%%%%%%%%%%%%%%%%%%%%%%%%%%%%%%%%%%%%%%%%%%%%%%%%%%%%%%%%%%%%%%%%%%%%%%%%%%%
		\begin{table}[H]
			\caption{ Errors presented in the fourth method }
			\label{table3}
			\centering
			\setlength{\tabcolsep}{8pt} % Default value: 6pt
			\renewcommand{\arraystretch}{1.5} % Default value: 1
			\begin{tabular}{llcccccc}
				\hline 
				\hline
				\multicolumn{2}{l}{\multirow{2}{*}{}}    & \multicolumn{2}{c}{$\boldsymbol{ \dfrac{ || u- \overline{u}||_{L^2(\partial\Omega\times K)}}{||  u||_{L^2(\partial\Omega\times K)}}}$}    & \multicolumn{2}{c}{$\boldsymbol{ \dfrac{ ||\rho - \overline{\rho }||_{L^2(\Omega)}}{|| \rho ||_{L^2(\Omega)}}}$}                                      & \multicolumn{2}{c}{$ \boldsymbol{||\nabla \mathcal{J}(\rho)||_{L^2(\Omega)}}$}    \\ \cmidrule(lr){3-4} \cmidrule(lr){5-6}  \cmidrule(lr){7-8}

				\multicolumn{2}{c}{}  & \multicolumn{1}{c}{Initial 1} & Initial 2 & \multicolumn{1}{c}{Initial 1}                      &          Initial 2             & \multicolumn{1}{c}{Initial 1} &  Initial 2 
				\\

				\multicolumn{1}{c}{}     & \textbf{Gradient Descent} & \multicolumn{1}{c}{0.011} & 0.004 & \multicolumn{1}{c}{0.060}                  &      0.05             & \multicolumn{1}{c}{0.005} &  0.002  
				\\ 	
				\noalign{\vskip -3mm} 
				
				\multicolumn{1}{c}{\textbf{Physics-Based}}          &  & \multicolumn{1}{c}{} &  & \multicolumn{1}{c}{}           &  & \multicolumn{1}{c}{} & 
				\\   
				\noalign{\vskip -3mm} 
				\multicolumn{1}{c}{}                 & \textbf{Truncated Gauss-Newton} & \multicolumn{1}{c}{0.00047} & 0.00015 & \multicolumn{1}{c}{0.062}                  &           0.042        & \multicolumn{1}{c}{0.0002} &  0.00007  
				
				\\ \hline \hline
			\end{tabular}
		\end{table}

		%%%%%%%%%%%%%%%%%%%%%%%%%%%%%%%%%%%%%%%%%%%%%%%%%%%%%%%%%%%%%%%%%%%%%%%%%%%%%%%%%%%%%%%%%%%%%%
		%%%%%%%%%%%%%%%%%%%%%%%%%%%%%%%%%%%%%%%%%%%%%%%%%%%%%%%%%%%%%%%%%%%%%%%%%%%%%%%%%%%%%%%%%%%%%%

		\begin{table}[H]
			\caption{Errors presented in the third method}  
			\label{table2}
			\centering
			\setlength{\tabcolsep}{4.0pt} % Default value: 6pt
			\renewcommand{\arraystretch}{1.6} % Default value: 1
			
			\begin{tabular}{ccccccccccc}
				\hline
				\hline
				\multirow{3}{*}{} & \multicolumn{4}{c}{$\mathbf{Loss_1} $}                                & \multicolumn{4}{c}{$\mathbf{Loss_2} $}                                                    & \multicolumn{2}{c}{\multirow{2}{*}{$\boldsymbol{ \dfrac{ ||\rho - \overline{\rho }||_{L^2(\Omega)}}{|| \rho ||_{L^2(\Omega)}}}$}}    \\ \cmidrule(lr){2-5} 
				\cmidrule(lr){6-9}

				& \multicolumn{2}{c}{$\mathbf{Reconstruction \,\, Loss}$}                         & \multicolumn{2}{c}{$\mathbf{\mathcal{D}}_{KL_1}$}    & \multicolumn{2}{c}{$\mathbf{Reconstruction \,\, Loss}$}                         & \multicolumn{2}{c}{$\mathbf{\mathcal{D}}_{KL_2}$}    & \multicolumn{2}{c}{}    \\ \cmidrule(lr){2-3} \cmidrule(lr){4-5}
				\cmidrule(lr){6-7} \cmidrule(lr){8-9}
				\cmidrule(lr){10-11}

				& \multicolumn{1}{c}{Training} & \multicolumn{1}{c}{Test} & \multicolumn{1}{c}{Training} & Test & \multicolumn{1}{c}{Training} & \multicolumn{1}{c}{Test} & \multicolumn{1}{c}{Training} & Test & \multicolumn{1}{c}{Training} &  Test
				\\

				\multicolumn{1}{l}{\textbf{New-VAE}}	& \multicolumn{1}{c}{55} & \multicolumn{1}{c}{90} & \multicolumn{1}{c}{81} & 81 & \multicolumn{1}{c}{50} & \multicolumn{1}{c}{60} & \multicolumn{1}{c}{83} & 83  & \multicolumn{1}{c}{0.026} & 0.030  \\ \hline \hline

			\end{tabular}
		\end{table}

		%%%%%%%%%%%%%%%%%%%%%%%%%%%%%%%%%%%%%%%%%%%%%%%%%%%%%%%%%%%%%%%%%%%%%%%%%%%%%%%%%%%%%%%%%%%%%%
		%%%%%%%%%%%%%%%%%%%%%%%%%%%%%%%%%%%%%%%%%%%%%%%%%%%%%%%%%%%%%%%%%%%%%%%%%%%%%%%%%%%%%%%%%%%%%%

		In the New-VAE row, in Tables {\ref{table1}} and {\ref{table2}}, the term $\dfrac{||\rho - \overline{\rho}||_{L^2(\Omega)}}{||\rho||_{L^2(\Omega)}}$  represents the relative error between the true density $\rho$ and mean of this probabilistic method which we denoted by $\overline{\rho}$. Notably, this mean is derived from samples generated from position 4 of Figure {\ref{VAE_New}}, using the Linear-to-Nonlinear results at position 2 of this Figure. Moreover, the values of $\alpha_1$ and $\alpha_2$ used to train the third method in this study are  $1$ and $0.2$, respectively.

		In this paper, we have tried to use the most optimal and accurate network structure to train all the networks by implementing different experiments. In training all the networks, we utilized batch sizes of 25, the Leaky ReLU activation function, and the Adam optimizer. Furthermore, each network has specific epochs and decay learning rates. The exact hyperparameters can be found in the provided code.

		You can visually compare the methods in Figures {\ref{Compare}} and {\ref{Physics}}, which display the results on the test dataset. Figure {\ref{Compare}} presents a comparison between the ground truth and data-driven methods, while Figure {\ref{Physics}}, compares the ground truth with physics-based methods, with two different initial guesses.
		To evaluate the third method, the probabilistic technique New-VAE, Figure {\ref{Probablistic}} is provided, showing the mean and pointwise standard deviation of the VAE method alongside the ground truth and the Linear-to-Nonlinear method. The Linear-to-Nonlinear method is included because its results is used in the optimization of the New-VAE method (step 2 in Figure {\ref{VAE_New}}) and it achieved the best performance among the methods presented.
		The mean and pointwise standard deviation is calculated using the Monte Carlo method, with 5,000 samples generated in step 4 of Figure  {\ref{VAE_New}}.
		Also, Figure {\ref{Histogram}} displays pointwise normalized histograms at four locations, along with the true values at those points for several density fields.
		Additionally, Figure \ref{Generator} illustrates the samples produced by the generator mentioned in the third method.

		%%%%%%%%%%%%%%%%%%%%%%%%%%%%%%%%%%%%%%%%%%%%%%%%% Compare
		%%%%%%%%%%%%%%%%%%%%%%%%%%%%%%%%%%%%%%%%%%%%%%%%%
		\begin{figure}[H]
			\centering
			\includegraphics[width=\linewidth]{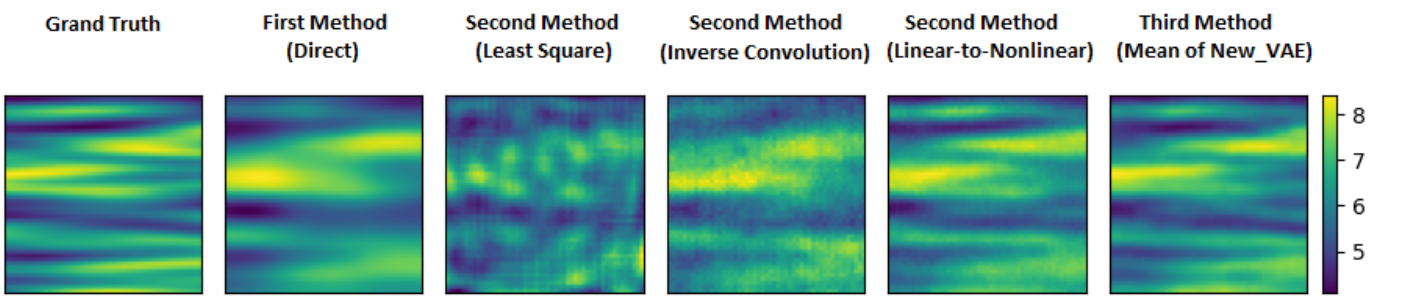}
		\end{figure}
		\vspace{-0.75 cm}
		\begin{figure}[H]
			\centering
			\includegraphics[width=\linewidth]{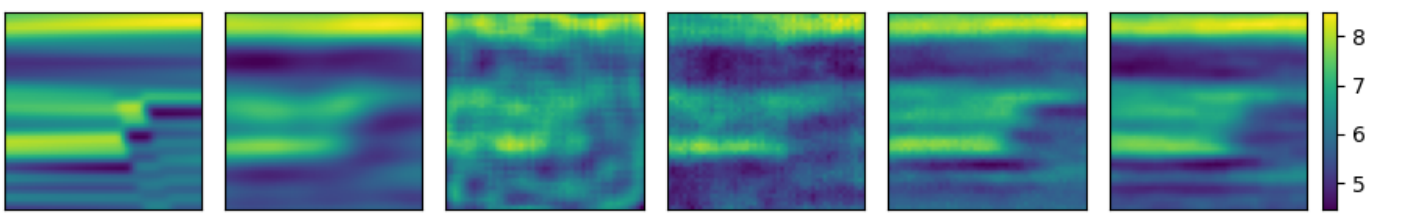}
		\end{figure}
		\vspace{-0.7 cm}	
		\begin{figure}[H]
			\centering
			\includegraphics[width=\linewidth]{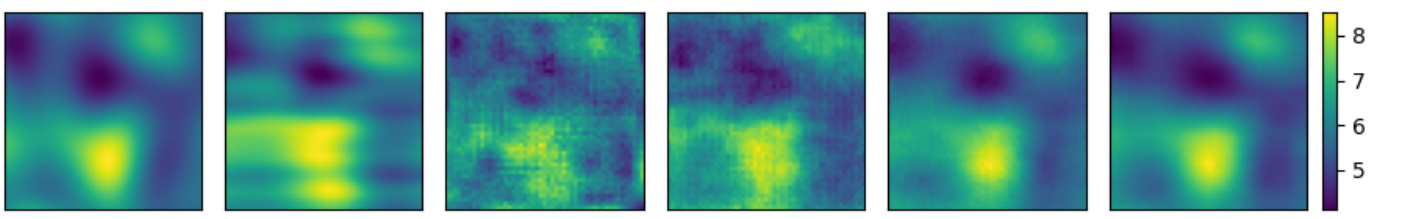}
		\end{figure}
		\vspace{-0.7 cm}
		\begin{figure}[H]
			\centering
			\includegraphics[width=\linewidth]{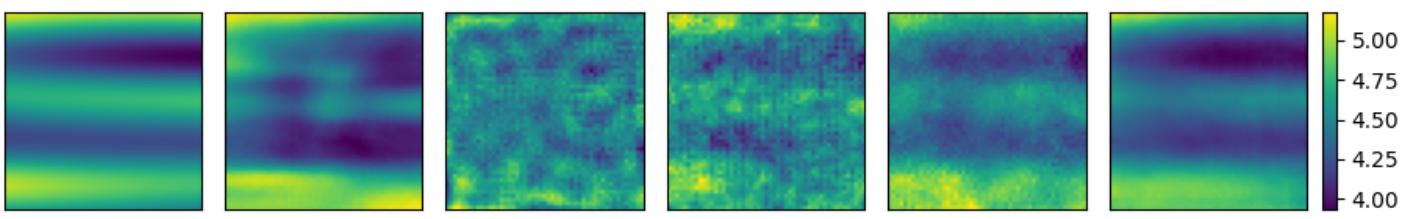}
		\end{figure}
		\vspace{-0.7 cm}		
		\begin{figure}[H]
			\centering
			\includegraphics[width=\linewidth]{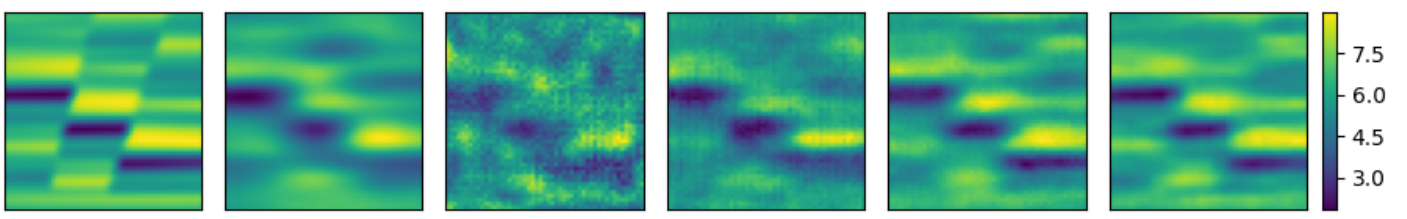}
		\end{figure}
		\vspace{-0.7 cm}
		\begin{figure}[H]
			\centering
			\includegraphics[width=\linewidth]{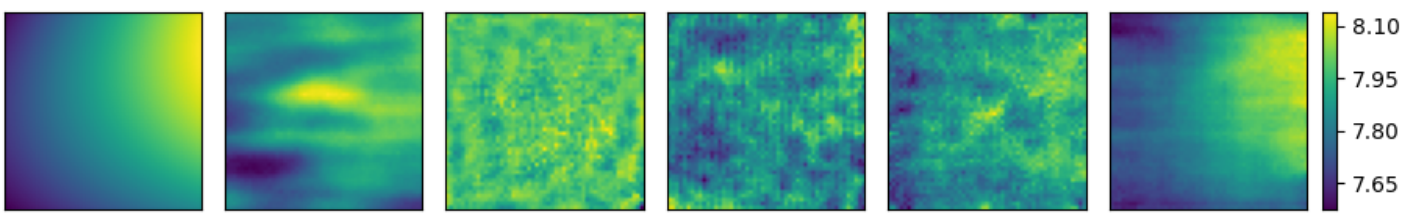}
		\end{figure}
		\vspace{-0.7 cm}
		\begin{figure}[H]
			\centering
			\includegraphics[width=\linewidth]{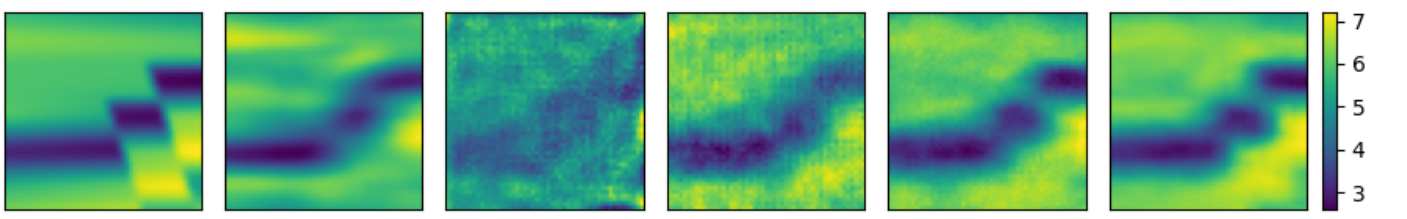}
		\end{figure}
		\vspace{-0.7 cm}
		\begin{figure}[H]
			\centering
			\includegraphics[width=\linewidth]{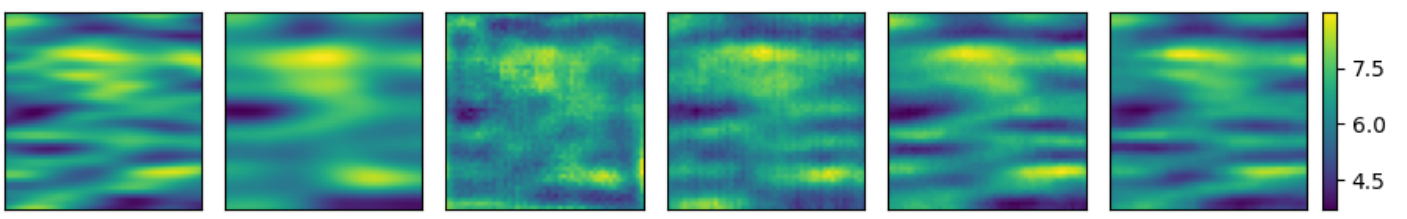}
		\end{figure}
		\vspace{-0.7 cm}
		\begin{figure}[H]
			\centering
			\includegraphics[width=\linewidth]{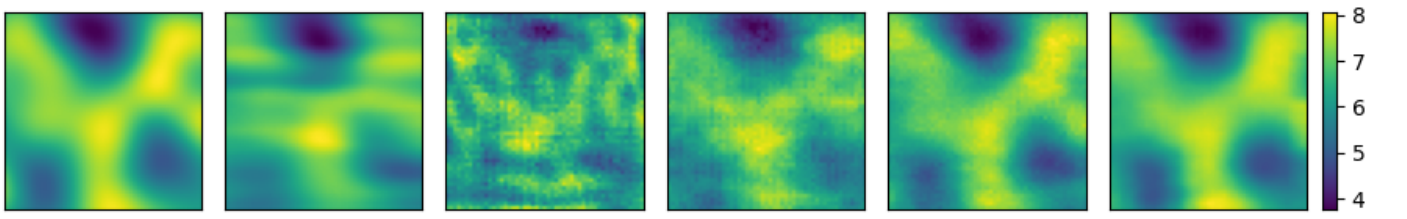}
		\end{figure}
		\vspace{-0.7 cm}
		\begin{figure}[H]
			\centering
			\includegraphics[width=\linewidth]{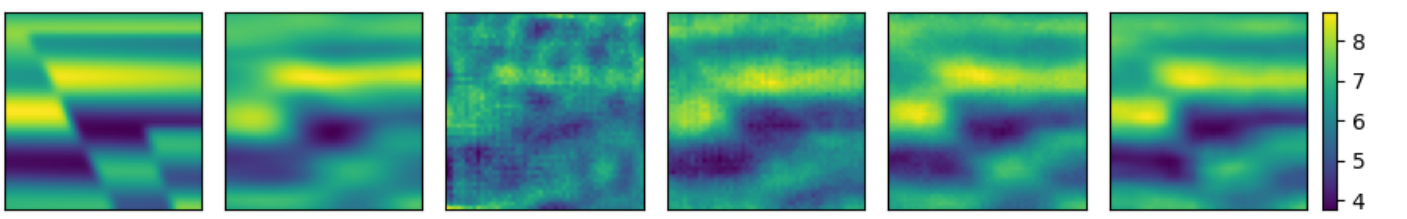}
		\end{figure}
		\vspace{-0.7 cm}
		\begin{figure}[H]
			\centering
			\includegraphics[width=\linewidth]{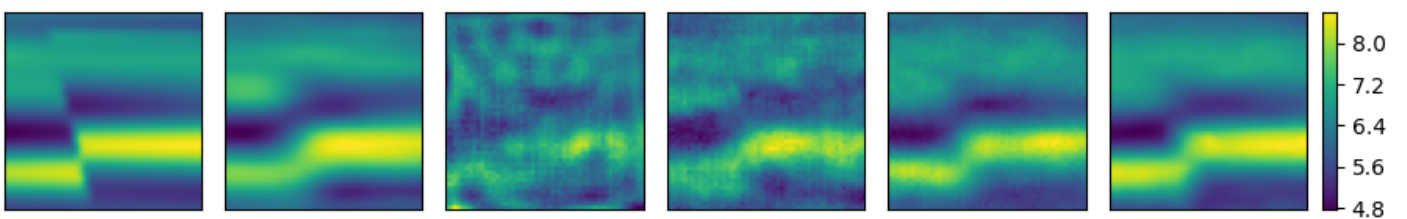}
		\end{figure}
		\vspace{-0.7 cm}
		\begin{figure}[H]
			\centering
			\includegraphics[width=\linewidth]{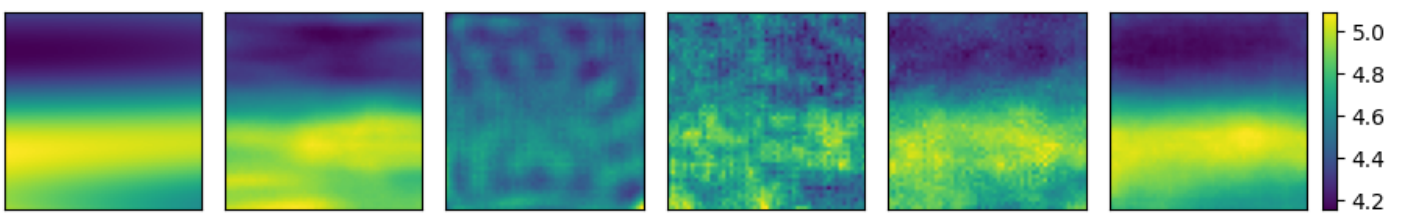}
			\caption{Comparison of data-driven methods' results with the true density} 
			\label{Compare}
		\end{figure}
		
		%%%%%%%%%%%%%%%%%%%%%%%%%%%%%%%%%%%%%%%%%%%%%%%%% Physics-Based
		%%%%%%%%%%%%%%%%%%%%%%%%%%%%%%%%%%%%%%%%%%%%%%%%%
		\newpage
		\begin{figure}[H]
			\centering
			\includegraphics[scale=0.5]{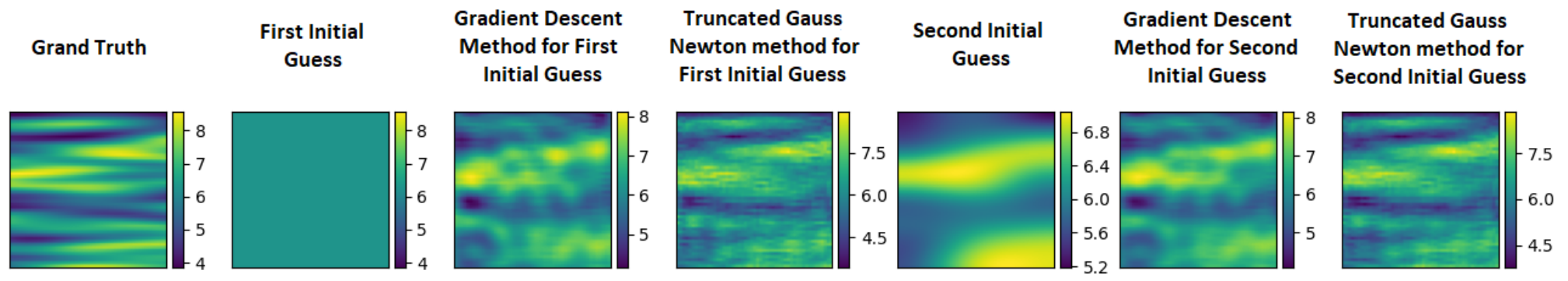}
		\end{figure}
		\vspace{-0.7 cm}
		\begin{figure}[H]
			\centering
			\includegraphics[scale=0.5]{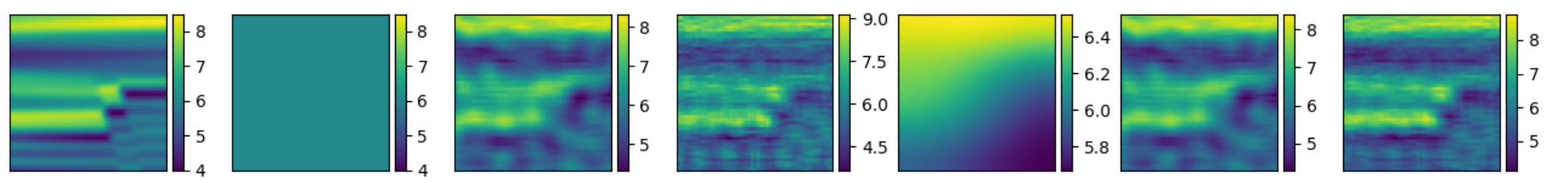}
		\end{figure}
		\vspace{-0.7 cm}	
		\begin{figure}[H]
			\centering
			\includegraphics[scale=0.5]{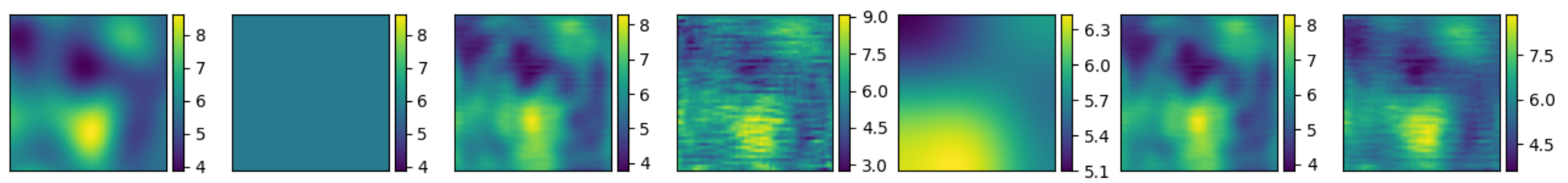}
		\end{figure}
		\vspace{-0.7 cm}
		\begin{figure}[H]
			\centering
			\includegraphics[scale=0.5]{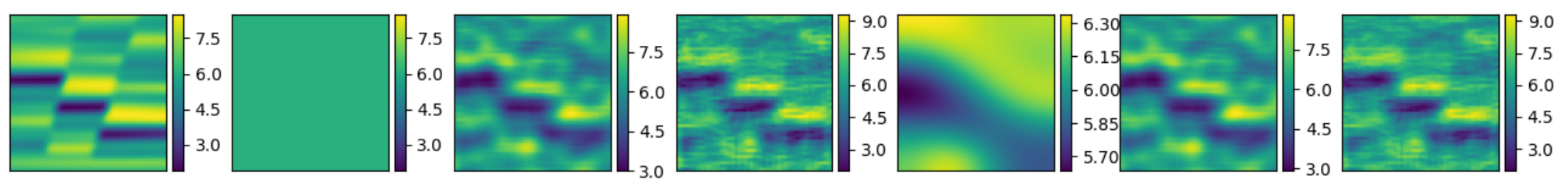}
		\end{figure}
		\vspace{-0.7 cm}		
		\begin{figure}[H]
			\centering
			\includegraphics[scale=0.5]{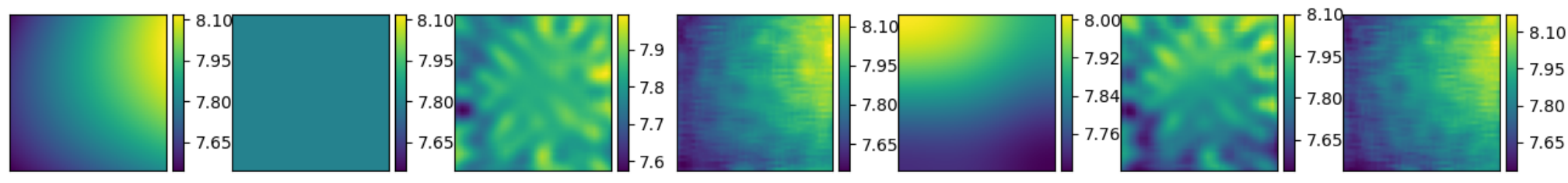}
		\end{figure}
		\vspace{-0.7 cm}
		\begin{figure}[H]
			\centering
			\includegraphics[scale=0.5]{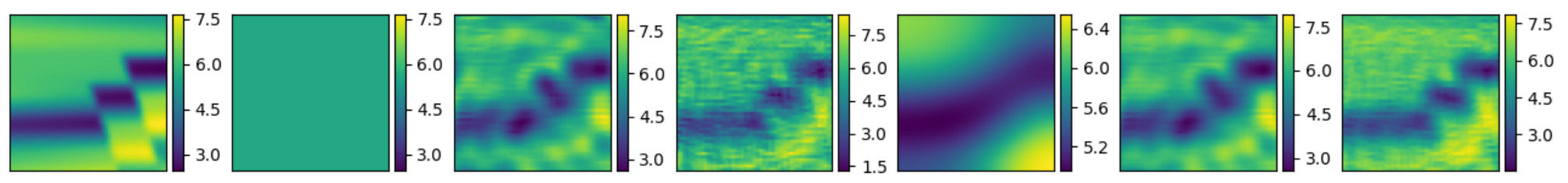}
		\end{figure}
		\vspace{-0.7 cm}
		\begin{figure}[H]
			\centering
			\includegraphics[scale=0.5]{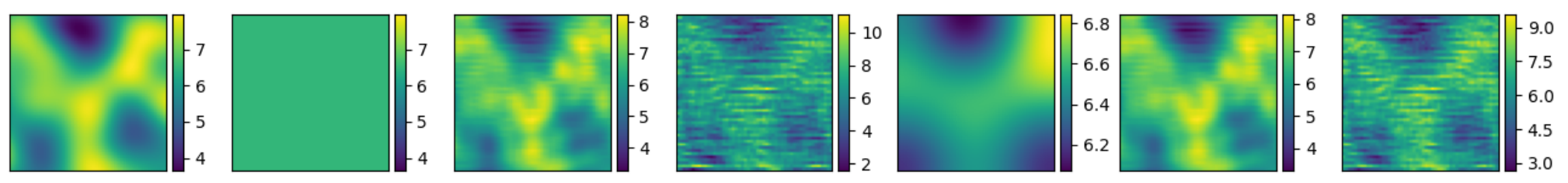}
		\end{figure}
		\vspace{-0.7 cm}
		\begin{figure}[H]
			\centering
			\includegraphics[scale=0.5]{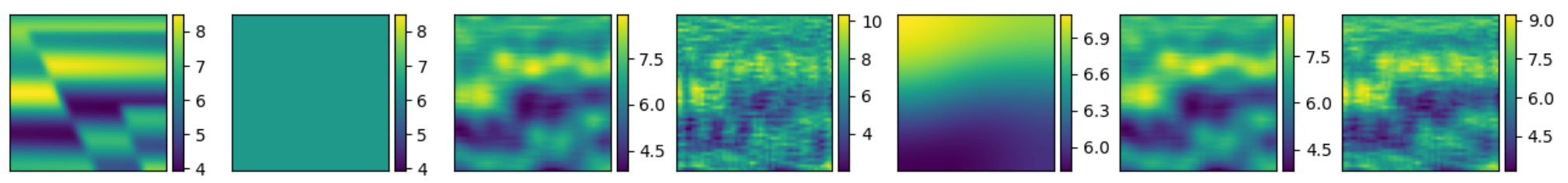}
		\end{figure}
		\vspace{-0.7 cm}
		\begin{figure}[H]
			\centering
			\includegraphics[scale=0.5]{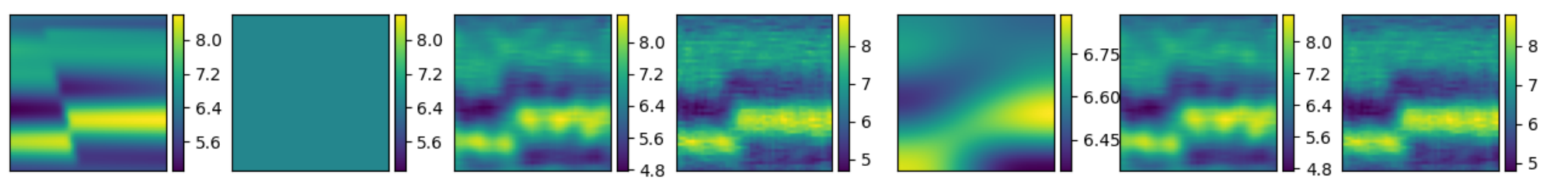}
			\caption{Comparison of physics-based methods' results, using two different initial guesses, with the true density } 
			\label{Physics}
		\end{figure}	
		
		%%%%%%%%%%%%%%%%%%%%%%%%%%%%%%%%%%%%%%%%%%%%%%%%% Probability
		%%%%%%%%%%%%%%%%%%%%%%%%%%%%%%%%%%%%%%%%%%%%%%%%%
		\newpage
		\begin{figure}[H]
			\centering
			\begin{subfigure}[b]{0.95\textwidth}
				\includegraphics[width=\linewidth]{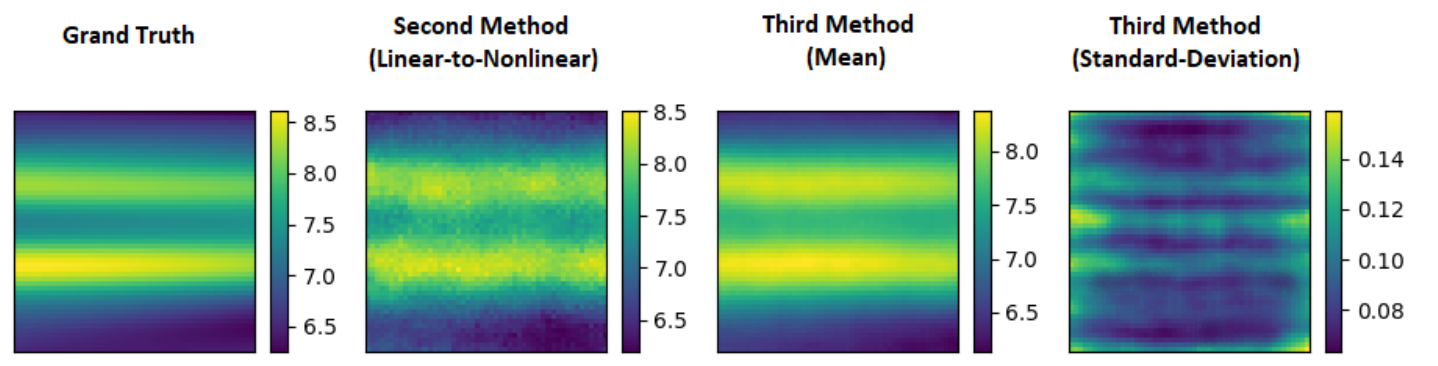}
			\end{subfigure}%
			
			\vspace{-0.0cm}
			\begin{subfigure}[b]{0.95\textwidth}
				\includegraphics[width=\linewidth]{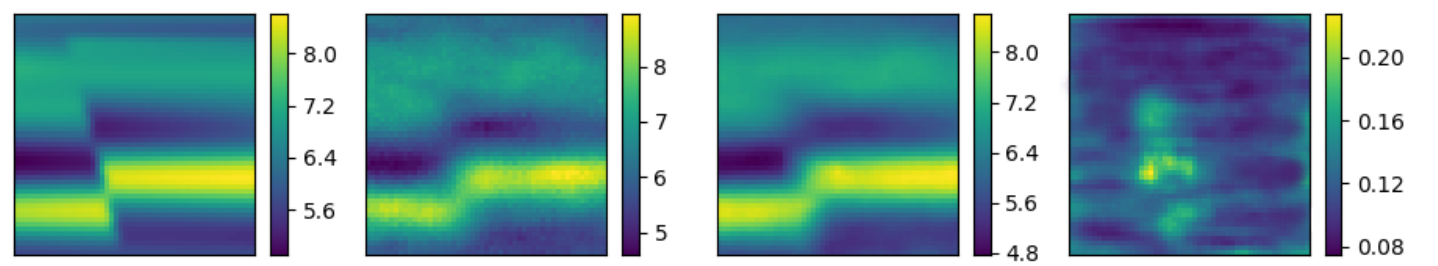}
			\end{subfigure}%
			
			\vspace{-0.0cm}
			\begin{subfigure}[b]{0.95\textwidth}
				\includegraphics[width=\linewidth]{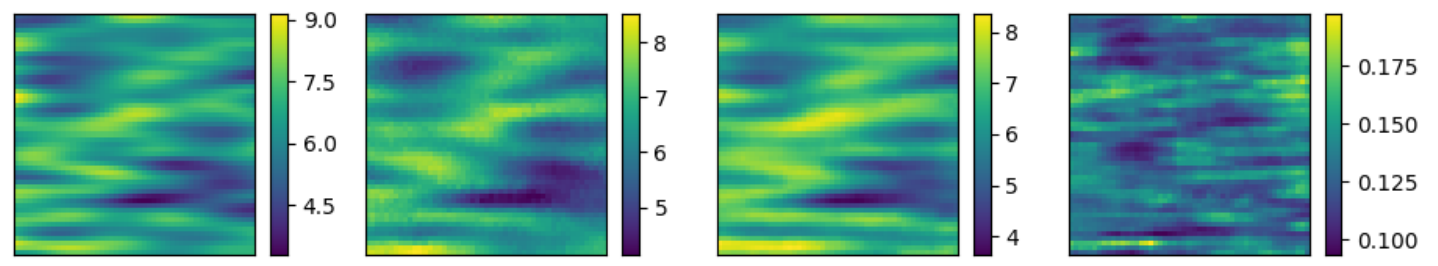}
			\end{subfigure}%
			
			\vspace{-0.0cm}
			\begin{subfigure}[b]{0.95\textwidth}
				\includegraphics[width=\linewidth]{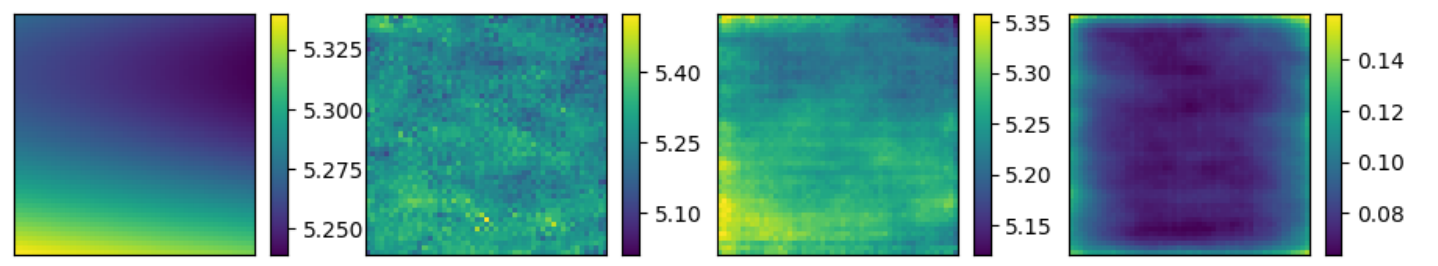}
			\end{subfigure}
			
			\vspace{-0.0cm}
			\begin{subfigure}[b]{0.95\textwidth}
				\includegraphics[width=\linewidth]{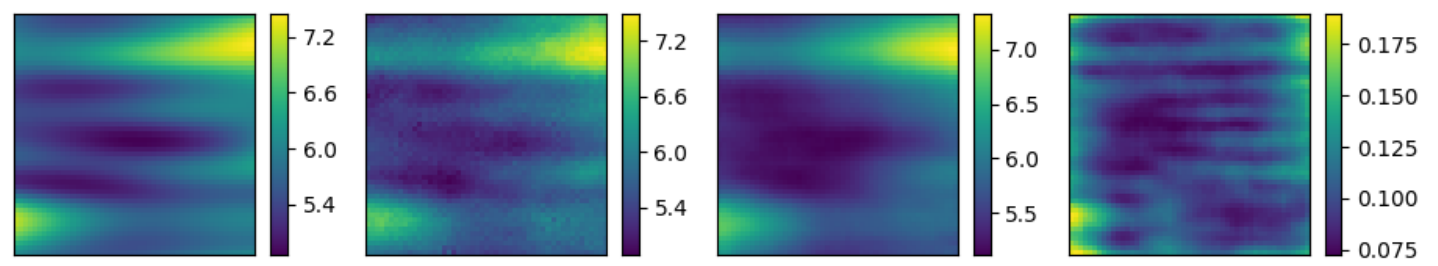}
			\end{subfigure}
			
			\vspace{-0.0cm}
			\begin{subfigure}[b]{0.95\textwidth}
				\includegraphics[width=\linewidth]{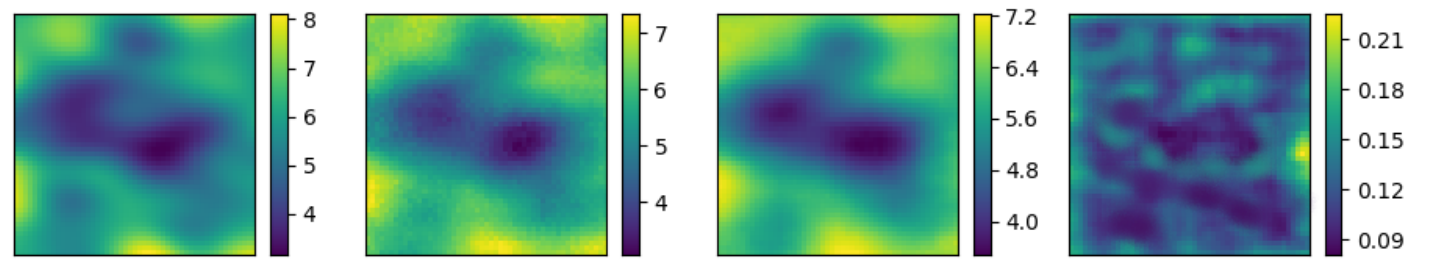}
			\end{subfigure}
			
			\vspace{-0.0cm}
			\begin{subfigure}[b]{0.95 \textwidth}
				\includegraphics[width=\linewidth]{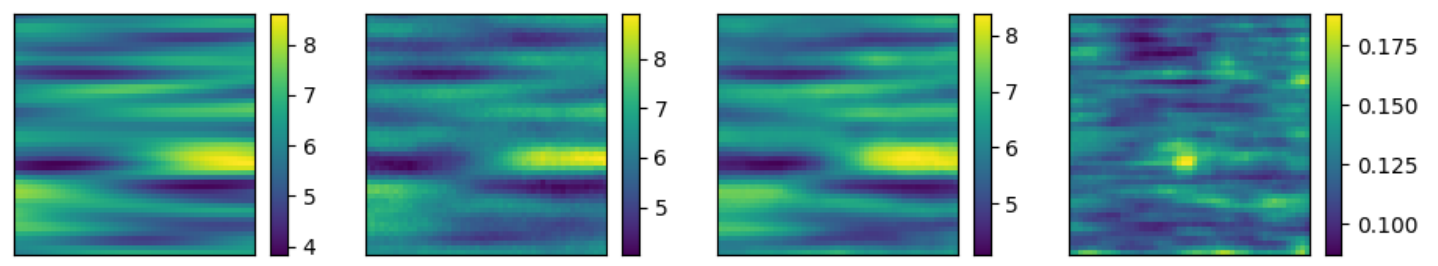}
			\end{subfigure}
			\caption{Mean and pointwise standard deviation of the third method, compared with true density and the results from the Linear-to-Nonlinear method}  
			\label{Probablistic}
		\end{figure}
		
		%%%%%%%%%%%%%%%%%%%%%%%%%%%%%%%%%%%%%%%%%%%%%%%%% Hist
		%%%%%%%%%%%%%%%%%%%%%%%%%%%%%%%%%%%%%%%%%%%%%%%%% 
		\begin{figure}[H]
			\centering
			\includegraphics[width=\linewidth]{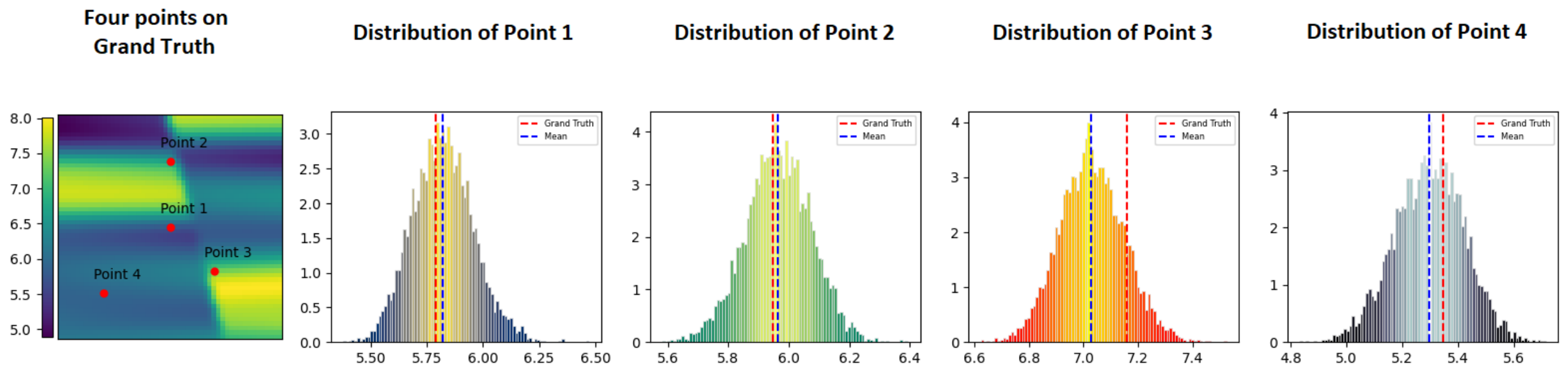}
		\end{figure}
		\vspace{-0.95 cm}
		\begin{figure}[H]
			\centering
			\includegraphics[width=\linewidth]{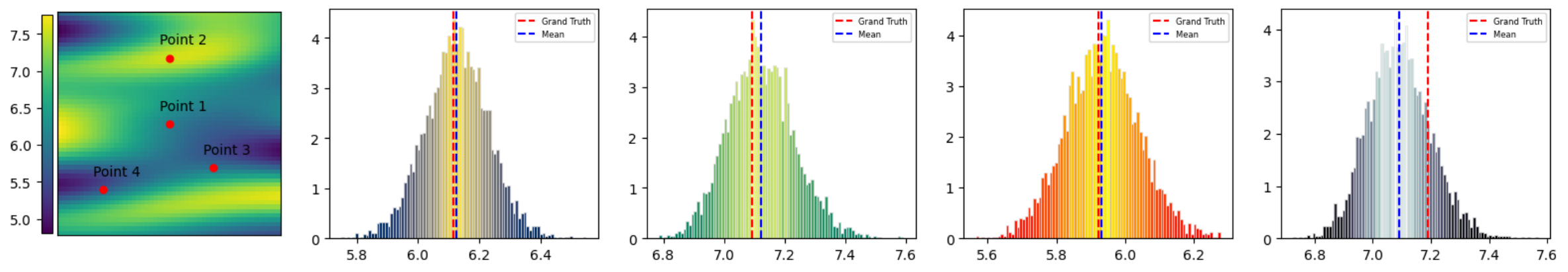}
		\end{figure}
		\vspace{-0.95 cm}	
		\begin{figure}[H]
			\centering
			\includegraphics[width=\linewidth]{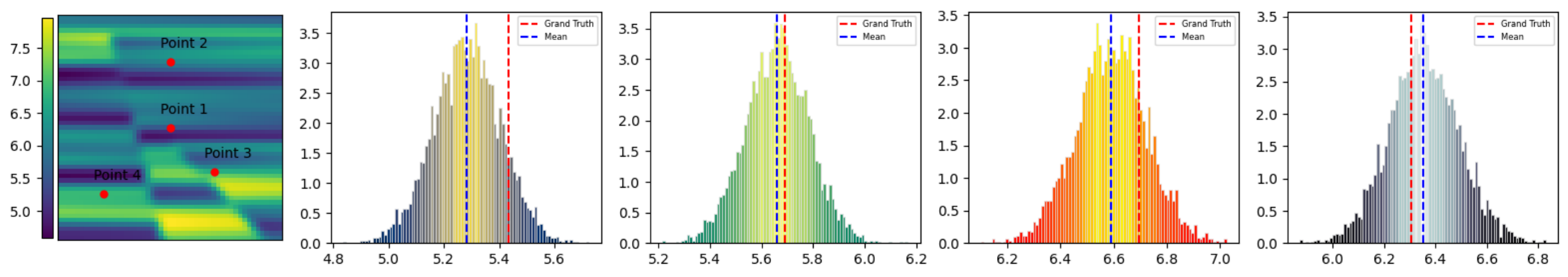}
		\end{figure}
		\vspace{-0.95 cm}
		\begin{figure}[H]
			\centering
			\includegraphics[width=\linewidth]{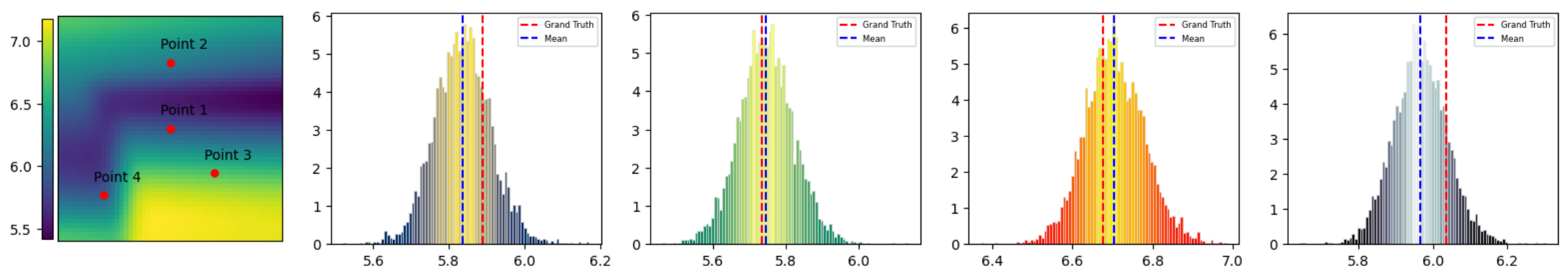}
		\end{figure}
		\vspace{-0.95 cm}		
		\begin{figure}[H]
			\centering
			\includegraphics[width=\linewidth]{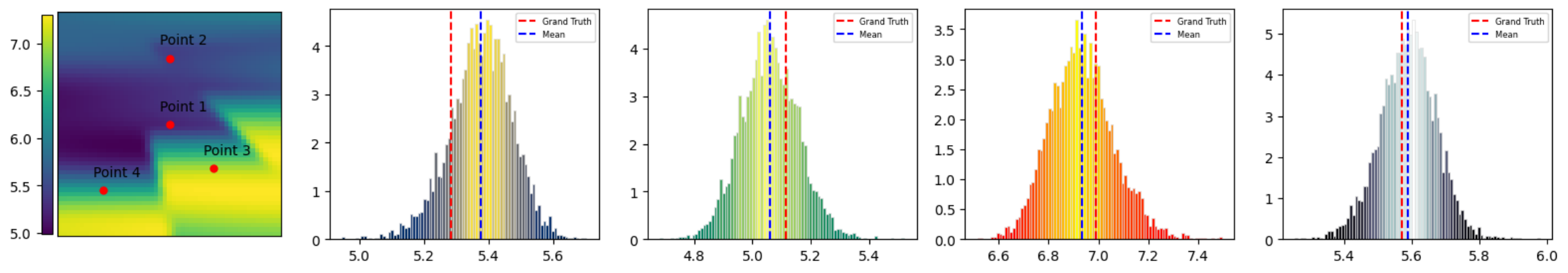}
		\end{figure}
		\vspace{-0.95 cm}
		\begin{figure}[H]
			\centering
			\includegraphics[width=\linewidth]{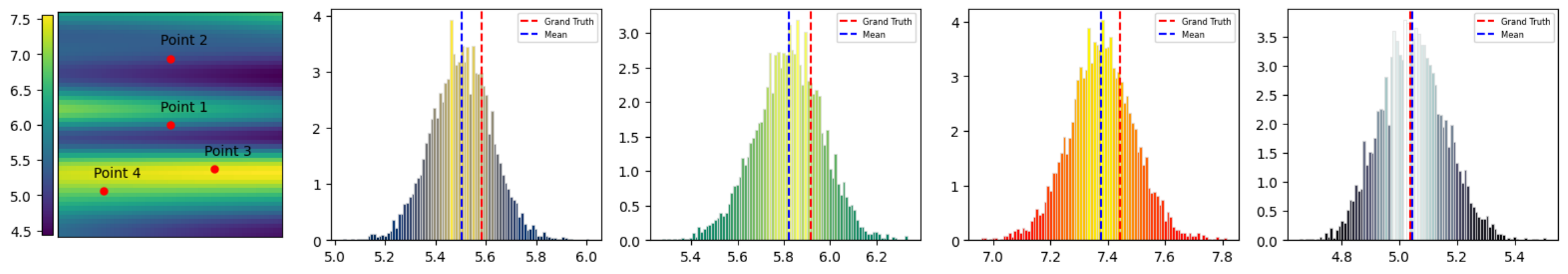}
		\end{figure}
		\vspace{-0.95 cm}
		\begin{figure}[H]
			\centering
			\includegraphics[width=\linewidth]{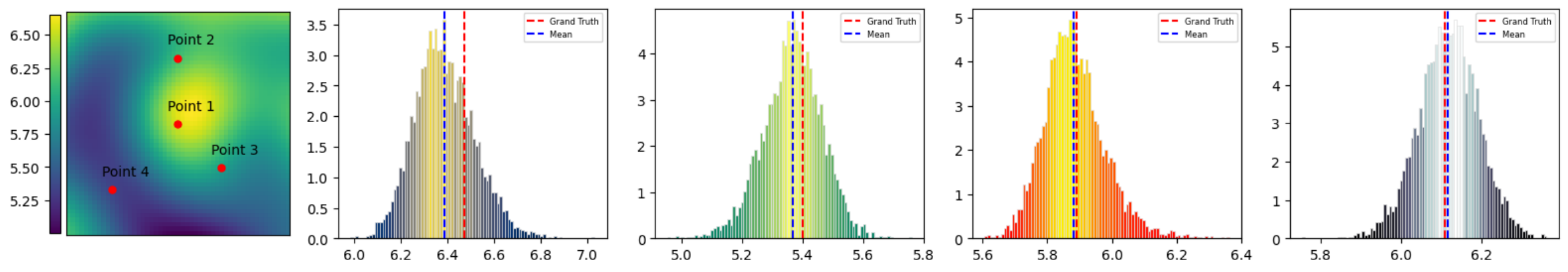}
			\caption{Pointwise normalized histogram at four locations, shown alongside the true values at these points for some density fields  }
			\label{Histogram}
		\end{figure}
		%%%%%%%%%%%%%%%%%%%%%%%%%%%%%%%%%%%%%%%%%%%%%%%%%
		%%%%%%%%%%%%%%%%%%%%%%%%%%%%%%%%%%%%%%%%%%%%%%%%%
		
		%%%%%%%%%%%%%%%%%%%%%%%%%%%%%%%%%%%%%%%%%%%%%%%%% Generator
		%%%%%%%%%%%%%%%%%%%%%%%%%%%%%%%%%%%%%%%%%%%%%%%%%
		
		\begin{figure}[H]
			\centering
			\includegraphics[scale=0.63]{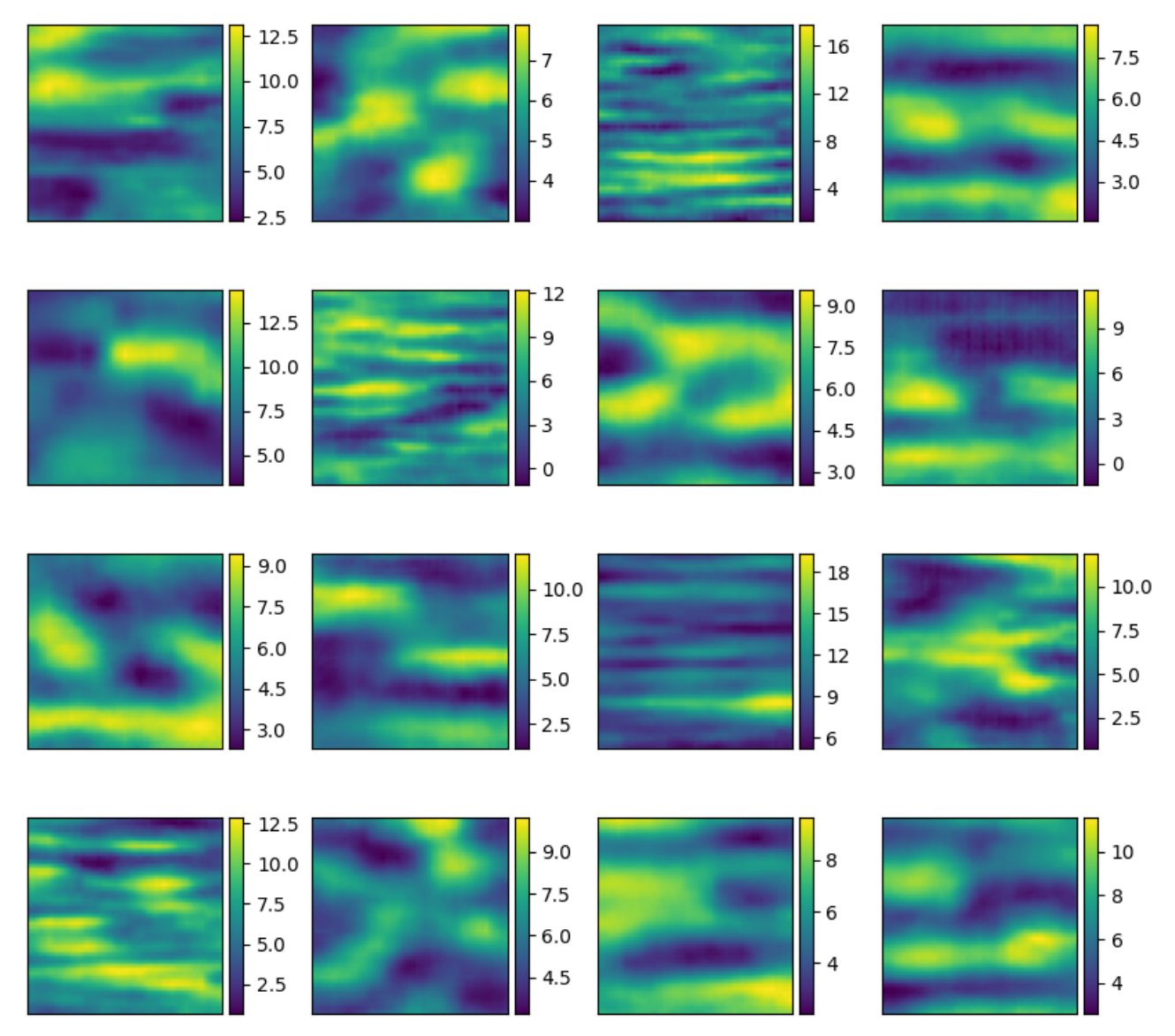}
			\caption{Generating some samples from New-VAE in the third method} 
			\label{Generator}
		\end{figure}	
		%%%%%%%%%%%%%%%%%%%%%%%%%%%%%%%%%%%%%%%%%%%%%%%%%%%%%%%%%%%%%%%%%%%%%%%%%%%%%%%%%%%%%%%%%%%%%%%%%%%%%%%%%%%%%%%%%%%%%%%%%%%%%%%%%%%%%%%%%%%%%%%%%%%%%%%%%%%%%%%%%%%%%%%%%%%%%%%%%%%%%%%%%%%%%%%%%%%%%%%%%%%%%%%%%%%%%%%%%%%%%%%%%%%%%%%%%%%%%%%%%%%%%%%%%%%%%%%%%%%%%%%%%%%%%%%%%%%%%%%%%%%%%%%%%%%%%%%%%%%%%%%%%%%%%%%%%%%%%%%%%%%%%%%%%%%%%%%%%%%%%%%%%%%%%%%%%%%%%%%%%%%%%%%%%%%%%%%%%%%%%%%%%%%%%%%%
		
		\section{Discussion}
		In this paper, we present and compare methods based on physics-based and data-driven techniques for the time-harmonic elastic FWI problem. Our study consists of several methods, including  physics-based, pure data-driven technique (deep learning based), the integrated application of machine learning and the physics underlying the problem, and a probabilistic deep learning technique that can also utilize the physics of the problem. Each of these methods possesses its own strengths and weaknesses. The aim of our comparative analysis is to offer guidance for researchers and experts in the field, including data scientists and geophysicists, helping them determine the most suitable method based on their specific requirements and limitations.
		
		Physics-based methods are comparable to data-driven techniques like deep learning in different ways. One of the main reasons that  deep learning techniques were developed was physical methods' weaknesses, such as noise sensitivity and the need for a good initial guess to start optimization. Notably, deep learning techniques typically allocate a substantial computational cost to dataset creation and training, while the inference phase incurs relatively low costs. It should be noted that since physics-based methods like FWI are performed once, their computational cost
		is much lower. However, for those applications that require inverting multiple seismic surveys at the same location, such as timelapse monitoring, our approach can be very cost-effective.
		
		A crucial phase  of the training process is having a comprehensive and suitable dataset for that task. One of the significant achievements of this paper is to create a dataset encompassing various geological patterns to assess the networks' ability to recover these diverse patterns. From the visual and numerical results, we observe that their performance is excellent and deemed acceptable. 
		
		In the following, we compare the methods presented in this paper on a case-by-case basis, according to results obtained from Tables {\ref{table1}}, {\ref{table3}}, {\ref{table2}}, and Figures  {\ref{Compare}}, {\ref{Physics}}, {\ref{Probablistic}} and {\ref{Histogram}}.
		
		\begin{enumerate}
			\item 
			The first method, \textbf{ Direct Deep Learning Inversion},  a traditional technique, employs a neural network to take surface seismic data into subsurface density. This process  exhibits relatively good numerical and visual performance, with a lower computational cost compared to other methods discussed in this paper. It is important to note that this technique does not leverage the physics of the problem; instead, it is presented for comparative purposes with other methods.
			
			\item
			In the second method, using machine learning and the physics of the problem (Lippman integral in the theory of elastic scattering), three techniques are presented to solve  elastic FWI. These techniques offer advantages over PINN as they do not require auto differentiation, which itself causes errors. Additionally, since PINNs are meshless, they are incapable of finding details and have excessive smoothness. However, they are more computationally optimal than all the methods presented in this paper and doesnt need to dataset.

			\begin{itemize}
				\item 
				In the  \textbf{Linear-to-Nonlinear} technique within this method, the simultaneous utilization of physics and machine learning demonstrates its superiority in the visual aspect. Numerically, this technique is more accurate than other deterministic methods presented in this paper.
				
				\item
				In the \textbf{Least Squares} technique within this method, accuracy is the lowest among all methods. However, as the number of incident waves and the accuracy of the estimation of unknown parameters increase, its better physical compatibility allows it to be potentially replaceable with other methods. A notable challenge with this technique lies in its relatively high computational cost and the requirement for significant memory.
				
				\item
				In the  \textbf{Inverse Convolution} technique within this method, which has relatively low accuracy, we attempt to estimate the inversion of the linear system in the form of convolution. However, due to the high ill-conditioning of the problem, the minor noise in the data (slight noise in estimating the displacement field over the area) destroys the accuracy of the solution. This technique, while not suitable for highly ill-conditioned inverse problems, can find application in other types of inverse problems.
			\end{itemize}
			
			\item
			In the \textbf{New-VAE} technique, which is a probabilistic method, we configure its deterministic component as a transcription of the Linear-to-Nonlinear method. By visually comparing the mean of this method with that of other approaches, it reveals that this method demonstrates resistance to noise. Moreover, it offers the advantage of performing uncertainty quantification analysis and proposing alternative solutions that may be more suitable than the primary one.
			This capability allows experts to compare these solutions with those obtained from alternative methods and make decisions with more certainty. Additionally, this method can  be used in Dataset Augmentation, as  described in  \ref{new-vae}. 
			
			\item 
			In the fourth method, \textbf{Physics-Based} approach, which is included for comparison with data-driven methods, we employ two techniques: \textbf{Gradient Descent}  and \textbf{Truncated Gauss Newton} method. For each technique, we use two different initial guesses, as shown in Table {\ref{table3}} and Figure {\ref{Physics}}. These results indicate that when the initial guess is close to the ground truth, the accuracy of these methods, particularly the Gauss method, improves significantly.
			This indicates that a suitable initial guess is essential for achieving high accuracy, which represents a limitation of this approach since a suitable initial guess may not always be available. Although the accuracy of this physics-based method is lower than that of methods like Linear-to-Nonlinear, it is expected that with better initial guesses, its accuracy can improve.
		\end{enumerate}

		Extending the presented methods from time-harmonic to the general time-domain state can be a potential way for future exploration. 
		Moreover, obtaining additional parameters in elastic FWI, such as P-wave and S-wave velocities, requires the use of a  Green's function for scenarios involving non-constant Lame parameters. This Green's function may be derived analytically or approximated numerically, offering a promising direction for future research.
		Furthermore, the presented methods in this paper can be compared with other physics-based \cite{metivier} and data-driven techniques \cite{review}, and can also be merged with them.

		\section*{Code availability}
		The paper's codes can be found on the GitHub repository. You can access them via the link: \url{https://github.com/Vahid-Negahdari/Elastic-Full-Waveforrm-Inversion}.
		
		\section*{Data availability}
		The dataset generated in this paper is openly available in Mendeley Data at \cite{Data}. %\href{https://data.mendeley.com/datasets/z2n2f23pxw/1}{Negahdari, Vahid (2024), “Elastic Full-Waveform-Inversion”, Mendeley Data, V1, doi: 10.17632/z2n2f23pxw.1}.

		\section*{Declaration of competing interest}
		The authors declare that they have no known competing financial interests or personal relationships that could have appeared to influence the work reported in this paper.

		\section*{Acknowledgments}
		The authors would like to thank the Sharif University of Technology for supporting this paper. Additionally, special thanks to Shirin Samadi Bahrami for her valuable efforts in editing and proofreading the text of this paper.

		\bibliography{mybibfile}
		
	\end{document}